\documentclass[aps,preprint,floats,epsf,epsfig,nofootinbib,letter]{revtex4}

\usepackage{epsfig}
\usepackage{graphicx}
\usepackage{dcolumn}
\usepackage{bm}
\usepackage{subfigure}

\def\DESepsf(#1 width #2){\epsfxsize=#2 \epsfbox{#1}}

\newcommand{\be}{\begin{eqnarray}}
\newcommand{\en}{\end{eqnarray}}
\newcommand{\ov}{\overline}
\newcommand{\A}{{\cal A}}
\newcommand{\B}{{\cal B}}
\newcommand{\Sc}{{\cal S}}
\newcommand{\T}{{\cal T}}
\newcommand{\U}{{\cal U}}
\newcommand{\V}{{\cal V}}
\newcommand{\la}{{\langle}}
\newcommand{\ra}{{\rangle}}
\newcommand{\non}{{\nonumber}}
\newcommand{\bold}{{\bf}}

\begin{document}

\title{Revisiting Final State Interaction in Charmless $B_q\to P P$ Decays}
%

\author{Chun-Khiang Chua}
\affiliation{ 
Department of Physics and Chung Yuan Center for High Energy Physics,
Chung Yuan Christian University,
Taoyuan, Taiwan 32023, Republic of China}

\date{\today}

\begin{abstract}

Various new measurements in charmless $B_{u,d,s}\to PP$ modes, where $P$ is a low lying pseudoscalar meson, are reported by Belle and LHCb. These include the rates of $B^0\to\pi^0\pi^0$, $\eta\pi^0$, $B_s\to\eta'\eta'$,  $B^0\to K^+K^-$ and $B^0_s\to\pi^+\pi^-$ decays. Some of these modes are highly suppressed and are among the rarest $B$ decays. Direct CP asymmetries on various modes are constantly updated. It is well known that direct CP asymmetries and rates of suppressed modes are sensitive to final state interaction (FSI). As new measurements are reported and more data will be collected, it is interesting and timely to revisit the rescattering effects in $B_{u,d,s}\to PP$ states. We perform a $\chi^2$ analysis with all available data on CP-averaged rates and CP asymmetries in $\overline B{}_{u,d,s}\to PP$ decays. Our numerical results are compared to data and those from factorization approach. The quality of the fit is improved significantly from the factorization results in the presence of rescattering. The relations on topological amplitudes and rescattering are explored and they help to provide a better understanding of the effects of FSI. As suggested by U(3) symmetry on topological amplitudes and FSI, a vanishing exchange rescattering scenario is considered. The exchange, annihilation, $u$-penguin, $u$-penguin annihilation and some electroweak penguin amplitudes are enhanced significantly via annihilation and total annihilation rescatterings. In particular, the $u$-penguin annihilation amplitude is sizably enhanced by the tree amplitude via total annihilation rescattering. These enhancements affect rates and CP asymmetries. Predictions can be checked in the near future.

 \end{abstract}


\maketitle

\section{Introduction}

In recent years there are some experimental progresses on measurements of the charmless $B_{u,d,s}\to PP$ decays.
In year 2015, Belle reported a $3.0\sigma$ significant measurement on $\ov B{}^0\to\eta\pi^0$ decay rate with
${\cal B}(B^0\to\eta\pi^0)=(4.1^{+1.7+0.5}_{-1.5-0.7})\times10^{-7}$~\cite{Pal:2015ewa}
and
${\cal B}(B_s^0\to K^0\ov K{}^0)=(19.6^{+5.8}_{-5.1}\pm 1.0\pm 2.0)\times 10^{-6}$~\cite{Pal:2015ghq} with $5.1~\sigma$ significance,
while LHCb observed $B_s\to\eta'\eta'$ decay at $(3.31\pm 0.64\pm 0.28\pm0.12)\times 10^{-5}$ at $6.4~\sigma$ significance~\cite{Aaij:2015qga}.
In year 2016, LHCb reported on the observation of annihilation modes with ${\cal B}(B^0\to K^+K^-)=(7.80\pm1.27\pm0.81\pm0.21)\times 10^{-8}$ and ${\cal B}(B^0_s\to\pi^+\pi^-)=(6.91\pm 0.54\pm0.63\pm0.19\pm0.40)\times 10^{-7}$~\cite{Aaij:2016elb}. 
Last year Belle reported the rate of $\overline B{}^0\to\pi^0\pi^0$ of ${\cal B}(\overline B{}^0\to\pi^0\pi^0)=(1.31\pm0.19\pm0.18)\times 10^{-6}$~\cite{Julius:2017jso}.
Some of these modes are highly suppressed and are among the rarest $B$ decays.
There were constant updates on other measurements, such as rates and  
direct CP asymmetries on $B_{(s)}\to K\pi, KK, \pi\pi$ modes~\cite{HFAG, PDG, Gershon}.

It is well known that direct CP asymmetries and rates of suppressed modes are sensitive to final state interaction (FSI)~\cite{Cheng:2004ru, Gronau}.
In a study on the effects of FSI on $B_{u,d,s}\to PP$ modes~\cite{Chua:2007cm}, 
the so called (too large) $\B(\pi^0\pi^0)/\B(\pi^+\pi^-)$ ratio and (non-vanishing) $\Delta\A\equiv\A(K^-\pi^+)-\A(K^-\pi^0)$ direct CP asymmetry puzzles in $\overline B_{u,d}$ decays can both be resolved by considering rescattering among $PP$ states.~\footnote{One is referred to \cite{Cheng:2014rfa,Kpi} for some recent analyses on these puzzles.} 
Several rates and CP asymmetries were predicted.
The newly observed $B_s^0\to K^0\ov K{}^0$ rate is consistent with the prediction.
However, there are some results that are in tension with the recent measurement. 
In particular, the predicted $B_s\to\eta'\eta'$ rate is too high compared to data. 
In fact, its central value is off by a factor of 3.
As new measurements are reported and more data will be collected in LHCb, and Belle II will be turned on in the very near future,
it is interesting and timely to revisit the subject.

It will be useful to give the physical picture.
From the time-invariant property of the Wilson operators in the weak Hamiltonian, one finds that the decay amplitude satisfies \cite{Suzuki:1999uc}~\footnote{See Appendix A for a derivation.} 
 \be
 A_i=\sum_{k=1}^N\Sc^{1/2}_{ik} A^0_k,
 \label{eq:master}
 \en
where  $A_i$ 
is a $\ov B_q\to PP$ decay amplitude with weak as well as strong phases, 
$A^0_k$ is a amplitude containing weak phase only, $i=1,\dots,n$, denotes all charmless $PP$ states and
$k=1,\dots,n,n+1,\dots,N,$ denotes {\it all} possible states that
can rescatter into the charmless $PP$ states
through the strong interacting $S$-matrix, $\Sc$.
Strong phases are encoded in the rescattering matrix. 
This is known as the Watson theorem \cite{Watson:1952ji}.
There are two points needed to be emphasised.
First, the above result is exact. Every $\ov B{}_q\to PP$ decay amplitude should satisfy it.
Second, for a typical $\overline B{}_q$
decay, since the $B$ mass is large there is a large number of kinematically allowed states involved in the above equation, i.e. $N$ in the above equation is large.
Consequently, the equation is hard to solve.

Although the largeness of the $B$ mass 
makes it difficult to solve the above equation, 
it is interesting that on the contrary it is precisely the largeness of $m_B$ that makes the problem somewhat trackable. 
According to the duality argument,
when the contributions from all hadronic
states at a large enough energy scale are summed over, one should
be able to understand the physics in terms of the quark and gluon
degrees of freedom.
Indeed, several quantum chromodynamics (QCD)-based factorization approaches, such
as pQCD~\cite{pQCD}, QCD factorization
(QCDF)~\cite{Beneke:2001ev,Beneke:2003zv} and soft collinear
effective theory (SCET)~\cite{SCET}
make use of the large $B$ mass
 and give predictions on
the facrorization amplitudes, $A^{\rm fac}$. 
In other words, using the largeness of $m_B$ comparing to $\Lambda_{QCD}$, 
the factorization approaches provide solutions to Eq.~(\ref{eq:master}), 
i.e. $ A^{\rm fac}_i=\sum_{k=1}^N\Sc^{1/2}_{ik} A^0_k$.

In the infinite $m_B$ limit, the above program may work perfectly.
However, in the physical $m_B$ case, power corrections can be important and may not be neglected.
In fact, 
the effects of power corrections are strongly hinted from some unexpected enhancements in rates of several color suppressed modes, such as $\overline B{}^0\to\pi^0\pi^0$ decay~\cite{PDG,HFAG}, and some unexpected signs of direct $CP$ asymmetries, as in the difference of direct CP asymmetries of $\ov B{}^0\to\ov K{}^-\pi^+$ and $B^-\to K^-\pi^0$ decays~\cite{Lin:2008zzaa}. 
These anomalies lead to the above mentioned $\pi\pi$ and $K\pi$ puzzles.
It is fair to say that the factorization approaches can reasonably produce rates of color allowed modes, 
but it encounters some difficulties in rates of color-suppressed states and CP asymmeties. 
It is to plausible to assume that factorization approaches do not give the full solution to Eq.~(\ref{eq:master}),
some residual rescattering or residual final state interaction is still allowed and needed in $\ov B{}_q\to PP$ decays.
Note that the group of charmless $PP$ states is unique
to $\ov B{}_q\to PP$ decays, as $P$ belongs to the same SU(3) multiplet
and
$PP$ states are well separated from all
other states, where the duality argument cannot be applied to
these limited number of states~\cite{quasielastic,quasielastic0}.
Note that residual rescattering among $PP$ modes only slightly affect the rates of color allowed modes, but it can easily change direct CP violation of most modes and the rates of color suppressed modes at the same time.
It can be a one stone two birds scenario. 
It can potentially solve two problems at the same time without affecting the successful results of factorization approach on color allowed rates. 
In fact, this approach is modest than the factorization approach
as it left some rooms for our ignorance on strong dynamics.
In the following text, unless indicated otherwise we use rescattering among $PP$ states or rescattering for short to denote this particular type of rescattering, while we assume that FSI contributions from all other states are contained in the factorization amplitudes. 

The quark diagram or the so-called topological
approach has been extensively used in mesonic modes~\cite{Zeppenfeld:1980ex,Chau:tk,Chau:1990ay,Gronau:1994rj,Gronau:1995hn,Cheng:2014rfa, Gronau}.
It will be useful and interesting to study the FSI effects on topological amplitudes.
For some early works in different approach, one is referred to ref.~\cite{Gronau}.  
The relation on topological amplitudes and rescattering will be explored 
and it can help to provide a better understanding on the effects of residual rescattering.

The layout of the present paper is as follows: In Sec. II we
give the formalism. Results and discussions are presented in
Sec. IV. Sec. V contains our conclusions. Some useful formulas and derivations are collected in
Appendices A and B.

\section{Formalism}

In this section we will give the rescattering (res.) formulas, topological amplitudes (TA) of $\ov B{}_q\to PP$ decays and the relations between res. and TA.

\subsection{Rescattering Formulas}

Most of the following formulas are from~\cite{Chua:2007cm}, but some are new.
As noted in the Introduction section, in the rescattering we have~(see Appendix A)
 \be
 A_i=\sum_{j=1}^n(\Sc_{res}^{1/2})_{ij} A^{fac}_j,
 \label{eq:master1}
 \en
where 
$i,j=1,\dots,n$ denote all charmless
$PP$ states. To apply the above formula, we need to specify the factorization
amplitudes. In this work, we use the factorization amplitudes
obtained in the QCD factorization approach~\cite{Beneke:2003zv}.

According to the quantum numbers of the final states, which can be mixed under FSI, $\overline B_q\to PP$ decays can be grouped into 4 groups.
Explicit formulas are collected in Appendix~A. 
Here we give an example for illustration. 
The $\overline B^0_d\to K^-\pi^+$ decay can rescatter with three other states, namely $\overline B^0_d\to \ov K{}^0\pi^0$, $\ov K{}^0\eta_8$ and $\ov K{}^0\eta_1$, via charge exchange, singlet exchange and annihilation rescatterings as denoted in Fig.~\ref{fig:r} (a)-(c). These states are the group-1 modes.
The relevant rescattering formula is given by
\begin{eqnarray}
\left(
\begin{array}{l}
 A_{\ov B {}^0_{d}\to K^-\pi^+}\\
 A_{\ov B {}^0_{d}\to \ov K {}^0 \pi^0}\\
 A_{\ov B {}^0_{d}\to \ov K {}^0 \eta_8}\\
 A_{\ov B {}^0_{d}\to \ov K {}^0 \eta_1}
\end{array}
\right)
 &=& \Sc_{res,1}^{1/2}
 \left(
\begin{array}{l}
 A^{fac}_{\ov B {}^0_{d}\to K^-\pi^+}\\
 A^{fac}_{\ov B {}^0_{d}\to \ov K {}^0 \pi^0}\\
 A^{fac}_{\ov B {}^0_{d}\to \ov K {}^0 \eta_8}\\
 A^{fac}_{\ov B {}^0_{d}\to \ov K {}^0 \eta_1}
\end{array}
\right),
 \label{eq:FSIB0Kpi}
\end{eqnarray}
with
${\cal S}^{1/2}_{res,1}=(1+i{\cal T}_1)^{1/2}$ and
\begin{eqnarray}
 {\cal T}_1 &=& \left(
\begin{array}{cccc}
r_0+r_a
      &\frac{-r_a+r_e}{\sqrt2}
      &\frac{-r_a+r_e}{\sqrt6}
      &\frac{2\bar r_a+\bar r_e}{\sqrt3}
      \\
\frac{-r_a+r_e}{\sqrt2}
      &r_0+\frac{r_a+r_e}{2}
      &\frac{r_a-r_e}{2\sqrt3}
      &-\frac{2\bar r_a+\bar r_e}{3\sqrt2}
      \\
\frac{-r_a+r_e}{\sqrt6}
      &\frac{r_a-r_e}{2\sqrt3}
      &r_0+\frac{r_a+5r_e}{6}
      &-\frac{2\bar r_a+\bar r_e}{3\sqrt2}
      \\
\frac{2\bar r_a+\bar r_e}{\sqrt3}
      &-\frac{2\bar r_a+\bar r_e}{\sqrt6}
      &-\frac{2\bar r_a+\bar r_e}{3\sqrt2}
      &\tilde r_0+\frac{4\tilde r_a+2\tilde r_e}{3}
\end{array}
\right).
\en
The rescattering parameters $r_{0,a,e,t}$, $\bar r_{0,a,e,t}$,
$\tilde r_{0,a,e,t}$, $\hat r_{0,a,e,t}$ and $\check r_{0,a,e,t}$
denote~\footnote{Note that $\hat r$ and $\check r$ do not appear in ${\cal T}_1$, but they will contribute to some other $PP$ modes.}
rescattering in $\Pi({\bf 8})\,\Pi({\bf 8})\to\Pi({\bf
8})\,\Pi({\bf 8})$, $\Pi({\bf 8})\,\Pi({\bf 8})\to\Pi({\bf
8})\,\eta_1$, $\Pi({\bf 8})\eta_1\to\Pi({\bf 8})\eta_1$ and
$\eta_1\eta_1\to\eta_1\eta_1$, respectively, with $\Pi(\bf 8)$ the SU(3) octet and
$\eta_1$ the singlet, and the subscripts
$0,a,e,t$ represent flavor singlet, annihilation, exchange and
total-annihilation rescatterings, respectively (see
Fig.~\ref{fig:r}). 

\begin{figure}[t]
\centering
\subfigure[]{
  \includegraphics[width=6cm]{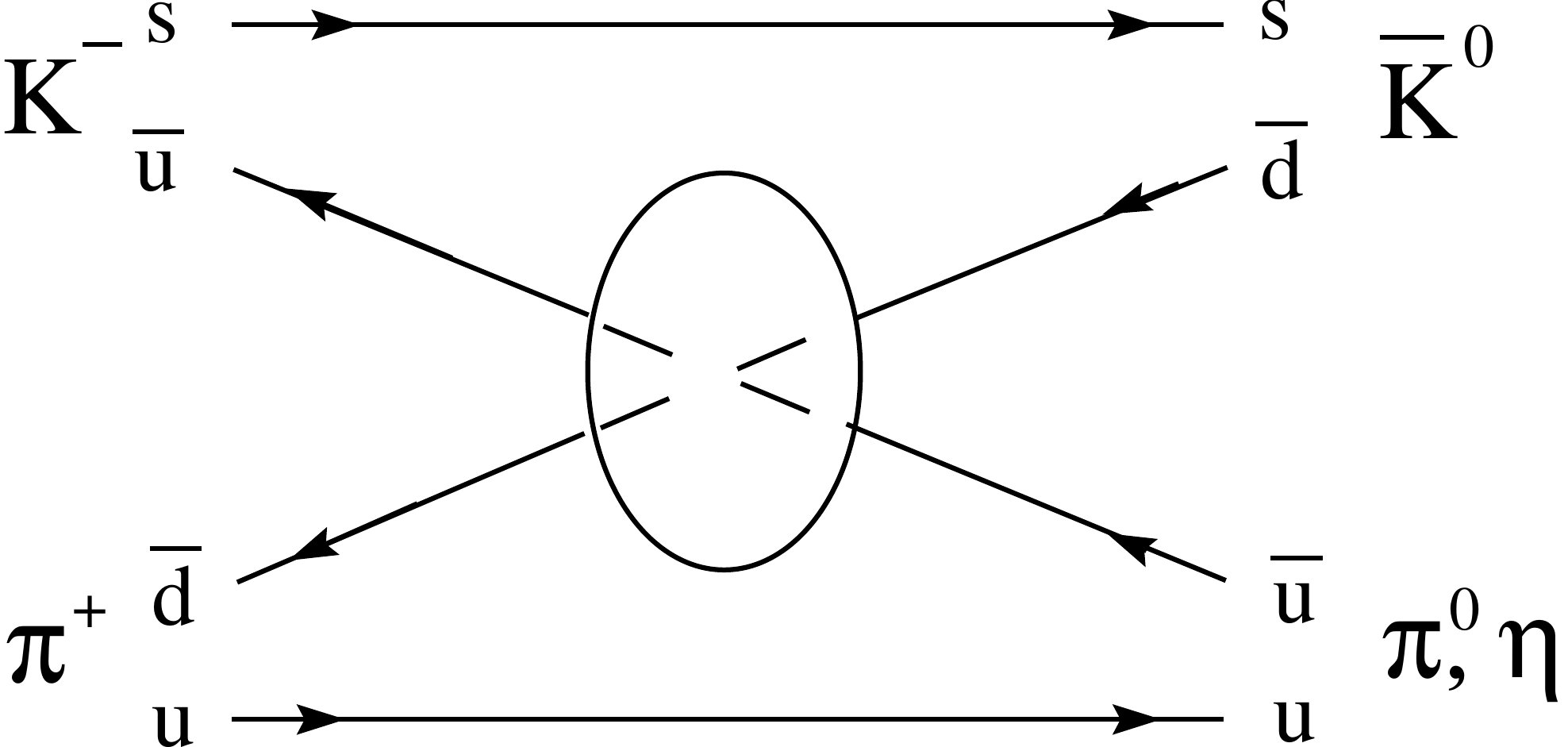}
 }
\subfigure[]{
  \includegraphics[width=6cm]{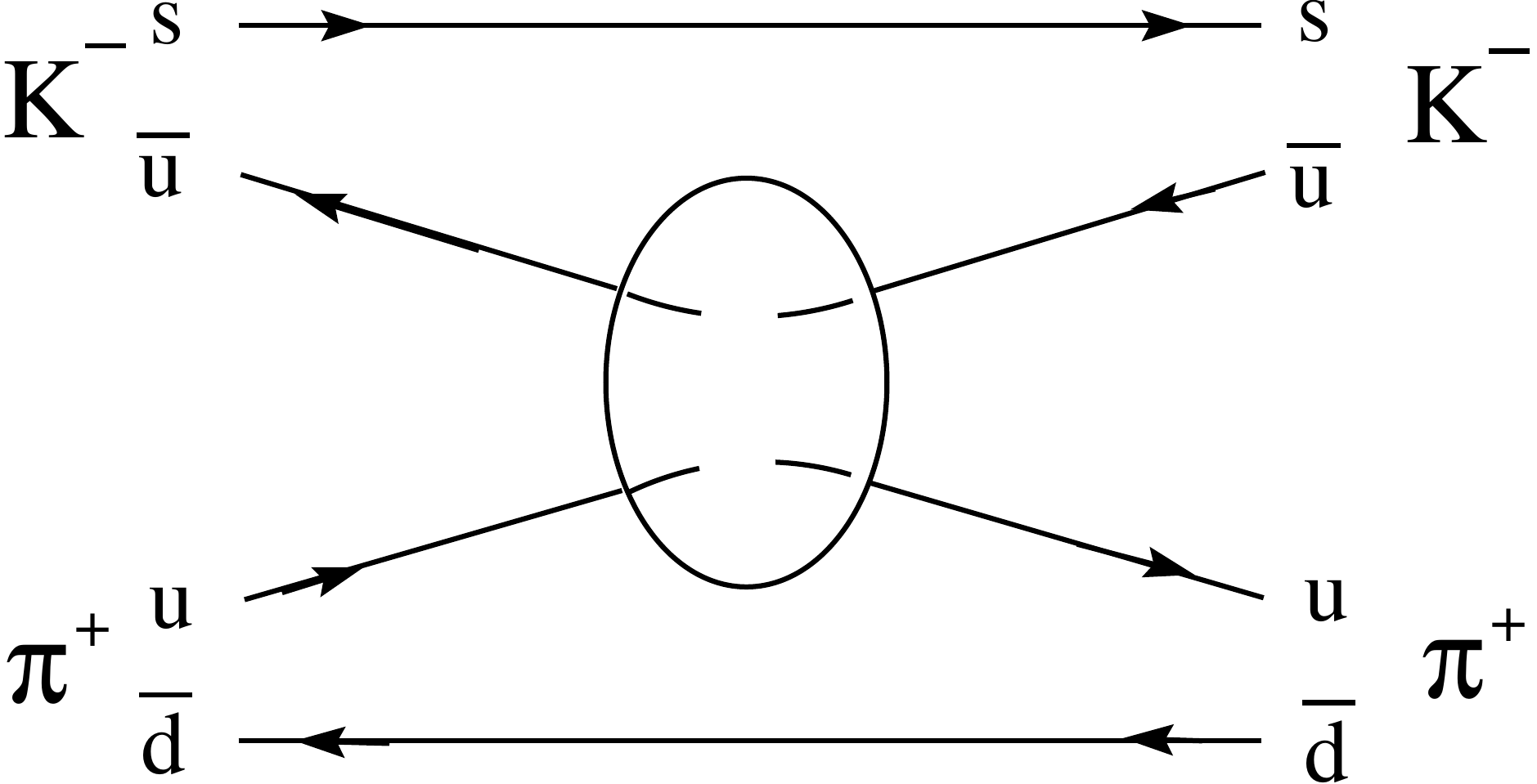}
}
\subfigure[]{
  \includegraphics[width=6cm]{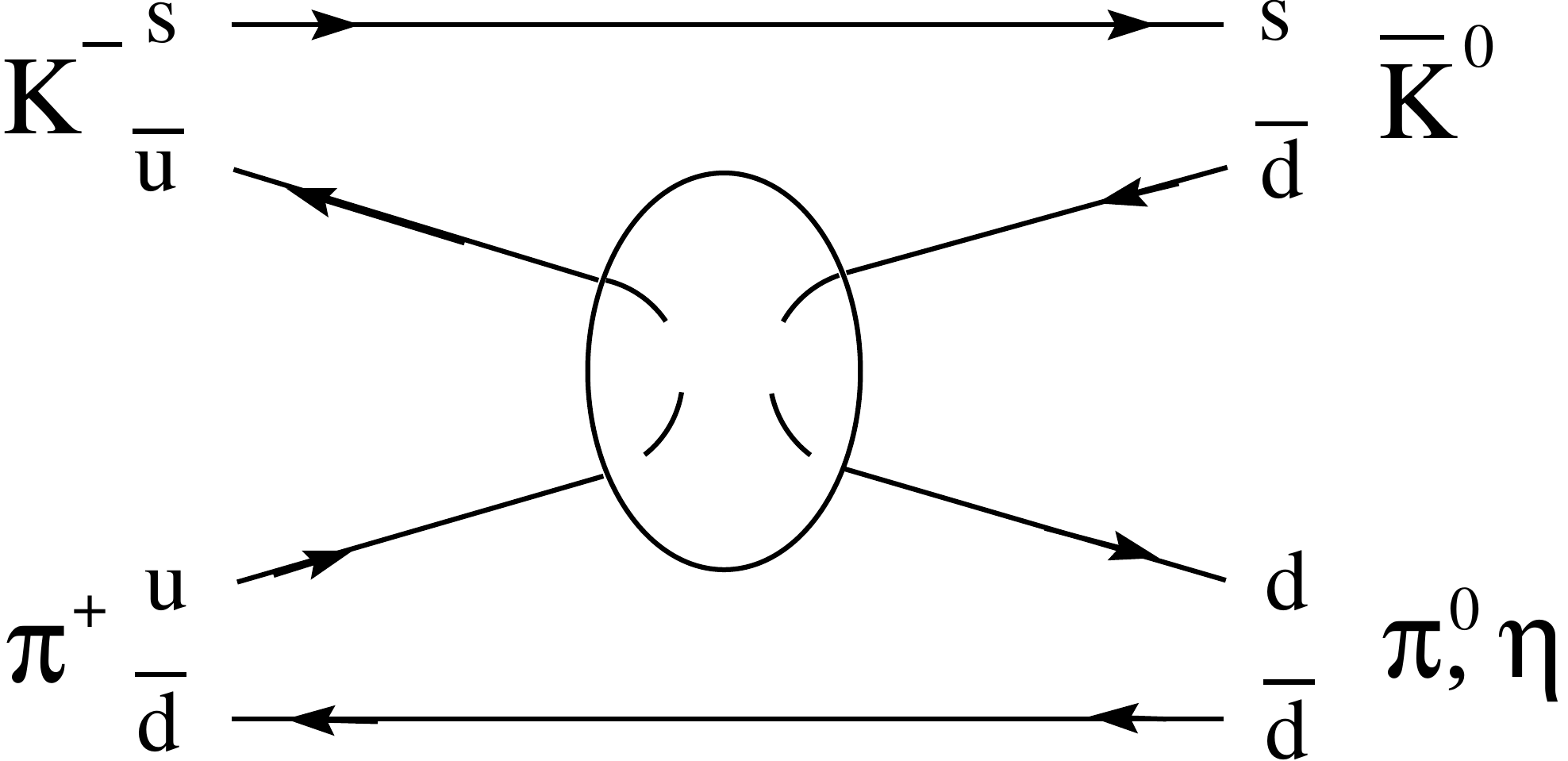}
}
\subfigure[]{
  \includegraphics[width=6cm]{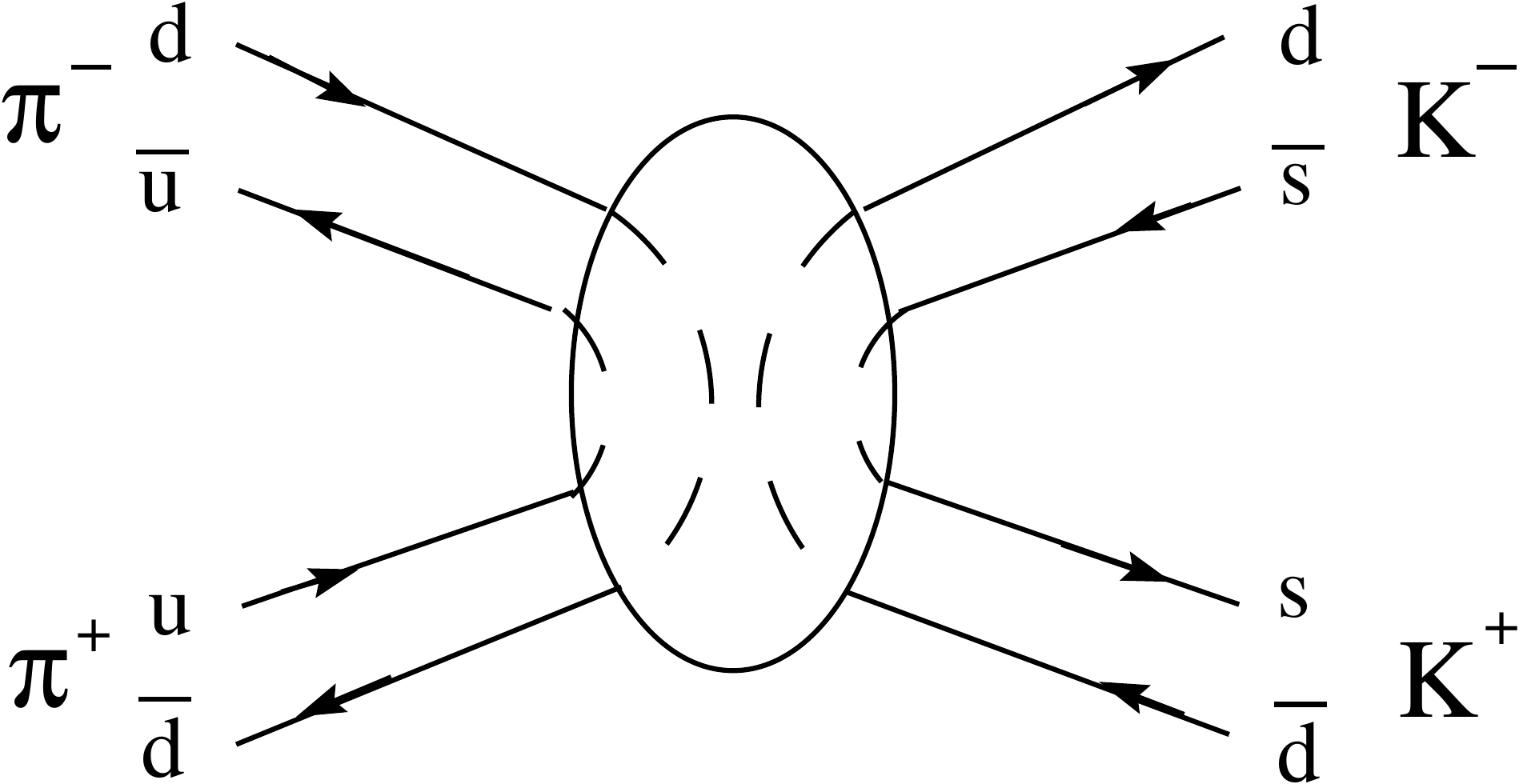}
}
\caption{Pictorial representation of
   (a) charge exchange $r_e$, (b) singlet exchange $r_0$,
   (c) annihilation $r_a$ and (d) total-annihilation $r_t$
   for $PP$ (re)scattering.} 
 \label{fig:r}
\end{figure}

Flavor symmetry requires that $(\Sc_{res})^m$ with an arbitrary
power of $m$ should also have the same form as $\Sc_{res}$. More
explicitly, from SU(3) symmetry, we should have
 \be
 (\Sc_{res})^m&=&(1+i\T)^m\equiv 1+i\T^{(m)},
 \label{eq:SSU3a}
 \en
where $\T^{(m)}$ is defined through the above equation and its
form is given by
 \be
 \T^{(m)}&=&\T\,\,\, {\rm with\,\,}(r_j,\bar r_j,\tilde r_j,\check r_j)
 \to (r^{(m)}_j,\bar r^{(m)}_j,\tilde r^{(m)}_j,\check r^{(m)}_j),
 \label{eq:SSU3b}
 \en
for $j=0,a,e,t$.

It is useful to note that we have $\bf 8\otimes 8$, $\bf 8\otimes 1$,
$\bf 1\otimes 8$ and $\bf 1\otimes 1$ SU(3) products for $P_1P_2$
final states, which has to be symmetric under the exchange of $P_1$ and
$P_2$ in the $\overline B\to PP$ decay as the meson pair is in $s$-wave configuration 
and they have to satisfy the Bose-Einstein statistics.
The allowed ones are the $\bf 27$, $\bf 8$ and the $\bf 1$
from $\bf 8\otimes 8$, the $\bf 8'$ from the symmetrized $\bf
8\otimes 1$+$\bf 1\otimes 8$, and $\bf 1'$ from $\bf 1\otimes1$
(see, for example \cite{TDLee}, for the decomposition). 
Hence, from SU(3) symmetry and
the Bose-Einstein statistics, we should have
 \be
 (\Sc_{res})^m
 =\sum_{a=1}^{27}|{\bf 27};a\rangle e^{2m\,i\delta_{27}}\langle {\bf 27};a|
  +\sum_{b=1}^{8}\sum_{p,q=8,8'}|p;b\rangle \U^{m}_{pq} \langle q;b|
  +\sum_{p,q=1,1'}|p;1\rangle \V^{m}_{pq} \langle q;1|,
 \label{eq:SU3decomposition}
 \en
where $a$ and $b$ are labels of states within multiplets,
and matrices $\U^{m}$ and $\V^{m}$ are given by
 \be
\U^{m}(\tau,\delta_8,\delta'_{8})
 &\equiv&\left(
\begin{array}{cc}
\cos\tau
       &\sin\tau
       \\
-\sin\tau
       &\cos\tau
\end{array}
\right) \left(
\begin{array}{cc}
e^{2m\,i\delta_8}
       &0
       \\
0
       &e^{2m\,i\delta_8'}
\end{array}
\right) \left(
\begin{array}{cc}
\cos\tau
       &-\sin\tau
       \\
\sin\tau
       &\cos\tau
\end{array}
\right),
 \nonumber\\
 \V^{m}(\nu,\delta_1,\delta'_{1})
 &\equiv&\left(
\begin{array}{cc}
\cos\nu
       &\sin\nu
       \\
-\sin\nu
       &\cos\nu
\end{array}
\right) \left(
\begin{array}{cc}
e^{2m\,i\delta_1}
       &0
       \\
0
       &e^{2m\,i\delta_1'}
\end{array}
\right) \left(
\begin{array}{cc}
\cos\nu
       &-\sin\nu
       \\
\sin\nu
       &\cos\nu
\end{array}
\right),
 \label{eq:UandV}
 \en
respectively.
Rescattering parameters $r_i$ as the solutions to Eqs.~(\ref{eq:SSU3a}) and
(\ref{eq:SSU3b}) can be expressed in terms of these angles and phases:
 \be
 1+i(r^{(m)}_0+r^{(m)}_a)
      &=&\frac{2 e^{2m\,i\delta_{27}}+3\U^{m}_{11}}{5},
      \nonumber\\
 i(r^{(m)}_e-r^{(m)}_a)
      &=&\frac{3 e^{2m\,i\delta_{27}}-3\U^{m}_{11}}{5},
      \nonumber\\
 i (r^{(m)}_a+r^{(m)}_t)
      &=&\frac{-e^{2m\,i\delta_{27}}- 4 \U^{m}_{11} + 5 \V^{m}_{11}}{20},
      \nonumber\\
 i(2\bar r^{(m)}_a+\bar r^{(m)}_e)
      &=&\frac{3}{\sqrt5} \U^{m}_{12},
      \nonumber\\
 1+i(\tilde r^{(m)}_0+\frac{4\tilde r^{(m)}_a+2\tilde r_e^{(m)}}{3})
      &=&\U_{22}^{m},
      \nonumber\\
 i(\hat r^{(m)}_t+\frac{4\hat r^{(m)}_a+2\hat r^{(m)}_e}{3})
      &=&\frac{1}{\sqrt2} \V^{m}_{12},
      \nonumber\\
  1+i(\check r^{(m)}_0+\frac{4\check r^{(m)}_a+2\check r_e^{(m)}+3\check r^{(m)}_t}{6})
      &=&\V_{22}^{m},
 \label{eq:solution}
 \en
with $\U^{m}_{ij}$ and $\V^{m}_{ij}$ given in Eq. (\ref{eq:UandV}).

It is interesting to see how the rescattring behaves in a U(3)
symmetric case. It is known that the U$_A(1)$ breaking is
responsible for the mass difference between $\eta$ and $\eta'$ and U(3)
symmetry is not a good symmetry for low-lying pseudoscalars.
However, U(3) symmetry may still be a reasonable one for
a system that rescatters at energies of order $m_B$. 
The mass difference between $\eta$
and $\eta'$, as an indicator of U(3) symmetry breaking effect,
does not lead to sizable energy difference of these particles in
charmless $B$ decays. 
Note that in the literature, some authors also make
use of U(3) symmetry in charmless $B$ decays (see, for
example~\cite{Pham:2007nt}).
We note that in the U(3) case, we have
 \be
 r_i=\bar r_i=\tilde r_i=\hat r_i=\check r_i.
 \label{eq: U3r}
 \en
Consequently, by requiring 
 \be
 \T^{(m)}&=&\T\,\,\, {\rm with\,\,}(r_j,\bar r_j,\tilde r_j,\check r_j)
 \to (r^{(m)}_j,r^{(m)}_j,r^{(m)}_j,r^{(m)}_j),
 \en
as required by Eq. (\ref{eq: U3r}), one must have~\cite{Chua:2007cm} 
\be
r^{(m)}_a r^{(m)}_e=0.
\label{eq: rare=0}
\en
There are two solutions, either $r^{(m)}_e=0$ or $r^{(m)}_a=0$~[see. Eqs.~(\ref{eq:solutionreU3ra}) and (\ref{eq:solutionreU3re})]. 
Note that in both solutions, we have
\be
 \delta_{27}=\delta_8'=\delta'_1.
\en
To reduce the number of the rescattering parameters and as a working assumption, the above relations will be used in this work, 
although we are not imposing the full U(3) symmetry to FSI.  

After imposing the above relation and factor out a over phase factor, say $\delta_{27}$,  
we are left with two mixing angles and two phase
differences:
 \be
 \tau,\quad
 &&\nu,
 \quad
 \delta\equiv\delta_8-\delta_{27},
 \quad
 \sigma\equiv\delta_1-\delta_{27},%
 \label{eq:FSIparameters}
 \en
in the scattering matrices. 
The rescattering formula Eq.~(\ref{eq:master1}) now becomes 
 \be
 A=\Sc_{res}^{1/2}(\tau,\nu;\delta,\sigma)\cdot
 A^{\rm fac},
 \label{eq:master2}
 \en
with the overall phase removed. 
In summary, 4 additional parameters from Res are introduced to the decay amplitudes.

We find that it is useful to incorporate SU(3) breaking effect in the scattering matrix. 
The idea is that we try to remove the SU(3) breaking effect in $A^{\rm fac}$ before recattering and put the SU(3) breaking effect back after the rescattering.
The underlying reason is as following. 
In the core of FSI, the rescattering processes are occuring at the $m_B$ energy scale, 
the SU(3) breaking effect cannot be very important at this stage. 
Hence the amplitudes to be rescattered are taken in the SU(3) limit, 
but after the rescattering, as the hadronization process takes place, SU(3) breaking cannot be neglected  and their effect needs to be included.
In practice we use ratio of decay constants to model the SU(3) breaking effect.
For example, the $B^-\to K^0K^-$ factorization amplitude is multiplied by $(f_\pi/f_K)^2$ before recattering with other states and is multiplied by $(f_K/f_\pi)^2$ after rescattering.
For convenient these two factors are absorbed in $\Sc_{res}^{1/2}$.
These are new to Ref.~\cite{Chua:2007cm}.

The rescattering matrices needed in this work are collected in Appendix A.
As we will see in the next section, including these four rescattering parameters will enhance the agreement of theory and data notably.


\subsection{Rescattering and Topological Amplitudes in the SU(3) limit}

Topological amplitude approach or flavor flow approach is based on SU(3) symmetry. 
The amplitudes can contain weak and strong phases. 
FSI will generate additional strong phases and can potentially mixed up different topological amplitudes. 
It is therefore interesting to investigate the relation of the FSI and topological amplitudes. 
We will take a closer look of this issue in the presence of the rescattering among $PP$ states.
We will consider the topological amplitudes in the SU(3) limit,
rescattering of topological amplitudes in the SU(3) limit and, finally, topological amplitudes and rescattering in the U(3) limit.
The discussion will be useful to provide a better understanding of the effect of FSI in $\ov B{}_q\to PP$ decays.
These are all new to Ref.~\cite{Chua:2007cm}.

\subsubsection{Topological Amplitudes in the SU(3) limit}

It is well-known
that the fields annihilating $B^-,\,\overline B^0_{d,s}$ and creating $\pi,\,K,\,\eta_8$ transform
respectively as $\overline {\bold 3}$ and $\bold 8$
under SU(3) (see, for example~\cite{TDLee}),
\begin{eqnarray}
{\overline B}&=& \left(
\begin{array}{ccc}
B^- &\overline B {}^0 &\overline B {}^0_s
\end{array}
\right),
\nonumber\\
\Pi&=& \left(
\begin{array}{ccc}
{{\pi^0}\over\sqrt2}+{{\eta_8}\over\sqrt6}&{\pi^+} &{K^+}\\
{\pi^-}&-{{\pi^0}\over\sqrt2}+{{\eta_8}\over\sqrt6}&{K^0}\\
{K^-}  &{\overline K {}^0}&-\sqrt{2\over3}{\eta_8}
\end{array}
\right).
\end{eqnarray}
For the $b\to u\bar u d$ and $b\to q\bar q d$
processes, the tree (${\cal O}_T$) and penguin (${\cal O}_P$)
operators respectively have the following flavor structure,
\begin{eqnarray}
&&{\cal O}_T\sim (\bar u b )(\bar d u )
 =H^{ik}_j (\bar q_i b) (\bar q_k q^j),
\quad
{\cal O}_P\sim(\bar d b) (\bar q_i  q^i )
 =H^k (\bar q_k b) (\bar q_i q^i) ,
\non\\
&&{\cal O}_{EWP}\sim Q_j(\bar d b) ( \bar q_j q^j)
 =H_{EW}{}^{ik}_j (\bar q_i b) (\bar q_k q^j), 
\end{eqnarray}
where we define
$H^k=\delta^k_2$, $H^{ik}_j=\delta^i_1\delta^1_j \delta^k_2$ and $(H_{EW})^{ik}_j=\delta^i_2 Q_j \delta^k_j$ 
(no sum in indices).
Note that it is easy to check that we have
$H^{ik}_i=H^k$, $H^{ik}_k=0$, $(H_{EW})^{ik}_k=0$, $(H_{EW})^{ik}_i=Q_2\delta^k_2=Q_2 H^k$.
The flavor
structures of $|\Delta S|=1$ tree and penguin operators can be
obtained by replacing $d$ to $s$, 
$H^k=\delta^k_3$, $H^{ik}_j=\delta^i_1\delta^1_j \delta^k_3$ and $(H_{EW})^{ik}_j=\delta^i_3 Q_j \delta^k_j$.

The effective Hamiltonian, in term of the meson degree of freedom,
for the $\overline B\to PP$ decay should have the same SU(3)
transform property of $H_{\rm W}$. Consequently, we
have
\begin{eqnarray}
H_{\rm eff}&=&T\, \overline B_m H^{ik}_j (\Pi^{\rm out})^j_k (\Pi^{\rm out})^m_i
              +C\,\overline B_m H^{ik}_j (\Pi^{\rm out})^j_i (\Pi^{\rm out})^m_k
\nonumber\\
              &&+E\,\overline B_k H^{ik}_j (\Pi^{\rm out})^j_l (\Pi^{\rm out})^l_i
              +A\,\overline B_i H^{ik}_j (\Pi^{\rm out})^j_l (\Pi^{\rm out})^l_k
\nonumber\\
            &&+P\,\overline B_m H^k (\Pi^{\rm out})^m_i (\Pi^{\rm out})^i_k
              +\frac{1}{2}PA\,\overline B_k H^k (\Pi^{\rm out})^l_m(\Pi^{\rm out})^m_l
\non\\
&&
+P_{EW} \overline B_m H_{EW}{}^{ik}_j(\Pi^{\rm out})^m_i(\Pi^{\rm out})^j_k  
+P^C_{EW}  \overline B_m H_{EW}{}^{ik}_j (\Pi^{\rm out})^m_k (\Pi^{\rm out})^j_i 
\non\\
&&+P^E_{EW}\,\overline B_k H_{EW}{}^{ik}_j (\Pi^{\rm out})^j_l (\Pi^{\rm out})^l_i
              +P^A_{EW}\,\overline B_i H_{EW}{}^{ik}_j (\Pi^{\rm out})^j_l (\Pi^{\rm out})^l_k
              \non\\
             && +(H_{\rm eff})_{\rm singlet},
 \label{eq:B2PP Heff}
\end{eqnarray}
where the $A$, $P$, $PA$ and $P_{EW}$ terms correspond to annihilation,
penguin, penguin annihilation and electroweak penguin amplitudes, respectively.
and $(H_{\rm eff})_{\rm singlet}$ is the hamiltonain involving $\eta_1$, given by
\begin{eqnarray}
(H_{\rm eff})_{\rm singlet}&=&\bar T\, \overline B_m H^{ik}_j (\Pi^{\rm out})^j_k (\tilde\Pi^{\rm out})^m_i
              +\bar C_1\,\overline B_m H^{ik}_j (\Pi^{\rm out})^j_i (\tilde \Pi^{\rm out})^m_k
\nonumber\\
             &&+\bar C_2\,\overline B_m H^{ik}_j (\tilde \Pi^{\rm out})^j_i (\Pi^{\rm out})^m_k
                 +\tilde C\,\overline B_m H^{ik}_j (\tilde\Pi^{\rm out})^j_i (\tilde\Pi^{\rm out})^m_k
\nonumber\\
              &&+\bar E_1\,\overline B_k H^{ik}_j (\tilde\Pi^{\rm out})^j_l (\Pi^{\rm out})^l_i
              +\bar E_2\,\overline B_k H^{ik}_j (\Pi^{\rm out})^j_l (\tilde\Pi^{\rm out})^l_i
 \nonumber\\
              &&+\tilde E\,\overline B_k H^{ik}_j (\tilde\Pi^{\rm out})^j_l (\tilde\Pi^{\rm out})^l_i            
              +\bar A_1\,\overline B_i H^{ik}_j (\tilde \Pi^{\rm out})^j_l (\Pi^{\rm out})^l_k
\nonumber\\
             &&+\bar A_2\,\overline B_i H^{ik}_j (\Pi^{\rm out})^j_l (\tilde \Pi^{\rm out})^l_k
             +\bar P_1\,\overline B_m H^k (\tilde\Pi^{\rm out})^m_i (\Pi^{\rm out})^i_k
\non\\
            &&
              +\bar P_2\,\overline B_m H^k (\Pi^{\rm out})^m_i (\tilde\Pi^{\rm out})^i_k
              +\tilde P\,\overline B_m H^k (\tilde\Pi^{\rm out})^m_i (\tilde\Pi^{\rm out})^i_k
\non\\
            &&
            +\frac{1}{2}\widetilde {PA}\,\overline B_k H^k (\tilde\Pi^{\rm out})^l_m(\tilde\Pi^{\rm out})^m_l
+\bar P_{EW} \overline B_m H_{EW}{}^{ik}_j(\tilde\Pi^{\rm out})^m_i(\Pi^{\rm out})^j_k 
\non\\
&& 
+\bar P^C_{EW,1}  \overline B_m H_{EW}{}^{ik}_j (\tilde\Pi^{\rm out})^m_k (\Pi^{\rm out})^j_i 
+\bar P^C_{EW,2}  \overline B_m H_{EW}{}^{ik}_j (\Pi^{\rm out})^m_k (\tilde\Pi^{\rm out})^j_i 
\non\\
&&
+\tilde P^C_{EW}  \overline B_m H_{EW}{}^{ik}_j (\tilde\Pi^{\rm out})^m_k (\tilde\Pi^{\rm out})^j_i 
+\bar P^E_{EW,1}\,\overline B_k H_{EW}{}^{ik}_j (\tilde\Pi^{\rm out})^j_l (\Pi^{\rm out})^l_i
\non\\
&&
+\bar P^E_{EW,2}\,\overline B_k H_{EW}{}^{ik}_j (\Pi^{\rm out})^j_l (\tilde\Pi^{\rm out})^l_i
              +\tilde P^E_{EW}\,\overline B_k H_{EW}{}^{ik}_j (\tilde\Pi^{\rm out})^j_l (\tilde\Pi^{\rm out})^l_i
\non\\
&&
                +\bar P^A_{EW,1}\,\overline B_i H_{EW}{}^{ik}_j (\tilde\Pi^{\rm out})^j_l (\Pi^{\rm out})^l_k
              +\bar P^A_{EW,2}\,\overline B_i H_{EW}{}^{ik}_j (\Pi^{\rm out})^j_l (\tilde\Pi^{\rm out})^l_k
\non\\
&&
              +\tilde P^A_{EW}\,\overline B_i H_{EW}{}^{ik}_j (\tilde\Pi^{\rm out})^j_l (\tilde\Pi^{\rm out})^l_k,
 \label{eq:B2PP Heff sing}
\end{eqnarray}
with $(\tilde \Pi^{\rm out})^i_j=\eta_1^{\rm out}\delta^i_j/\sqrt3$.
Note that we introduce $P^E_{EW}$ and $P^A_{EW}$, namely the electroweak penguin exchange and electroweak penguin annihilation terms for completeness. The above $H_{\rm eff}$ contains all possible SU(3) invariant combinations in first order of $H^{ik}_j$, $H^k$ and $H_{EW}{}^{ik}_j$.
It should be emphasize that the effective Hamiltonian in Eq. (\ref{eq:B2PP Heff}) is obtained using flavor SU(3) symmetry argument only.
The TA amplitude can contain all possible FSI contributions, while the expressions of the decay amplitude in term of these TA will remain the same.

With redefinition of the following amplitudes:
\be
2\bar A\equiv \bar A_1+\bar A_2,
\quad
2\bar E\equiv\bar E_1+\bar E_2,
\quad
2\bar P\equiv\bar P_1+\bar P_2,
\non\\
2\bar P^A_{EW}\equiv\bar P^A_{EW,1}+\bar P^A_{EW,2},
\quad
2\bar P^E_{EW}\equiv \bar P^E_{EW,1}+\bar P^E_{EW,2},
\en
$(H_{\rm eff})_{\rm singlet}$ can be expressed in a more compact form,
\begin{eqnarray}
(H_{\rm eff})_{\rm singlet}&=&(\bar T+2\bar A)\, \overline B_i H^{ik}_j (\Pi^{\rm out})^j_k\eta_1^{\rm out}/\sqrt3
\non\\
&&+(\bar C_1+2\bar E)\,\overline B_k H^{ik}_j (\Pi^{\rm out})^j_i\eta_1^{\rm out}/\sqrt3
\non\\
&&+(\bar C_2+2\bar P-\frac{1}{3}\bar P^C_{EW,2})  \overline B_m H^k (\Pi^{\rm out})^m_k   \eta_1^{\rm out}/\sqrt3
\non\\
&&+(\bar P_{EW} +2\bar P^A_{EW})\,\overline B_i H_{EW}{}^{ik}_j (\Pi^{\rm out})^j_k\eta_1^{\rm out}/\sqrt3
\non\\
&& +(\bar P^C_{EW,1}+2\bar P^E_{EW})\,\overline B_k H_{EW}{}^{ik}_j (\Pi^{\rm out})^j_i\eta_1^{\rm out}/\sqrt3
\non\\
&&+(\tilde C+\tilde E +\tilde P+\frac{3}{2}\tilde {PA}-\frac{1}{3}\tilde P^C_{EW}-\frac{1}{3}\tilde P^E_{EW})\,\overline B_k H^k  \eta_1^{\rm out}\eta_1^{\rm out}/3.        
\label{eq:B2PP Heff0}
\end{eqnarray}



Using the above approach we can
reproduce familiar expressions of decay amplitudes in terms of TA~\cite{Gronau:1994rj,Gronau:1995hn}.~\footnote{See also \cite{He:2018php} for a recent discussion.}
Explicitly, we have the following amplitudes:
\begin{eqnarray}
 A_{\ov B {}^0_{d}\to K^-\pi^+}&=&T'+P'+\frac{1}{3}(2 P_{EW}^{\prime\,C}-P_{EW}^{\prime\,E}),
 \non\\
 A_{\ov B {}^0_{d}\to \ov K {}^0 \pi^0}&=&\frac{1}{3\sqrt2}(3 C'-3P'+3P'_{EW}+P_{EW}^{\prime\,C}+P_{EW}^{\prime\,E}),
 \non\\
 A_{\ov B {}^0_{d}\to \ov K {}^0 \eta_8}&=&\frac{1}{3\sqrt6}(3 C'-3P'+3P'_{EW}+P_{EW}^{\prime\,C}+P_{EW}^{\prime\,E}),
 \non\\
 A_{\ov B {}^0_{d}\to \ov K {}^0 \eta_1}&=&\frac{1}{3\sqrt3}(3 \bar C'_2+6\bar P'-\bar P_{EW,1}^{\prime\,C}-\bar P_{EW,2}^{\prime\,C}-2\bar P_{EW}^{\prime\,E}),
 \label{eq: TAgroup1}
\end{eqnarray}
for group-1 modes, 
\begin{eqnarray}
 A_{B^-\to \ov K^0\pi^-}&=&A'+P'+\frac{1}{3}(-P_{EW}^{\prime\,C}+2P_{EW}^{\prime\, E}),
 \non\\
 A_{B^-\to K^- \pi^0}&=&\frac{1}{3\sqrt2}(3T'+3C'+3A'+3P'+3 P'_{EW}+2 P_{EW}^{\prime\,C}+2 P_{EW}^{\prime\, E}),
 \non\\
 A_{B^-\to K^- \eta_8}&=&\frac{1}{3\sqrt6}(3T'+3C'-3A'-3P'+3 P'_{EW}+4 P_{EW}^{\prime\,C}-2 P_{EW}^{\prime\, E}),
 \non\\
 A_{B^-\to K^- \eta_1}&=&\frac{1}{3\sqrt3}(3\bar T'+3 \bar C'_2+6 \bar A'+6\bar P'+2\bar P_{EW,1}^{\prime\,C}-\bar P_{EW,2}^{\prime\,C}+4\bar P_{EW}^{\prime\,E}),
 \label{eq: TAgroup2}
\end{eqnarray}
for group-2 modes,
\begin{eqnarray}
 A_{B^-\to \pi^-\pi^0}&=&\frac{1}{\sqrt2}(T+C+P_{EW}+P_{EW}^C),
 \non\\
 A_{B^-\to K^0 K^-}&=&A+P+\frac{1}{3}(-P^C_{EW}+2 P_{EW}^E),
 \non\\
 A_{B^-\to \pi^- \eta_8}&=&\frac{1}{3\sqrt6}(3 T+3 C+6 A+6P+3 P_{EW}+P_{EW}^C+4 P_{EW}^E),
 \non\\
 A_{B^-\to \pi^- \eta_1}&=&\frac{1}{3\sqrt3}(3 \bar T+3\bar C_2+6 \bar A+6\bar P+2 \bar P_{EW,1}^C-\bar P_{EW,2}^C+4\bar P_{EW}^E),
 \label{eq: TAgroup3}
\end{eqnarray}
for group-3 modes,
\be
 A_{\ov B {}^0_{d}\to \pi^+\pi^-}&=&T+E+P+PA+\frac{1}{3}(2P_{EW}^C+P_{EW}^A-P_{EW}^E),
 \non\\
 A_{\ov B {}^0_{d}\to \pi^0 \pi^0}&=&\frac{1}{\sqrt2} (-C+E+P+ PA-P_{EW}-\frac{1}{3}P_{EW}^C+\frac{1}{3}P_{EW}^A-\frac{1}{3}P_{EW}^E),
 \non\\
 A_{\ov B {}^0_{d}\to \eta_8 \eta_8}&=&\frac{1}{9\sqrt2} (3 C+3 E+3 P+ 9PA+3P_{EW}-P_{EW}^C-3P_{EW}^A-P_{EW}^E),
 \non\\
 A_{\ov B {}^0_{d}\to \eta_8 \eta_1}&=&\frac{1}{9\sqrt2}(3 \bar C_1+3 \bar C_2+6 \bar E+6\bar P+3\bar P_{EW}+6 \bar P_{EW}^A-\bar P_{EW,1}^C-\bar P_{EW,2}^C-2 \bar P_{EW}^E),
 \non\\
 A_{\ov B {}^0_{d}\to \eta_1 \eta_1}&=&\frac{1}{9\sqrt2}(6\tilde C+6 \tilde E+6\tilde P+9\tilde{PA}-2\tilde P^C_{EW}-2 \tilde P_{EW}^E),
 \non\\
 A_{\ov B {}^0_{d}\to K^+ K^-}&=&E+PA+\frac{1}{3} P_{EW}^A,
 \non\\
 A_{\ov B {}^0_{d}\to K^0 \ov K {}^0}&=&P+PA-\frac{1}{3}(P_{EW}^C+2 P_{EW}^A+P_{EW}^E),
 \non\\
 A_{\ov B {}^0_{d}\to \pi^0 \eta_8}&=&\frac{1}{3\sqrt3}(3 E-3 P+P_{EW}^C+3 P_{EW}^A+P_{EW}^E),
 \non\\
 A_{\ov B {}^0_{d}\to \pi^0 \eta_1}&=&\frac{1}{3\sqrt6}(3 \bar C_1-3 \bar C_2+6 \bar E-6\bar P+3\bar P_{EW}+6 \bar P_{EW}^A+\bar P_{EW,1}^C+\bar P_{EW,2}^C+2 \bar P_{EW}^E),
 \label{eq: TAgroup4}
 \non\\
\en
for group-4 modes, 
and the following amplitudes: 
\begin{eqnarray}
 A_{\ov B {}^0_{s}\to K^+\pi^-}&=&T+P+\frac{1}{3}(2 P_{EW}^C-P_{EW}^E),
 \non\\
 A_{\ov B {}^0_{s}\to K {}^0 \pi^0}&=&\frac{1}{3\sqrt2}(3 C-3P+3 P_{EW}+P_{EW}^C+P_{EW}^E),
 \non\\
 A_{\ov B {}^0_{s}\to K {}^0 \eta_8}&=&\frac{1}{3\sqrt6}(3 C-3P+3 P_{EW}+P_{EW}^C+P_{EW}^E),
 \non\\
 A_{\ov B {}^0_{s}\to K {}^0 \eta_1}&=&\frac{1}{3\sqrt3}(3 \bar C_2+6\bar P-\bar P_{EW,1}^C-\bar P_{EW,2}^C-2 \bar P_{EW}^E),
 \label{eq: TABsgroup1}
\end{eqnarray}
and
\be
 A_{\ov B {}^0_{s}\to \pi^+\pi^-}&=&E'+PA'+\frac{1}{3}P_{EW}^{\prime \,A},
 \non\\
 A_{\ov B {}^0_{s}\to \pi^0 \pi^0}&=&\frac{1}{\sqrt2} (E'+ PA'+\frac{1}{3}P_{EW}^{\prime\,A}),
 \non\\
 A_{\ov B {}^0_{s}\to \eta_8 \eta_8}&=&\frac{1}{9\sqrt2} (-6 C'+3 E'+12 P'+ 9PA'-6P'_{EW}-4 P_{EW}^{\prime\,C}-3P_{EW}^{\prime\,A}-4P_{EW}^{\prime\,E}),
 \non\\
 A_{\ov B {}^0_{s}\to \eta_8 \eta_1}&=&\frac{1}{9\sqrt2}(3 \bar C'_1-6 \bar C'_2+6 \bar E'-12\bar P'+3\bar P'_{EW}+6 \bar P_{EW}^{\prime\,A}+2\bar P_{EW,1}^{\prime\,C}+2\bar P_{EW,2}^{\prime\,C}+4 \bar P_{EW}^{\prime\,E}),
 \non\\
 A_{\ov B {}^0_{s}\to \eta_1 \eta_1}&=&\frac{1}{9\sqrt2}(6\tilde C'+6 \tilde E'+6\tilde P'+9\tilde{PA}'-2\tilde P^{\prime\,C}_{EW}-2 \tilde P_{EW}^{\prime\,E}),
 \non\\
 A_{\ov B {}^0_{s}\to K^+ K^-}&=&T'+E'+P'+PA'+\frac{1}{3} (P_{EW}^{\prime\,A}+2P_{EW}^{\prime\,C}-P_{EW}^{\prime\,E}),
 \non\\
 A_{\ov B {}^0_{s}\to K^0 \ov K {}^0}&=&P'+PA'-\frac{1}{3}(P_{EW}^{\prime\,C}+2 P_{EW}^{\prime\,A}+P_{EW}^{\prime\,E}),
 \non\\
 A_{\ov B {}^0_{s}\to \pi^0 \eta_8}&=&\frac{1}{\sqrt3}(-C'+E'-P'_{EW}+P_{EW}^{\prime\,A}),
 \non\\
 A_{\ov B {}^0_{s}\to \pi^0 \eta_1}&=&\frac{1}{\sqrt6}(\bar C'_1+2 \bar E'+\bar P'_{EW}+2 \bar P_{EW}^{\prime\,A}),
 \label{eq: TABsgroup4}
\en
for $\overline B_s\to PP$ decays,
where the $T$, $C$, $A$, $P$, $PA$ and $P_{EW}$ terms correspond to color-allowed tree, color-suppressed tree, annihilation,
penguin, penguin annihilation and electroweak penguin amplitudes, respectively. 
Note that $P^E_{EW}$ and $P^A_{EW}$, namely the electroweak penguin exchange and electroweak penguin annihilation terms, are introduced for completeness. See Appendix B for details.
Those with (without) prime are for $\Delta S=-1$(0) transition. 

The one-to-one correspondence of the SU(3) parameters and the
topological amplitudes is not a coincidence. It can be understood
by using a flavor flow analysis. We take the first term of $H_{\rm
eff}$ for illustration. In $H_{\rm W}$ the decays are governed by
the 
${\cal O}_T\sim (\bar u b )(\bar d u )
 =H^{ik}_j (\bar q_i b) (\bar q_k q^j),$
$b\to q_i\, \bar q^j\, q_k $ transition with the corresponded
$H^{ik}_j$ coupling. The first term of $H_{\rm eff}$ in
Eq.~(\ref{eq:B2PP Heff}) is $T\, \overline B_m H^{ik}_j (\Pi^{\rm out})^j_k (\Pi^{\rm out})^m_i$. 
Note that we use subscript and superscript according to the field convention. 
For example, we assign a subscript (superscript) to the initial
(final) state anti-quark $\bar q_m$~($\bar q^m$). 
The $\overline B_m (\Pi^{\rm out})^m_i$ part in $T\, \overline B_m H^{ik}_j (\Pi^{\rm out})^j_k (\Pi^{\rm out})^m_i$ can be
interpreted as a $\overline B_m$ to $(\Pi^{\rm out})^m_i$ transition with the same
spectator anti-quark $\bar q_m$ from $\overline B_m$ becoming the final state spectator anti-quark $\bar q^m$, which ends up in $(\Pi^{\rm out})^m_i$. 
The quark $q_i$ from $b\to q_i$ transition also ends up in $(\Pi^{\rm out})^m_i$, while
the $(\Pi^{\rm out})^j_k$ part is responsible for the creation of the meson
where the $W$-emitted $\bar q^j q_k$ pair ends up with. 
The above
picture clearly corresponds to the external $W$-emission topology.
Similarly, the identification of the other topological amplitudes can be understood similarly.

One can check that all of the above amplitudes can be expressed in terms of the following combinations:
\be
T^{(\prime)}+C^{(\prime)},
\quad
C^{(\prime)}-E^{(\prime)},
\quad
A^{(\prime)}+C^{(\prime)},
\non\\
P^{(\prime)}-C^{(\prime)}+\frac{1}{3}P_{EW}^{(\prime)C},
\quad
PA^{(\prime)}-\frac{4}{9} C^{(\prime)}+\frac{13}{9} E^{(\prime)}-\frac{1}{3} P_{EW}^{(\prime)C},
\non\\
P^{(\prime)}_{EW}+P_{EW}^{(\prime)C},
\quad
P_{EW}^{(\prime)C}-P_{EW}^{(\prime)E},
\quad
P_{EW}^{(\prime)A}+P_{EW}^{(\prime)C},
\non\\
\bar T^{(\prime)}+2\bar A^{(\prime)},
\quad
\bar C^{(\prime)}_1+2\bar E^{(\prime)},
\quad
\bar C^{(\prime)}_2+2\bar P^{(\prime)}-\frac{1}{3}\bar P^{(\prime)C}_{EW,2},
\quad
\bar P^{(\prime)}_{EW} +2\bar P^{(\prime)A}_{EW},
\non\\
\bar P^{(\prime)C}_{EW,1}+2\bar P^{(\prime)E}_{EW},
\quad
\tilde C^{(\prime)}+\tilde E^{(\prime)} +\tilde P^{(\prime)}+\frac{3}{2}\tilde {PA}^{(\prime)}-\frac{1}{3}\tilde P^{(\prime)C}_{EW}-\frac{1}{3}\tilde P^{(\prime)E}_{EW}.
\label{eq: combinations}
\en
For example, we can express the decay amplitude of $\overline B^0\to K^-\pi^+$ in the following combinations:
\be
A(\overline B^0\to K^-\pi^+)=(T'+C') + (P'-C'+\frac{1}{3}P_{EW}^{C\prime}) +\frac{1}{3}(P^{C\prime}_{EW}-P^{E\prime}_{EW}).
\en


It is interesting to compare the amplitudes expressed in terms of the topological amplitudes with the those in the QCDF calculation.
We can obtain the following relations in the SU(3) limit: (using formulas in \cite{Beneke:2003zv} but taking the SU(3) limit)
\be
T^{(\prime)0}&=&A_{PP}\lambda^{(\prime)}_p\delta_{pu} \alpha_1,
\non\\
C^{(\prime)0}&=&A_{PP}\lambda^{(\prime)}_p\delta_{pu} \alpha_2,
\non\\
E^{(\prime)0}&=&A_{PP}\lambda^{(\prime)}_p\delta_{pu}\beta_1,
\non\\
A^{(\prime)0}&=&A_{PP}\lambda^{(\prime)}_p\delta_{pu}\beta_2,
\non\\
P^{(\prime)0}&=&A_{PP}\lambda^{(\prime)}_p(\alpha_4^p+\beta_3^p),
\non\\
PA^{(\prime)0}&=&2 A_{PP}\lambda^{(\prime)}_p\beta^p_4,
\non\\
P^{(\prime)0}_{EW}&=& \frac{3}{2} A_{PP}\lambda^{(\prime)}_p\alpha^p_{3,EW},
\non\\
P^{(\prime)C0}_{EW}&=& \frac{3}{2} A_{PP}\lambda^{(\prime)}_p\alpha^p_{4,EW},
\non\\
P^{(\prime)E0}_{EW}&=& \frac{3}{2} A_{PP}\lambda^{(\prime)}_p\beta^p_{3,EW},
\non\\
P^{(\prime)A0}_{EW}&=& \frac{3}{2} A_{PP}\lambda^{(\prime)}_p\beta^p_{4,EW},
\label{eq: TAQCDF1}
\en
where we use $\lambda^{(\prime)}_p\equiv V_{pb}V^*_{pd(s)}$, $p=u,c$ with $V_{pb, pd(s)}$ the Cabibbo-Kobayashi-Maskawa (CKM) matrix elements and summation over $p$ is implied. One can find detail definitions of $A_{PP}$, $\alpha$ and $\beta$ in \cite{Beneke:2003zv}. 
Note that $A_{PP}$ involves a $\overline B_q\to P$ transition and a $P$ decay constant:
\be
A_{PP}=\frac{G_F}{\sqrt2}F^{BP}_0(m_P^2) f_P m_B^2. 
\en
It should be note that we have removed an overall $i$ in the definition of $A_{PP}$.
The superscript $0$ on TA is denoting the fact that rescattering among $PP$ states has not taken place.
In the SU(3) limit, we will use $F^{BP}_0(m_P^2)=F^{B\pi}_0(0)$ and $f_P=f_\pi$ in later discussion.

For $B$ decays to a final state with $\eta_1$, things are more complicated.
For example, $A_{P\eta_1}$ is in principle different from $A_{\eta_1 P}$. 
We have in the SU(3) limit:~\cite{Beneke:2003zv} 
\be
\bar T^{(\prime)0}&=&A_{\eta_1P}\lambda^{(\prime)}_p\delta_{pu} \alpha_1,
\non\\
\bar C^{(\prime)0}_1&=&A_{\eta_1P}\lambda^{(\prime)}_p\delta_{pu} \alpha_2,
\non\\
\bar C^{(\prime)0}_2&=&A_{P\eta_1}\lambda^{(\prime)}_p\delta_{pu} \alpha_2,
\non\\
2\bar E^{(\prime)0}&\equiv&\bar E^{(\prime)0}_1+\bar E^{(\prime)0}_2
=A_{\eta_1P}\lambda^{(\prime)}_p\delta_{pu}\beta_1+A_{P\eta_1}\lambda^{(\prime)}_p\delta_{pu}(\beta_1+3\beta_{S1}),
\non\\
2\bar A^{(\prime)0}&\equiv&\bar A^{(\prime)0}_1+\bar A^{(\prime)0}_2
=A_{\eta_1P}\lambda^{(\prime)}_p\delta_{pu}\beta_2+A_{P\eta_1}\lambda^{(\prime)}_p\delta_{pu}(\beta_2+3\beta_{S2}),
\non\\
2\bar P^{(\prime)0}&\equiv&\bar P^{(\prime)0}_1+\bar P^{(\prime)0}_2
=A_{\eta_1P}\lambda^{(\prime)}_p(\alpha_4^p+\beta_3^p)+A_{P\eta_1}\lambda^{(\prime)}_p(\alpha_4^p+3\alpha^p_3+\beta_3^p+3\beta^p_{S3}),
\non\\
\bar P^{(\prime)0}_{EW}&=& \frac{3}{2} A_{\eta_1P}\lambda^{(\prime)}_p\alpha^p_{3,EW},
\non\\
\bar P^{(\prime)C0}_{EW,1}&=& \frac{3}{2} A_{\eta_1P}\lambda^{(\prime)}_p\alpha^p_{4,EW},
\non\\
\bar P^{(\prime)C0}_{EW,2}&=& \frac{3}{2} A_{P\eta_1}\lambda^{(\prime)}_p\alpha^p_{4,EW},
\non\\
2\bar P^{(\prime)E0}_{EW}&\equiv&\bar P^{(\prime)E0}_{EW,1}+\bar P^{(\prime)E0}_{EW,2}
= \frac{3}{2} [A_{\eta_1P}\lambda^{(\prime)}_p\beta^p_{3,EW}+A_{P\eta_1}\lambda^{(\prime)}_p(\beta^p_{3,EW}+3\beta^p_{S3,EW})],
\non\\
2\bar P^{(\prime)A0}_{EW}&\equiv&\bar P^{(\prime)A0}_{EW,1}+\bar P^{(\prime)A0}_{EW,2}
= \frac{3}{2} [A_{\eta_1P}\lambda^{(\prime)}_p\beta^p_{4,EW}+A_{P\eta_1}\lambda^{(\prime)}_p(\beta^p_{4,EW}+3\beta^p_{4S,EW})].
\en
Note that $A_{P\eta_1}$ involves a $\overline B\to P$ transition, while $A_{\eta_1P}$ involves a $\overline B\to \eta_1$ transition:
\be
A_{P\eta_1}=\frac{G_F}{\sqrt2}F^{\bar B\to P}_0 f_{\eta_1} m_B^2,
\quad
A_{\eta_1P}=\frac{G_F}{\sqrt2}F^{\bar B\to \eta_1}_0 f_P m_B^2\simeq\frac{G_F}{\sqrt2}F^{\bar B\to P}_0\frac{f_{\eta_1}}{f_P} f_P m_B^2,
\label{eq: TAQCDF2}
\en
where in the second equation, we have made use of the approximation from ~\cite{Beneke:2003zv}.
In fact we have $A_{P\eta_1}\simeq A_{\eta_1 P}\simeq A_{PP}(f_{\eta_1}/f_P)$.


Finally comparing our expressions and those in Ref.~\cite{Beneke:2003zv}, we have
\be
\tilde C^{(\prime)0}&=&A_{\eta_1\eta_1}\lambda^{(\prime)}_p\delta_{pu} \alpha_2,
\non\\
\tilde E^{(\prime)0}&=&A_{\eta_1\eta_1}\lambda^{(\prime)}_p\delta_{pu}(\beta_1+3\beta_{S1}),
\non\\
\tilde P^{(\prime)0}&=&A_{\eta_1\eta_1}\lambda^{(\prime)}_p(\alpha_4^p+3\alpha^p_3+\beta_3^p+3\beta^p_{S3}),
\non\\
\widetilde {PA}^{(\prime)}&=&2 A_{\eta_1\eta_1}\lambda^{(\prime)}_p(\beta^p_4+3\beta^p_{S4}),
\non\\
\tilde P^{(\prime)C0}_{EW}&=& \frac{3}{2} A_{\eta_1\eta_1}\lambda^{(\prime)}_p\alpha^p_{4,EW},
\non\\
\tilde P^{(\prime)E0}_{EW}&=& \frac{3}{2} A_{\eta_1\eta_1}\lambda^{(\prime)}_p(\beta^p_{3,EW}+3\beta^p_{S3,EW}),
\label{eq: TAQCDF3}
\en
with
\be
A_{\eta_1\eta_1}=\frac{G_F}{\sqrt2}F^{\bar B\to \eta_1}_0 f_{\eta_1} m_B^2
\simeq \frac{G_F}{\sqrt2}F^{\bar B\to P}_0 \frac{f_{\eta_1}}{f_P}f_{\eta_1} m_B^2=A_{PP}\bigg(\frac{f_{\eta_1}}{f_P}\bigg)^2.
\en
In the later discussion, we take $f_{\eta_1}=f_P=f_\pi$.

\subsubsection{Rescattering of Topological Amplitudes in the SU(3) limit}

We now turn to the rescattering part.
The matrices $\T_{1,2,3,4}$ can be obtained through a diagrammatic
method by matching the Clebsh-Gordan coefficients of scattering
mesons (see Fig.~\ref{fig:r}) or by using an operator method. We
have 
\be
O_e&=&Tr(\Pi^{\rm in}\Pi^{\rm out}\Pi^{\rm in}\Pi^{\rm out})/2,
\quad
O_a=Tr(\Pi^{\rm in}\Pi^{\rm in}\Pi^{\rm out}\Pi^{\rm out}),
\non\\ 
O_0&=&Tr(\Pi^{\rm in}\Pi^{\rm out})Tr(\Pi^{\rm in}\Pi^{\rm out})/2,
\quad
O_t=Tr(\Pi^{\rm in}\Pi^{\rm in})Tr(\Pi^{\rm out}\Pi^{\rm out})/4,
\label{eq:Oi}
\en
corresponding to $r_e$, $r_a$, $r_0$ and $r_t$ contributions, in the combination of
\be
{\cal T}^{(m)}=r^{(m)}_0 O_0+r^{(m)}_e O_e+r^{(m)}_a O_a+r^{(m)}_t O_t+\cdots,
\en
where the remaining terms will be specified in below.
The
above terms exhaust all possible combinations for $\Pi({\bf
8})\,\Pi({\bf 8})\to\Pi({\bf 8})\,\Pi({\bf 8})$ scatterings.

To obtain operators involving $\eta_1$, we simply replace $\Pi$ in the above operators to
$\Pi+\eta_1 1_{3\times 3}/\sqrt3$ and collect terms with different number of $\eta_1$ as
\be
\sqrt3 \bar O_e&=&\frac{\sqrt3}{2} \bar O_a 
=Tr(\Pi^{\rm in}\Pi^{\rm out}\Pi^{\rm in})\eta^{\rm out}_1+Tr(\Pi^{\rm out}\Pi^{\rm in}\Pi^{\rm out})\eta^{\rm in}_1,
\non\\
\tilde O_0&=&\frac{3}{4}\tilde O_a=\frac{3}{2}\tilde O_e=Tr(\Pi^{\rm in}\Pi^{\rm out})\eta_1^{\rm in}\eta_1^{\rm out},
\non\\
4\hat O_t&=&3\hat O_a=6\hat O_e=\eta^{\rm in}_1\eta^{\rm in}_1Tr(\Pi^{\rm out}\Pi^{\rm out})
+\eta^{\rm out}_1\eta^{\rm out}_1Tr(\Pi^{\rm in}\Pi^{\rm in}),
\non\\
2\check O_0&=&4\check O_t=3\check O_a=6\check O_e=\eta^{\rm in}_1\eta^{\rm out}_1\eta^{\rm in}_1\eta^{\rm out}_1.
\label{eq:Oieta1}
\en
Note that it is impossible to obtain a term containing three $\eta_1$ as is prohibited from $SU(3)$ symmetry.
We now have
\be
{\cal T}^{(m)}&=&
r^{(m)}_0 O_0+r^{(m)}_e O_e+r^{(m)}_a O_a+r^{(m)}_t O_t
+\bar r^{(m)}_e \bar O_e+\bar r^{(m)}_a \bar O_a
\non\\
&&+\tilde r^{(m)}_0 \tilde O_0+\tilde r^{(m)}_a \tilde O_a+\tilde r^{(m)}_e \tilde O_e
+\hat r^{(m)}_t \hat O_t+\hat r^{(m)}_a \hat O_a+\hat r^{(m)}_e \hat O_e
\non\\
&&+\check r^{(m)}_0 \check O_0+\check r^{(m)}_t \check O_t+\check r^{(m)}_a \check O_a+\check r^{(m)}_e \check O_e.
\en
Using Eq. (\ref{eq:Oieta1}), the above equation can be simplified into
\be
{\cal T}^{(m)}&=&
r^{(m)}_0 O_0+r^{(m)}_e O_e+r^{(m)}_a O_a+r^{(m)}_t O_t
+(\bar r^{(m)}_e+2\bar r^{(m)}_a) \bar O_e
\non\\
&&+\left(\tilde r^{(m)}_0 +\frac{4\tilde r^{(m)}_a +2\tilde r^{(m)}_e}{3} \right)\tilde O_0
+\left(\hat r^{(m)}_t +\frac{4\hat r^{(m)}_a +2\hat r^{(m)}_e}{3} \right)\hat O_t
\non\\
&&+\left(\check r^{(m)}_0+\frac{4\check r^{(m)}_a+2\check r^{(m)}_e+3\check r^{(m)}_t}{6}\right) \check O_0.
\label{eq:Tm}
\en
Note that various $\bar r^{(m)}_i$, $\tilde r^{(m)}_i$, $\hat r^{(m)}_i$ and $\check r^{(m)}_i$ occur in ${\cal T}^{(m)}$ only through some very specific combinations.
We still preserve the subscripts ($i=0,t,a,e$), since these $\bar r^{(m)}_i$, $\tilde r^{(m)}_i$, $\hat r^{(m)}_i$ and $\check r^{(m)}_i$ for different $i$ correspond to different flavor flow patterns in rescattering diagrams (see Fig.~\ref{fig:r}) and they will, in fact, reduce to $r^{(m)}_i$ in the $U$(3) limit.

It is straightforward to obtain the rescattering effects on topological amplitudes. 
In analogy to Eq. (\ref{eq:master2}):
 \be
 A=\Sc_{res}^{1/2}\cdot
 A^{\rm fac}=(1+i\T^{1/2})\cdot A^{\rm fac},
 \label{eq:master3}
 \en
we have
\be
H_{\rm eff}=(1+i\T^{1/2})\cdot H^0_{\rm eff}=H^0_{\rm eff}+i\T^{1/2}\cdot H^0_{\rm eff},
\label{eq:H=(1+iT')H0}
\en
where $H_{\rm eff}$ is given in Eq. (\ref{eq:B2PP Heff}), $\T^{1/2}$ in Eq. (\ref{eq:Tm}) but with $m=1/2$, $H^0_{\rm eff}$ is the un-scattered effective Hamiltonian with all $TA$ in $H_{\rm eff}$ replaced by $TA^0$ and the dot in the above equation implies all possible pairing of the $P^{\rm out}P^{\rm out}$ fields in $H_{\rm eff}^0$ to the $P^{\rm in}P^{\rm in}$ fields in $\T^{1/2}$ (the $P^{\rm out}P^{\rm out}$ in $\T^{1/2}$ remains unpaired). 
As noted previously since the effective Hamiltonian in Eq. (\ref{eq:B2PP Heff}) is obtained using flavor SU(3) symmetry argument only,
its flavor structure will not be changed in the presence of rescattering, i.e. Eq. (\ref{eq:H=(1+iT')H0}) will not modify the flavor structure of $H_{\rm eff}$.
This feature is indeed verified in the explicit computation. 
Therefore the expressions of the decay amplitude in term of the TA will remain the same, but now the these TA contain rescattering contributions.

The effect of rescattering on TA can be obtained using the above equation. 
The computation is straightforward, but tedious. 
Here we only give the final results, some derivations using the above equation can be found in Appendix B for illustration.
We obtain, in the presence of the rescattering, TA will receive corrections in the following ways:
\be
\delta T^{(\prime)}&=&ir'_0 T^{(\prime)0}+i r'_e C^{(\prime)0},
\non\\
\delta C^{(\prime)}&=&ir'_0C^{(\prime)0}+i r'_e T^{(\prime)0},
\non\\
\delta E^{(\prime)}&=&ir'_0 E^{(\prime)0}+ ir'_a T^{(\prime)0}-\frac{1}{3} i (r'_e+2r'_a) C^{(\prime)0}+\frac{1}{3}i(-2 r'_e+5 r'_a) E^{(\prime)0}
\non\\
         &&           +\frac{1}{3}i(\bar r'_e+2\bar r'_a)(\bar C^{(\prime)0}_1+2\bar E^{(\prime)0}),
\non\\
\delta A^{(\prime)}&=&\frac{1}{3}i(3 r'_0-2 r'_e +5 r'_a  )A^{(\prime)0}
-\frac{1}{3}i(r'_e+2r'_a) T^{(\prime)0}+ir'_a C^{(\prime)0}
\non\\
&&+\frac{1}{3}i(\bar r'_e+2\bar r'_a)(\bar T^{(\prime)0}+2\bar A^{(\prime)0}),
\non\\
\delta P^{(\prime)}&=&ir'_0 P^{(\prime)0}
+i r'_a T^{(\prime)0}
-\frac{1}{3}i( r'_e +2 r'_a) C^{(\prime)0}
+\frac{1}{3}i(-2 r'_e +5 r'_a) P^{(\prime)0}
\non\\
&&-\frac{1}{3} i r'_a P^{(\prime)0}_{EW}
+\frac{1}{9}i(r'_e +2 r'_a) P_{EW}^{(\prime)C0}
+ \frac{1}{3}i(\bar r'_e+2\bar r'_a)(\bar C^{(\prime)0}_2+2\bar P^{(\prime)0}-\frac{1}{3}\bar P^{(\prime)C0}_{EW,2}),
\non\\
\delta PA^{(\prime)}&=& 
\frac{1}{3}i(3r'_0
- r'_e +16 r'_a +12 r'_t) PA^{(\prime)0}
+ i r'_t T^{(\prime)0}
+\frac{1}{9}(2 i r'_e +4 i r'_a -3 i r'_t) C^{(\prime)0}
\non\\
&&
+\frac{2}{9}(i r'_e+11 ir'_a+12  ir'_t) E^{(\prime)0}
+\frac{2}{9}( ir'_e+11 ir'_a +12 ir'_t ) P^{(\prime)0}
-\frac{1}{3} i r'_t P^{(\prime)0}_{EW}
\non\\
&&
+\frac{1}{27}i(-2  r'_e -4 r'_a +3 r'_t) P^{(\prime)C}_{EW}
-\frac{2}{27} i (r'_e +11 r'_a +12 r'_t )P_{EW}^{(\prime)E0}
\non\\
&&-\frac{2}{9}i(\bar r'_e+2\bar r'_a)
\bigg(\bar C^{(\prime)0}_1+2\bar E^{(\prime)0}+\bar C^{(\prime)0}_2+2\bar P^{(\prime)0}-\frac{1}{3}\bar P^{(\prime)C0}_{EW,2}
-\frac{1}{3}\bar P^{(\prime)C0}_{EW,1} -\frac{2}{3}\bar P^{(\prime)E0}_{EW}\bigg)
\non\\
&&+\frac{1}{3}i(\hat r'_t +\frac{4\hat r'_a +2\hat r'_e}{3})
\bigg(\tilde C^{(\prime)0}+\tilde E^{(\prime)0}+\tilde P^{(\prime)0} +\frac{3}{2}\widetilde {PA}^{(\prime)0}-\frac{1}{3}\tilde P^{(\prime)C0}_{EW} -\frac{1}{3}\tilde P^{(\prime)E0}_{EW}\bigg),
\non\\
\delta P^{(\prime)}_{EW}&=&ir'_0 P^{(\prime)0}_{EW}+i r'_e P_{EW}^{(\prime)C0},
\non\\
\delta P_{EW}^{(\prime)C}&=&ir'_0 P_{EW}^{(\prime)C0}+i r'_e P_{EW}^{(\prime)0},
\non\\
\delta P_{EW}^{(\prime)E}&=&ir'_0 P_{EW}^{(\prime)E0}+ ir'_a P^{(\prime)0}_{EW}
-\frac{1}{3} i (r'_e+2 r'_a) P_{EW}^{(\prime)C0}
+\frac{1}{3}i(-2r'_e +5 r'_a) P_{EW}^{(\prime)E0}
\non\\
&&+\frac{1}{3}i(\bar r'_e+2\bar r'_a)(\bar P^{(\prime)C0}_{EW,1}  +2\bar P^{(\prime)E0}_{EW}),
\non\\
\delta P^{(\prime)A}_{EW}&=&
\frac{1}{3} i (3r'_0-2 r'_e +5r'_a) P_{EW}^{(\prime)A0}
-\frac{1}{3}i(r'_e +2r'_a) P^{(\prime)0}_{EW}
+ir'_a P^{(\prime)C0}_{EW}
\non\\
&&+\frac{1}{3}i(\bar r'_e+2\bar r'_a)(\bar P^{(\prime)0}_{EW} +2\bar P^{(\prime)A0}_{EW}),
\label{eq: delta TA1}
\en
\be
\delta (\bar T^{(\prime)}+2\bar A^{(\prime)})
&=&i(\bar r'_e+2\bar r'_a) (-\frac{2}{3}T^{(\prime)0}+C^{(\prime)0}+\frac{5}{3} A^{(\prime)0})
\non\\
&&
+i(\tilde r'_0 +\frac{4\tilde r'_a +2\tilde r'_e}{3})(\bar T^{(\prime)0}+2\bar A^{(\prime)0}),
\non\\
\delta (\bar C^{(\prime)}_1+2\bar E^{(\prime)})
&=&i(\bar r'_e+2\bar r'_a) (T^{(\prime)0} -\frac{2}{3} C^{(\prime)0}+\frac{5}{3} E^{(\prime)0})
\non\\
&&
+i(\tilde r'_0 +\frac{4\tilde r'_a +2\tilde r'_e}{3})(\bar C^{(\prime)0}_1+2\bar E^{(\prime)0}),
\non\\
\delta (\bar C^{(\prime)}_2+2\bar P^{(\prime)}-\frac{1}{3}\bar P^{(\prime)C}_{EW,2})
&=&i(\bar r'_e+2\bar r'_a) (T^{(\prime)0}-\frac{2}{3}  C^{(\prime)0}+\frac{5}{3} P^{(\prime)0}-\frac{1}{3} P^{(\prime)0}_{EW}+\frac{2}{9} P^{(\prime)C0}_{EW}
\non\\
&&+i(\tilde r'_0 +\frac{4\tilde r'_a +2\tilde r'_e}{3})
(\bar C^{(\prime)0}_2+2\bar P^{(\prime)0}-\frac{1}{3}\bar P^{(\prime)C0}_{EW,2}),
\non\\
\delta (\bar P^{(\prime)}_{EW} +2\bar P^{(\prime)A}_{EW})&=& 
i(\bar r'_e+2\bar r'_a)(-\frac{2}{3} P^{(\prime)0}_{EW}+ P^{(\prime)C0}_{EW}+\frac{5}{3} P^{(\prime)A0}_{EW})
\non\\
&&+i(\tilde r'_0 +\frac{4\tilde r'_a +2\tilde r'_e}{3})
(\bar P^{(\prime)0}_{EW} +2\bar P^{(\prime)A0}_{EW}),
\non\\
\delta (\bar P^{(\prime)C}_{EW,1}+2\bar P^{(\prime)E}_{EW})
&=&i(\bar r'_e+2\bar r'_a)  (P^{(\prime)0}_{EW}-\frac{2}{3} P^{(\prime)C0}_{EW}+\frac{5}{3} P^{(\prime)E0}_{EW})
\non\\
&&+i(\tilde r'_0 +\frac{4\tilde r'_a +2\tilde r'_e}{3})
(\bar P^{(\prime)C0}_{EW,1}  +2\bar P^{(\prime)E0}_{EW}),
\label{eq: delta TA2}
\en
and
\be
&&\delta(\tilde C^{(\prime)}+\tilde E^{(\prime)}+\tilde P+\frac{3}{2}\widetilde {PA}^{(\prime)}-\frac{1}{3}\tilde P^{(\prime)C}_{EW} -\frac{1}{3}\tilde P^{(\prime)E}_{EW})
\non\\
&&\quad
=i\left(\hat r'_t +\frac{4\hat r'_a +2\hat r'_e}{3}\right) 
\bigg(\frac{3}{2}T^{(\prime)0}-\frac{1}{2} C^{(\prime)0}+4 E^{(\prime)0}+4 P^{(\prime)0}+6 PA^{(\prime)0}
\non\\
&&\qquad
-\frac{1}{2} P^0_{EW}
+\frac{1}{6} (P^C_{EW})^0
-\frac{4}{3} (P_{EW}^E)^0\bigg)
\non\\
&&\qquad
+i\left(\check r'_0+\frac{4\check r'_a+2\check r'_e+3\check r'_t}{6}\right)
\bigg(\tilde C^{(\prime)0}+\tilde E^{(\prime)0}+\tilde P^{(\prime)0}
+\frac{3}{2}\widetilde {PA}^{(\prime)0}
-\frac{1}{3}\tilde P^{(\prime)C0}_{EW} -\frac{1}{3}\tilde P^{(\prime)E0}_{EW}\bigg),
\non\\
\label{eq: delta TA3}
\en
where the superscript 0 denote un-scattered amplitudes
and we define
$r'_i\equiv r^{(1/2)}_i$, 
$\bar r'_i\equiv \bar r^{(1/2)}_i$, 
$\hat r'_i\equiv \hat r^{(1/2)}_i$, 
$\check r'_i\equiv \check r^{(1/2)}_i$, 
$\tilde r'_i\equiv \tilde r^{(1/2)}_i$.

The full topological amplitudes contain the un-scattered and the contribution from the scattering. 
For example, for the tree amplitude the full amplitude is $T^{(\prime)}$, the un-scattered tree amplitude is $T^{(\prime)0}$. 
After scattering we have 
\be
T^{(\prime)}=T^{(\prime)0}+\delta T^{(\prime)}=T^{(\prime)0}+ir'_0 T^{(\prime)0}+i r'_e C^{(\prime)0}.
\en
One can check that the above equations are consistent with the topological amplitude expressions
Eqs.~(\ref{eq: TAgroup1}), (\ref{eq: TAgroup2}), (\ref{eq: TAgroup3}), (\ref{eq: TAgroup4}), (\ref{eq: TABsgroup1}) and (\ref{eq: TABsgroup4}), and the rescattering formulas, Eqs.~(\ref{eq:FSIB0Kpi}), (\ref{eq:FSIBKpi0}), (\ref{eq:FSIBpipi0}), (\ref{eq:FSIB0pipi}) and those of $B_s$ decays.
It should be pointed out that this is a non-trivial check, as one can see that Eqs.~(\ref{eq: TAgroup1}), (\ref{eq: TAgroup2}), (\ref{eq: TAgroup3}), (\ref{eq: TAgroup4}), (\ref{eq: TABsgroup1}) and (\ref{eq: TABsgroup4}) are rather complicate and a single error in them can easily spoil the consistency check.

Note that decay amplitudes can be expressed in terms of several combinations of topological amplitudes, such as $T+C$, $C-E$ and so on, 
and FSI affects these combinations only through,
\be
 1+i(r'_0+r'_a),
 \quad
 i(r'_e-r'_a),
  \quad
 i (r'_a+r'_t),
 \quad
 i(2\bar r'_a+\bar r'_e),
     \nonumber\\
 1+i(\tilde r'_0+\frac{4\tilde r'_a+2\tilde r'_e}{3}),
 \quad
 i(\hat r'_t+\frac{4\hat r'_a+2\hat r'_e}{3}).
\label{eq: combinations of r}  
\en
We have
\be
&&T^{(\prime)}+C^{(\prime)}
=[(1+ir'_0+ir'_a)+i(r'_e-r'_a)](T^{(\prime)0}+C^{(\prime)0})=e^{i\delta_{27}}(T^{(\prime)0}+C^{(\prime)0}),
\non\\
&&C^{(\prime)}-E^{(\prime)}
=(1+ir'_0+ir'_a)(C^{(\prime)0}-E^{(\prime)0})+\frac{1}{3}i(r'_e-r'_a)[3(T^{(\prime)0}+C^{(\prime)0})-2(C^{(\prime)0}-E^{(\prime)0})]
\non\\
&&\qquad\qquad\qquad
-\frac{1}{3}i(\bar r'_e+2\bar r'_a)(\bar C_1^{(\prime)0}+2\bar E^{(\prime)0}),
\non\\
&&A^{(\prime)}+C^{(\prime)}
=(1+ir'_0+ir'_a)(A^{(\prime)0}+C^{(\prime)0})+\frac{2}{3}i(r'_e-r'_a)[(T^{(\prime)0}+C^{(\prime)0})-(A^{(\prime)0}+C^{(\prime)0})]
\non\\
&&\qquad\qquad\qquad
+\frac{1}{3}i(\bar r'_e+2\bar r'_a)(\bar T^{(\prime)0}+2\bar A^{(\prime)0}),
\non\\
&&P^{(\prime)}-C^{(\prime)}+\frac{1}{3}P^{(\prime)C}_{EW}
=\bigg[(1+ir'_0+ir'_a)-\frac{2}{3}i(r'_e-r'_a)\bigg] \bigg(P^{(\prime)0}-C^{(\prime)0}+\frac{1}{3}P^{(\prime)C0}_{EW}\bigg)
\non\\
&&\qquad\qquad\qquad\qquad\qquad
+\frac{1}{3}i(r'_e-r'_a)
[-3 (T^{(\prime)0} + C^{(\prime)0})+ (P_{EW}^{(\prime)0} + P_{EW}^{(\prime)C0})]
\non\\
&&\qquad\qquad\qquad\qquad\qquad
+\frac{1}{3}i(\bar r'_e+2\bar r'_a)(\bar C^{(\prime)0}_2+2\bar P^{(\prime)0}-\frac{1}{3}\bar P^{(\prime)C0}_{EW,2}),
\non\\
&&PA^{(\prime)}-\frac{4}{9} C^{(\prime)}+\frac{13}{9} E^{(\prime)}-\frac{1}{3} P_{EW}^{(\prime)C}
\non\\
&&\quad
=(1+ir'_0+ir'_a-\frac{1}{3} i(r'_e-r'_a)+4 i(r'_a+r'_t) ) \bigg(PA^{(\prime)0}-\frac{4}{9} C^{(\prime)0}+\frac{13}{9} E^{(\prime)0}-\frac{1}{3} P_{EW}^{(\prime)C}\bigg)
\non\\
&&\qquad
+\bigg[-\frac{4}{9} i(r'_e-r'_a)+i(r'_a+r'_t)\bigg] (T^{(\prime)0} + C^{(\prime)0}) 
\non\\
&&\qquad
+\frac{7}{9}\bigg[\frac{1}{3} i(r'_e-r'_a)+ 4 i(r'_a+r'_t) \bigg](C^{(\prime)0} - E^{(\prime)0})
\non\\
&&\qquad
+\bigg[\frac{2}{9} i(r'_e-r'_a)+\frac{8}{3} i(r'_a+r'_t) \bigg]
\bigg(P^{(\prime)0} - C^{(\prime)0} + \frac{1}{3} P_{EW}^{(\prime)C0}\bigg) 
\non\\
&&\qquad
-\frac{1}{3} [i(r'_e-r'_a)+i(r'_a+r'_t)] 
(P_{EW}^{(\prime)0} + P_{EW}^{(\prime)C0})
\non\\
&&\qquad
+\bigg[\frac{2}{27} i(r'_e-r'_a)+\frac{8}{9} i(r'_a+r'_t) \bigg]
(P_{EW}^{(\prime)C0} - P_{EW}^{(\prime)E0})
\non\\
&&\qquad
+\frac{1}{27} i(\bar r'_e+2\bar r'_a) \bigg[7 (\bar C_1^{(\prime)0} + 2 \bar E^{(\prime)0}) - 
   6 \bigg( \bar C_2^{(\prime)0} + 2\bar P^{(\prime)0} - \frac{1}{3} \bar P_{EW,2}^{(\prime)C0}\bigg)
   + 2 (\bar P_{EW,1}^{(\prime)C0} + 2 \bar P_{EW}^{(\prime)E0})\bigg]
\non\\   
&&\qquad
+\frac{1}{3}i(\hat r'_t +\frac{4\hat r'_a +2\hat r'_e}{3})
(\tilde C^{(\prime)0}+\tilde E^{(\prime)0}+\tilde P^{(\prime)0} +\frac{3}{2}\widetilde {PA}^{(\prime)0}-\frac{1}{3}\tilde P^{(\prime)C0}_{EW} -\frac{1}{3}\tilde P^{(\prime)E0}_{EW}),   
\non\\ 
&&P^{(\prime)}_{EW}+P^{(\prime)C}_{EW}
=[(1+ir'_0+ir'_a)+i(r'_e-r'_a)](P^{(\prime)0}_{EW}+P^{(\prime)C0}_{EW})=e^{i\delta_{27}}(P^{(\prime)0}_{EW}+P^{(\prime)C0}_{EW}),
\non\\
&&P_{EW}^{(\prime)C}-P_{EW}^{(\prime)E}
=(1+ir'_0+ir'_a)(P_{EW}^{(\prime)C0}-P_{EW}^{(\prime)E0})
\non\\
&&\qquad\qquad\qquad\quad
+\frac{1}{3}i(r'_e-r'_a)[3(P_{EW}^{(\prime)0}+P_{EW}^{(\prime)C0})
-2(P_{EW}^{(\prime)C0}-P_{EW}^{(\prime)E0})]
\non\\
&&\qquad\qquad\qquad\quad
-\frac{1}{3}i(\bar r'_e+2\bar r'_a)(\bar P_{EW,1}^{(\prime)C0}+2\bar P_{EW}^{(\prime)E0}),
\non\\
&&P_{EW}^{(\prime)A}+P_{EW}^{(\prime)C}
=(1+ir'_0+ir'_a)(P_{EW}^{(\prime)A0}+P_{EW}^{(\prime)C0})
\non\\
&&\qquad\qquad\qquad\quad
+\frac{2}{3}i(r'_e-r'_a)[(P_{EW}^{(\prime)0}+P_{EW}^{(\prime)C0})-(P_{EW}^{(\prime)A0}+P_{EW}^{(\prime)C0})]
\non\\
&&\qquad\qquad\qquad\quad
+\frac{1}{3}i(\bar r'_e+2\bar r'_a)(\bar P_{EW}^{(\prime)0}+2\bar P_{EW}^{(\prime)A0}),
\label{eq: rcombinations1}
\en
\be
&&(\bar T^{(\prime)}+2\bar A^{(\prime)})
=
\bigg[1+i\bigg(\tilde r'_0 +\frac{4\tilde r'_a +2\tilde r'_e}{3}\bigg)\bigg](\bar T^{(\prime)0}+2\bar A^{(\prime)0})
\non\\
&&\qquad\qquad\qquad\quad
+i(\bar r'_e+2\bar r'_a) \bigg[-\frac{2}{3}(T^{(\prime)0}+C^{(\prime)0})+\frac{5}{3} (A^{(\prime)0}+C^{(\prime)0})\bigg],
\non\\
&&(\bar C^{(\prime)}_1+2\bar E^{(\prime)})
=\bigg[1+i\bigg(\tilde r'_0 +\frac{4\tilde r'_a +2\tilde r'_e}{3}\bigg)\bigg](\bar C^{(\prime)0}_1+2\bar E^{(\prime)0})
\non\\
&&\qquad\qquad\qquad\quad
+i(\bar r'_e+2\bar r'_a) \bigg[(T^{(\prime)0}+C^{(\prime)0})-\frac{5}{3}(C^{(\prime)0}- E^{(\prime)0})\bigg],
\non\\
&&
\bar C^{(\prime)}_2+2\bar P^{(\prime)}-\frac{1}{3}\bar P^{(\prime)C}_{EW,2}
=\bigg[1+i(\tilde r'_0 +\frac{4\tilde r'_a +2\tilde r'_e}{3}\bigg)\bigg]
(\bar C^{(\prime)0}_2+2\bar P^{(\prime)0}-\frac{1}{3}\bar P^{(\prime)C0}_{EW,2})
\non\\
&&\qquad\qquad\qquad\qquad\qquad\quad
+i(\bar r'_e+2\bar r'_a) \bigg[(T^{(\prime)0}+C^{(\prime)0})]
+\frac{5}{3} \bigg(P^{(\prime)0}-C^{(\prime)0}+\frac{1}{3}P^{(\prime)C0}_{EW}\bigg)
\non\\
&&\qquad\qquad\qquad\qquad\qquad\quad
-\frac{1}{3} (P^{(\prime)0}_{EW}+P^{(\prime)C0}_{EW})\bigg],
\non\\
&&(\bar P^{(\prime)}_{EW} +2\bar P^{(\prime)A}_{EW})
=
\bigg[1+i\bigg(\tilde r'_0 +\frac{4\tilde r'_a +2\tilde r'_e}{3}\bigg)\bigg]
(\bar P^{(\prime)0}_{EW} +2\bar P^{(\prime)A0}_{EW})
\non\\
&&\qquad\qquad\qquad\qquad
+i(\bar r'_e+2\bar r'_a)\bigg[-\frac{2}{3} (P^{(\prime)0}_{EW}+P^{(\prime)C0}_{EW})+\frac{5}{3} (P^{(\prime)A0}_{EW}+P^{(\prime)C0}_{EW})\bigg],
\non\\
&&(\bar P^{(\prime)C}_{EW,1}+2\bar P^{(\prime)E}_{EW})
=
\bigg[1+i\bigg(\tilde r'_0 +\frac{4\tilde r'_a +2\tilde r'_e}{3}\bigg)\bigg]
(\bar P^{(\prime)C0}_{EW,1}  +2\bar P^{(\prime)E0}_{EW})
\non\\
&&\qquad\qquad\qquad\qquad
+
i(\bar r'_e+2\bar r'_a) \bigg[(P^{(\prime)0}_{EW}+P^{(\prime)C0}_{EW})-\frac{5}{3} (P^{(\prime)C0}_{EW}-P^{(\prime)E0}_{EW})\bigg],
\label{eq: rcombinations2}
\en
and
\be
&&(\tilde C^{(\prime)}+\tilde E^{(\prime)}+\tilde P^{(\prime)}+\frac{3}{2}\widetilde {PA}^{(\prime)}-\frac{1}{3}\tilde P^{(\prime)C}_{EW} -\frac{1}{3}\tilde P^{(\prime)E}_{EW})
\non\\
&&\quad
=i\left(\hat r'_t +\frac{4\hat r'_a +2\hat r'_e}{3}\right) 
\bigg[\frac{3}{2}(T^{(\prime)0}+C^{(\prime)0})+\frac{14}{3}(C^{(\prime)0}-E^{(\prime)0})+4 \bigg(P^{(\prime)0}-C^{(\prime)0}+\frac{1}{3}P_{EW}^{(\prime)C0}\bigg)
\non\\
&&\qquad
+6 \bigg(PA^{(\prime)0}-\frac{4}{9} C^{(\prime)0}+\frac{13}{9} E^{(\prime)0}-\frac{1}{3} P_{EW}^{(\prime)C0}\bigg)
-\frac{1}{2} (P^{(\prime)0}_{EW}+P_{EW}^{(\prime)C0})
+\frac{4}{3}(P^{(\prime)C0}_{EW}-P_{EW}^{(\prime)E0})\bigg]
\non\\
&&\qquad
\left[1+i\left(\check r'_0+\frac{4\check r'_a+2\check r'_e+3\check r'_t}{6}\right)\right]
\bigg(\tilde C^{(\prime)0}+\tilde E^{(\prime)0}+\tilde P^{(\prime)0}
+\frac{3}{2}\widetilde {PA}^{(\prime)0}
-\frac{1}{3}\tilde P^{(\prime)C0}_{EW} -\frac{1}{3}\tilde P^{(\prime)E0}_{EW}\bigg).
\non\\
\label{eq: rcombinations3}
\en
With the help of Eq. (\ref{eq:solution}) (with $m=1/2$)
we will be able to study the effect of rescattering to the above combinations 
and give a clearer picture.
Note that the above transformation formulas of the combined topological amplitudes,  in Eqs. (\ref{eq: rcombinations1}), (\ref{eq: rcombinations2}) and (\ref{eq: rcombinations3}) are not as powerful compared to transformation formulas of the individual topological amplitudes, Eqs.~(\ref{eq: delta TA1}), (\ref{eq: delta TA2}) and (\ref{eq: delta TA3}). 
They are, however, the ones that can have in terms of the combinations of $r'_i$ [Eq. (\ref{eq: combinations of r})] and hence the rescattering angles and phases, $\tau$, $\nu$, $\sigma$ and $\delta$ [see Eqs. (\ref{eq:solution}) and (\ref{eq:FSIparameters})], without introducing additional assumption.

\subsubsection{Topological Amplitudes and rescattering in the U(3) limit}

It is interesting to investigate the above relations in the U(3) limit, where we take Eq.~(\ref{eq: U3r}) and
\be
\bar T=\tilde T=T,
\quad
\bar C_{1}=\bar C_{2}=\tilde C=C,
\quad
\bar E=\tilde E=E,
\quad
\bar A=\tilde A=A,
\non\\
\bar P=\tilde P=P,
\quad
\bar P_{EW}=\tilde P_{EW}=P_{EW},
\quad
\bar P^C_{EW,1}=\bar P^C_{EW,2}=\tilde P^C_{EW}=P^C_{EW},
\non\\
\bar P^E_{EW}=\tilde P^E_{EW}=P^E_{EW},
\quad
\bar P^A_{EW}=\tilde P^A_{EW}=P^A_{EW},
\quad
\quad
\bar {PA}=\widetilde {PA}=PA.
\label{eq: U3TA}
\en
Using Eq. (\ref{eq: U3r}) and Eqs. (\ref{eq: delta TA1}), (\ref{eq: delta TA2}) and (\ref{eq: delta TA3}), we find that
\be
\delta (\bar T+2\bar A)-\delta (T+2A)
&=&3i r'_e A^0,
\non\\
\delta (\bar C_1+2\bar E)-\delta (C+2E)
&=&3i r'_e E^0,
\non\\
\delta (\bar C_2+2\bar P-\frac{1}{3}\bar P^C_{EW,2})
-\delta (C+2P-\frac{1}{3}P^C_{EW})
&=&3i r'_e P^0,
\non\\
\delta (\bar P_{EW} +2\bar P^A_{EW})
-\delta (P_{EW} +2P^A_{EW})
&=& 3i r'_e P^{A0}_{EW},
\non\\
\delta (\bar P^C_{EW,1}+2\bar P^E_{EW})
-\delta (P^C_{EW}+2P^E_{EW})
&=& 3i r'_e P^{E0}_{EW},
\label{eq: U3 cal1}
\en
and
\be
&&\delta(\tilde C+\tilde E+\tilde P +\frac{3}{2}\widetilde {PA}-\frac{1}{3}\tilde P^C_{EW} -\frac{1}{3}\tilde P^E_{EW})
-\delta(C+E+P +\frac{3}{2}{PA}-\frac{1}{3}P^C_{EW}-\frac{1}{3}P^E_{EW})
\non\\
&&\quad
=\frac{1}{2} i r'_e (6 E^0 + 6 P^0 + 9 PA^0 - 2 P_{EW}^{E0}).
\label{eq: U3 cal2}
\en
The above relations can be consistent with the relations in the U(3) limit, Eq.~(\ref{eq: U3TA}), only if we take
\be
r'_e=0.
\label{eq: re=0}
\en
It is useful to recall that by requiring U(3) symmetry to the rescattering matrix $\T$ [Eq. (\ref{eq: U3r})] one only leads to $r'_e r'_a=0$ [see Eq. (\ref{eq: rare=0})], which can either be $r'_e=0$ or $r'_a=0$. Now we can select out the $r'_e=0$ solution. 
The reason of being more specify here is that we now apply U(3) symmetry to both rescattering matrix $\T$ [Eq. (\ref{eq: U3r})] and to the topological amplitudes [Eq. (\ref{eq: U3TA})].
Hence it leads to a more specify relation.

\section{Numerical Results}

In this section, we will present our numerical results. 
First, we will give an overview of the results of the fits. 
We will then discuss the rescattering effects on topological amplitudes. 
Finally, numerical results for decay rates and CP asymmetries will be shown.

\subsection{Overview of the Results of the Fits}

Before present our numerical results, we specify the inputs used 
in the following numerical study.
Masses of all particles and total widths of $B_{u,d,s}$ mesons are taken from the
review of the Particle Data Group (PDG)~\cite{PDG}
and the branching ratios of $B$ to
charmless meson decays are taken
from the latest averages in~\cite{HFAG}. 

For theoretical inputs, 
we use $f_\pi=$ 130.2 MeV, $f_{K}=$ 155.6 MeV and
$f_{B_{(s)}}=$ 187.1 (227.2) MeV for decay constants 
and $m_s(2{\rm GeV})=93.5$ MeV for the strange quark mass, which is taken from the central value of the PDG averaged value, $93.5\pm 2$ MeV~\cite{PDG}.~\footnote{Note that in the previous study~\cite{Chua:2007cm} $m_s$ is taken as a fit parameter in the range of $100\pm30$ MeV, but now as the value becomes more precisely known it is better to use the present central value as an input parameter.~\label{footnote4}}
The values of CKM matrix elements, except $\gamma/\phi_3$, are also taken from the central
values of the latest PDG's results~\cite{PDG}.
We use the QCD factorization calculated amplitudes
\cite{Beneke:2003zv} for the factorization amplitudes in the
right-hand-side of Eq.~(\ref{eq:master1}). We take the
renormalization scale as $\mu=4.2$~GeV and the power correction
parameters $X_{A,H}=\ln(m_B/\Lambda_h)(1+\rho_{A,H}
e^{i\phi_{A,H}} )$. 
For meson wave functions, we use the following Gegenbauer moments: $\alpha_1^{\bar K}=-\alpha_1^K=0.2$, $\alpha_2^{\bar K}=\alpha_2^K=0.1$,
$\alpha_1^\pi=0$, $\alpha_2^\pi=0.2$ and $\alpha_{1,2}^{\eta,\eta'}=0$~\cite{Beneke:2003zv}.
Several hadronic parameters, in additional to the CKM phase $\gamma/\phi_3$, $\rho_{A,H}$ and $\phi_{A,H}$, in factorization
amplitudes are fit parameters and
are allowed to vary in the following ranges:
 \be
 &&F^{B\pi}_0(0)=0.25\pm0.05,
 \quad
 F^{BK}_0(0)=0.35\pm0.08,
 \quad
 F^{B_sK}(0)=0.31\pm0.08,
 \nonumber\\
 &&\gamma/\phi_3=(73.2\pm 10)^\circ,
 \quad
 \lambda_B=0.35\pm 0.25\, {\rm GeV},
 \quad
 \lambda_{B_s}=0.35\pm 0.25\, {\rm GeV}.
 \label{eq:QCDFHparameters}
 \en
These
estimations agree with those in \cite{Beneke:2003zv,LF,MS,Ball:2004ye,Duplancic:2008ix,Khodjamirian:2011ub,Bharucha:2012wy}, while
the ranges of form factors and $\gamma/\phi_3$ are slightly enlarged.
For example, the above $F^{B\pi}_0(0)$ can be compared to the following reported values for the quantity: 
$0.28\pm0.05$ \cite{Beneke:2003zv},
$0.25$ \cite{LF},
$0.29$ \cite{MS},
$0.258\pm0.031$ \cite{Ball:2004ye},
$0.26^{+0.04}_{-0.03}$ \cite{Duplancic:2008ix},
$0.281^{+0.027}_{-0.029}$ \cite{Khodjamirian:2011ub}
and
$0.261^{+0.020}_{-0.023}$ \cite{Bharucha:2012wy}.~\footnote{It is preferable to use the form factors as inputs instead of variables in the fit, but in the present situation no definite values for these form factors can be found (see for example the collected $F^{B\pi}_0(0)$ values from \cite{Beneke:2003zv,LF,MS,Ball:2004ye,Duplancic:2008ix,Khodjamirian:2011ub,Bharucha:2012wy}) and we therefore treat them as fitting variables to avoid bias in this work. Hopefully the situation can be improved in future. See also Footnote~\ref{footnote4}.} 

It is known that semileptonic $B\to\pi l\nu$ decays are related to the $B\to \pi$ form factor and the determination of $|V_{ub}|$~\cite{PDG}.
Using data from BaBar~\cite{delAmoSanchez:2010af,Lees:2012vv} and Belle \cite{Ha:2010rf,Sibidanov:2013rkk}, HFLAG obtain the following result in 2014:~\cite{HFAG}
\be
F^{B\pi}_0(0)|V_{ub}|=(9.23\pm0.24)\times 10^{-4},
\label{eq: SL}
\en
We will use this in our $\chi^2$ analysis.

In summary, 9 hadronic parameters,
$\rho_{A,H}$, $\phi_{A,H}$, $F^{B\pi}_0(0)$, $F^{BK}_0(0)$, $F^{B_sK}(0)$, $ \lambda_B$, $ \lambda_{B_s}$,
and one CKM phase, $\gamma/\phi_3$, involved in the QCDF amplitudes will be fitted from data.
The residue rescattering part add 4 more parameters,
$ \tau$, $\nu$, $\delta$ and $\sigma$, giving 14 parameters in total.
Note that the majority of the fitted parameters are from the factorization part.

In this analysis there are totally 93 measurable  quantities, including 34 rates, 34 direct CP asymmetries, 24 mixing induced CP asymmetries and one measurement from semileptonic $B$ decay [Eq. (\ref{eq: SL})]. 
Among them we will fit to all available data, including 26 rates, 16 direct CP asymmetries, 5 mixing induced CP asymmetries and 1 semileptonic decay data, 
giving 48 in total, 
and will have prediction on 8 rates, 18 direct CP asymmetries and 19 mixing induced CP asymmetries. 
The explicit list of these 48 items will be shown later.
The total numbers of data in fit and in predictions are roughly the same.
The summary of these numbers is shown in Table~\ref{tab:overview}.


\begin{table}[t!]
\caption{ 
Numbers of rates $\B$, direct CP asymmetries $\A$ and mixing induced CP asymmetries $S$ of $\overline B_q\to PP$ decays involved in this study.
 \label{tab:overview}
}
\begin{ruledtabular}
{\footnotesize
\begin{tabular}{cccccc}
 
 &number of $\B$
 &number of $\A$
 &number of $S$
 & number of $SL$
 & Total number
 \\
  \hline
   All 
  & 34
  & 34
  & 24
  & 1
  &93
 \\
   Fitted
  & 26
  & 16
  & 5
  & 1
  & 48
  \\
  Predicted
  & 8
  & 18
  & 19
  & 0
  & 45
  \end{tabular}
  }
\end{ruledtabular}
\end{table}

We perform a $\chi^2$ analysis with all available data on
CP-averaged rates and CP asymmetries in $\ov B{}_{u,d,s}\to PP$
decays.
In the following study we use two different scenarios: Fac and Res. 
For the formal we use only factorization amplitudes (i.e. $A_i=A_i^{\rm fac}$), 
while for the latter we add residue FSI effect as well (i.e. $A_i=\sum_{j=1}^n(\Sc_{res}^{1/2})_{ij} A^{\rm fac}_j$). 
Both are fitted to data. 
The confidence levels and $\chi^2$s for the best fitted cases in both senarios are shown
in Table~\ref{tab:chisquare}. 
Contributions to $\chi^2_{\rm min.}$
from various sub-sets of data are also given. 
Modes that are related through the Res are grouped together [see
Eq.~(\ref{eq:FSIB0Kpi}), and see Eqs.
(\ref{eq:FSIBKpi0})--(\ref{eq:FSIB0pipi}) for other groups]. 
Off course only those with data can contribute to $\chi^2$.
Numbers of data used are shown in parentheses.
Explicitly,
$\chi^2_{\{\B(\ov B{}^0\to K\pi),\dots\}}$ and $\chi^2_{\{\A(\ov B{}^0\to K\pi),\dots\}}$ in the table denote
the $\chi^2$ contribution obtained from 4 CP-average rates and 3 direct CP asymmetries, respectively, of the group-1 modes consisting of $\ov
B{}^0\to K^-\pi^+,\, \ov K {}^0\pi^0,\,\ov K{}^0\eta,\,\ov
K{}^0\eta'$ decays  (except $\A(\ov B{}^0\to \ov K{}^0\eta$)); 
$\chi^2_{\{\B(B^-\to K\pi),\dots\}}$ and $\chi^2_{\{\A(B^-\to K\pi),\dots\}}$ 
are contributed from the group-2 modes: $B^-\to \ov K{}^0\pi^-,\, K^-\pi^0,\,K^-\eta,\,K^-\eta'$ decays; 
$\chi^2_{\{\B(B^-\to \pi\pi),\dots\}}$ and $\chi^2_{\{\A(B^-\to \pi\pi),\dots\}}$ are contributed from the group-3 modes: 
$B^-\to \pi^-\pi^0,\, K^-K^0,\,\pi^-\eta,\,\pi^-\eta'$ decays;
$\chi^2_{\{\B(\ov B{}^0\to \pi\pi),\dots\}}$ is contributed from the group-4 modes: 
$\ov B{}^0\to \pi^+\pi^-,\,\pi^0\pi^0,\, \eta\eta,\,\eta\eta',\,\eta'\eta',K^+K^-,\,\ov K{}^0 K^0,\,\pi^0\eta,\,\pi^0\eta'$
decays, while $\chi^2_{\{\A(\ov B{}^0\to \pi\pi),\dots\}}$ only contributed from 3 of the above modes, $\ov B{}^0\to \pi^+\pi^-,\,\pi^0\pi^0,\, \ov K{}^0 K^0$ decays;
$\chi^2_{\{\B(\ov B_s),\A(\ov B_s)\}}$ is contributed from 5 CP-averaged rates in 
$\ov B{}^0_s\to K^+\pi^-,\,\pi^+\pi^-,\,\eta'\eta',\,K^+K^-,\, K^0\ov K{}^0$ decays and 
from 2 direct CP asymmetries in 
$\ov B{}^0_s\to K^+\pi^-,\,K^+K^-$ decays;
$\chi^2_{\{S(\ov B {}^0)),\,S(\ov B {}^0_s))\}}$ is contributed from mixing induced CP asymmetries 
of $\ov B{}^0\to \ov K{}^0\pi^0, \,\ov K{}^0\eta',\,\pi^+\pi^-,\,K_S\ov K_S$ and $\ov B{}_s^0\to K^+K^-$ decays. 
The semiloptonic data, Eq. (\ref{eq: SL}) is also included in the fit.
The above lists are the 26 rates, 16 direct CP asymmetries, 5 mixing induced asymmetries and 1 semileptonic data [Eq. (\ref{eq: SL})], 48 in totally, that go into the fit.

\begin{table}[t!]
\caption{ Confidence level (C.L.), $\chi^2_{\rm min}/{\rm d.o.f.}$
and various contributions to $\chi^2_{\rm min}$ for the best
fitted solutions. The $p$-value of the rescattering (Res) case is 5.5\%.
Numbers of data used are shown in parentheses.
 \label{tab:chisquare}
}
\begin{ruledtabular}
{\footnotesize
\begin{tabular}{cccccc}
 
 &$\chi^2_{\rm min.}/{\rm d.o.f.}$
 &$\chi^2_{\{\B(\ov B{}^0\to K\pi),\dots\}}$
 &$\chi^2_{\{\A(\ov B{}^0 \to K\pi),\dots\}}$
 &$\chi^2_{\{\B(B^-\to K\pi),\dots\}}$
 &$\chi^2_{\{\A(B^- \to K\pi),\dots\}}$
 \\
  \hline
   Fac 
  & 213.4/38 (48)
  & 10.1 (4)
  & 1.8 (3)
  & 24.7 (4)
  & 5.2 (4)
 \\
   \hline
   Res
  & 48.1/34 (48)
  & 7.2 (4)
  & 1.1 (3)
  & 6.3 (4)
  & 0.6 (4)
 \\
 \hline
  & $\chi^2_{\{\B(B^-\to \pi\pi),\dots\}}$
  &$\chi^2_{\{\A(B^- \to \pi\pi),\dots\}}$
  &$\chi^2_{\{B(\ov B{}^0\to \pi\pi),\dots\}}$
  &$\chi^2_{\{\A(\ov B{}^0 \to \pi\pi),\dots\}}$
  &$\chi^2_{\{\B(\ov B_s),\A(\ov B_s)\}}$
   \\
   \hline
   Fac
  & 10.6 (4)
  & 6.5 (4)
  & 55.3 (9)
  & 15.7 (3)
  & 64.0 (7)
  \\
    \hline
    Res
  & 6.4 (4)
  & 7.5 (4)
  & 7.8 (9)
  & 4.7 (3)
  & 0.6 (7)
  \\
  \hline
  &$\chi^2_{\{S(\ov B {}^0)),\,S(\ov B {}^0_s))\}}$
  &$\chi^2_{SL}$
  \\
   \hline
   Fac
  & 12.9 (5)
  & 8.0 (1)
  \\
    \hline
    Res
  & 5.2 (5)  
  & 0.7 (1)
  \end{tabular}
  }
\end{ruledtabular}
\end{table}

Table~\ref{tab:chisquare} shows the overall performances of the fits.
We discuss the factorization case first.
The $\chi^2$ per degree of freedom of Fac is $213.4/(48-10)$.
One can compare the $\chi^2$ values and the numbers of data used in
the corresponding groups. When the ratio of $\chi^2$ and the number of data is smaller than one,
the fit in the group is reasonably well.
By inspecting the table, we see that Fac gives a good fit in the direct CP asymmetries of group-1 modes ($\ov B{}^0\to K^-\pi^+,\cdots$), and produces reasonable fits in the direct CP asymmetries of group-2 modes ($B^-\to \ov K{}^0\pi^-,\cdots$) and of group-3 modes ($B^-\to\pi^-\pi^0,\cdots$),
but the fits in rates and mixing induced CP asymmetries of all modes (including $B_s$ decay modes) and direct CP asymmetries of group-4 modes are poor. 
In particular, the ratios of $\chi^2$ per number of data used in rates of the group-2 modes ($B^-\to \ov K{}^0\pi^-,\cdots$), group-4 modes ($\ov B{}^0\to\pi^+\pi^-,\cdots$), in the rates and direct CP asymmetries of $B_s$ modes and in the semileptonic quantity
are as large as $24.7/4$, $55.3/9$, $64.0/7$ and $8.0/1$, respectively, indicating the badness of the fit in these sectors.

The fit is significant improved when the rescattering is added. 
In the best fitted case, the $\chi^2$ per degree of freedom of the fit is $48.1/(48-14)$ giving the $p$-value of $5.5\%$.
It should be noted that with 4 additional parameters the quality of the fit is improved significantly.
All $\chi^2$, except the direct CP of group-3 modes ($B^-\to\pi^-\pi^0,\cdots$), which is slightly enhanced,  are reduced. 
In particular, the $\chi^2$ per number of data of rates of the group-2 modes ($B^-\to \ov K{}^0\pi^-,\cdots$), group-4 modes ($\ov B{}^0\to\pi^+\pi^-,\cdots$), 
the rates and direct CP asymmetries of $B_s$ modes and in the semileptonic quantity are $6.3/4$, $7.8/9$, $0.6/7$ and $0.7/1$, respectively. 
The performance of the fit in these sector is improved significantly.
Detail results will be shown later.

\begin{table}[t!]
\caption{ Fitted hadronic and FSI parameters. Upper table contains
fitted parameters in factorization amplitudes (Fac), while the lower ones are parameters
in the rescattering (Res) case. 
 \label{tab:parameters1}
}
\begin{ruledtabular}
\begin{tabular}{ccccccccc}
 & $\rho_{A}$
 &$\rho_{H}$ 
 &$\phi_A({}^\circ)$
 &$\phi_H({}^\circ)$
 &$F_0^{B\pi}(0)$
 &$F_0^{BK}(0)$
 &$F_0^{B_sK}(0)$
 \\
  \hline
  Fac
  & $0.97_{-0.02}^{+0.01}$
  & $2.82_{-0.61}^{+0.20}$
  & $-28.4_{-0.1}^{+0.3}$
  & $-111.5_{-13.6}^{+4.4}$
  & $0.239\pm0.002$
  & $0.27_{-0.00}^{+0.00}$
  & $0.23_{-0.00}^{+0.00}$
  \\
    \hline
  Res
  & $2.87_{-0.03}^{+0.02}$
  & $2.33\pm0.63$
  & $165.1\pm0.9$ 
  & $-111.7\pm20.6$ 
  & $0.253\pm0.002$
  & $0.28\pm0.01$
  & $0.24\pm0.01$
  \\
  \hline
  &$\lambda_B$(GeV)
  &$\lambda_{B_s}$(GeV)
  &$\gamma/\phi_3({}^\circ)$
  &$\tau({}^\circ)$
  &$\nu({}^\circ)$
  &$\delta({}^\circ)$
  &$\sigma({}^\circ)$
  \\
  \hline
  Fac
  & $0.19_{-0.05}^{+0.02}$
  & $0.60_{-0.04}^{+0.00}$
  & $75.4_{-1.6}^{+1.7}$
  & --
  & --
  & --
  & --
  \\
   \hline
  Res
  & $0.22\pm0.06$
  & $0.45_{-0.34}^{+0.15}$
  & $68.9\pm1.8$
  & $22.2\pm2.2$
  & $78.1\pm2.9$
  & $23.3\pm4.0$
  & $120.7\pm22.3$
\end{tabular}
\end{ruledtabular}
\end{table}

The fitted parameters are shown in Table~\ref{tab:parameters1}.
Uncertainties are obtained by scanning the parameter space with
$\chi^2\leq\chi^2_{\rm min}+1$. The parameters consist of those in
factorization amplitude and of Res. 
The Fac fit gives $F^{B\pi}=0.239\pm0.002$, while the Res fit gives $F^{B\pi}=0.253\pm0.002$.
They correspond to 
$F^{B\pi}|V_{ud}|=(8.55_{-0.05}^{+0.08})\times 10^{-4}$ and 
$F^{B\pi}|V_{ud}|=(9.03\pm0.09)\times 10^{-4}$ for $|V_{ub}|= 35.76\times 10^{-4}$ employed in the numerical study, 
respectively, and they can be compared 
the HFLAG average,
$F^{B\pi}_0(0)|V_{ub}|=(9.23\pm0.24)\times 10^{-4}$.
The Res result agrees better with the data.

Both fits prefer large $\lambda_{B_s}$. 
Except $\rho_A$ and $\phi_A$, most common parameters in Fac and Res have similar fitted values.
A closer look reveals that Fac prefers $\gamma/\phi_3$ close to its center value [see Eq.~(\ref{eq:QCDFHparameters})], 
while Res prefers a lower $\gamma/\phi_3$.
Comparing the fitted phases to those in the U(3) exchange-type solution~[see Eq.~(\ref{eq:solutionreU3re})] 
$\tau=24.1^\circ$, $\nu=35.3^\circ$ and $\sigma-\delta=0$ and in
the U(3) annihilation-type solution~[see Eq.~(\ref{eq:solutionreU3ra})]
$\tau=-41.8^\circ$, $\nu=-19.5^\circ$ and $\sigma-\delta\neq 0$, 
we see that the fitted $\tau\simeq 22^\circ$ and $\nu\simeq 78^\circ$ seem to prefer the exchange-type solution, 
while the fitted $\sigma-\delta\simeq 97.4^\circ$ supports the annihilation-type solution.

\subsection{Rescattering effects on Topological Amplitudes}\label{subsec: FSITA}

In this part, we will show the rescattering effects on topological amplitudes in certain combinations and on some individual topological amplitudes of interest.
Note that the discussion in the first part is generic, while we need to impose further assumption in the second part.

\subsubsection{Rescattering effects on the Combinations of Topological Amplitudes}

It is useful to show the fitted results on residual rescattering parameters $r'_i$ (or $r^{(1/2)}_i$):
\be
 1+i(r'_0+r'_a)
      &=&
       (0.979_{-0.008}^{+0.007})\exp[i(11.98_{-1.81}^{+1.66})^\circ+i\delta_{27}], 
      \nonumber\\
 i(r'_e-r'_a)
      &=&
      (0.208_{-0.031}^{+0.028})\exp[i(-78.36\pm2.02)^\circ+i\delta_{27}], 
      \nonumber\\
 i (r'_a+r'_t)
      &=&
      (0.059\pm 0.009)\exp[i(-92.06_{-13.21}^{+9.09})^\circ+i\delta_{27}],
      \nonumber\\
 i(2\bar r'_a+\bar r'_e)
      &=&
       (0.189_{-0.044}^{+0.048})\exp[i(-78.36\pm2.02)^\circ+i\delta_{27}],
      \nonumber\\
 1+i(\tilde r'_0+\frac{4\tilde r'_a+2\tilde r'_e}{3})
      &=&
       (0.990_{-0.006}^{+0.004})\exp[i(3.27_{-1.01}^{+1.24})^\circ+i\delta_{27}],
      \nonumber\\
i \hat r'_t+i\frac{4\hat r'_a+2\hat r'_e}{3}
      &=&
       (0.248_{-0.068}^{+0.067})\exp[i(-29.66_{-11.13}^{+11.13})^\circ+i\delta_{27}], 
      \nonumber\\
  1+i(\check r'_0+\frac{4\check r'_a+2\check r'_e+3\check r'_t}{6})
      &=&
       (0.936_{-0.041}^{+0.031})\exp[i(118.43_{-21.73}^{+22.24})^\circ+i\delta_{27}]. 
 \label{eq:rfit}
 \en
From the above equation, we see that most of these parameters have large phases (with respect to $\delta_{27}$). 
Note that $i \hat r'_t+i({4\hat r'_a+2\hat r'_e})/3$, $i(r'_e-r'_a)$ and  $i(2\bar r'_a+\bar r'_e)$ are three most sizable combinations and they are close to 
$\lambda$, $-i\lambda$ and $-i\lambda$ (taking the overall phase $\delta_{27}=0$), respectively, where $\lambda$ is the Wolfenstein parameter.

\begin{table}[t!]
\caption{Combinations of topological amplitudes of $\Delta S=0$, $\bar B_q\to PP$ and $B_q\to PP$ decays before rescattering ($A^0$) and after rescattering ($A_{FSI}$) in the unit of $10^{-8}$ GeV. 
These results are obtained using the best fitted solution and Eqs.~(\ref{eq: TAQCDF1}), (\ref{eq: TAQCDF2}), (\ref{eq: TAQCDF3}), (\ref{eq: rcombinations1}), (\ref{eq: rcombinations2}) and (\ref{eq: rcombinations3}). 
Without lost of generality the overall phase ($\delta_{27}$) for $A_{FSI}$ is set to 0 for simplicity.
 \label{tab:TADS0}
}
{\footnotesize
\begin{ruledtabular}
\begin{tabular}{ccccccccc}
 & $A^0(\overline B)$
 & $A_{FSI}(\overline B)$ 
 & $A_{FSI}/A^0(\overline B)$
 & $A^0(B)$
 & $A_{FSI}(B)$ 
 & $A_{FSI}/A^0(B)$
  \\ 
  \hline
$T+C$
  &$ 3.23 e^{-i 79.8^\circ}$ 
  &$ 3.23 e^{-i 79.8^\circ}$ 
  & $1$
  &$ 3.23 e^{i 57.9^\circ}$ 
  &$ 3.23 e^{i 57.9^\circ}$ 
  & $1$
  \\
    \hline
$C-E$
  & $1.13 e^{-i119.5^\circ}$ 
  & $1.58 e^{-i118.8^\circ}$ 
  & $1.40 e^{i0.7^\circ}$
  & $1.18 e^{i18.2^\circ}$ 
  & $1.58 e^{i19.0^\circ}$ 
  & $1.40 e^{i0.7^\circ}$  
  \\
  \hline
$A+C$  
  & $1.07 e^{-i122.4^\circ}$ 
  & $1.52 e^{-i120.7^\circ}$ 
  & $1.42 e^{i1.7^\circ}$
  & $1.07 e^{i15.3^\circ}$ 
  & $1.52 e^{i17.0^\circ}$ 
  & $1.42 e^{i1.7^\circ}$  
  \\
  \hline
{\footnotesize
$P-C+\frac{1}{3}P_{EW}^{C}$ 
} 
  & $1.77 e^{i34.1^\circ}$ 
  & $2.23 e^{i44.1^\circ}$ 
  & $1.26 e^{i10.0^\circ}$
  & $0.80 e^{-i102.1^\circ}$ 
  & $0.94 e^{-i128.3^\circ}$ 
  & $1.17 e^{-i26.2^\circ}$
  \\
  \hline
$PA-\frac{4}{9} C$
  & $0.56 e^{i75.4^\circ}$ 
  & $0.45 e^{i69.7^\circ}$ 
  & $0.81 e^{-i5.7^\circ}$
  & $0.64 e^{-i160.6^\circ}$ 
  & $0.80 e^{-i141.5^\circ}$ 
  & $1.25 e^{i19.1^\circ}$
  \\
 $+\frac{13}{9} E-\frac{1}{3} P_{EW}^{C}$
  &   
  \\
  \hline
$P_{EW}+P_{EW}^{C}$
  & $0.10 e^{i11.9^\circ}$
  & $0.10 e^{i11.9^\circ}$
  & $1$
  & $0.10 e^{-i32.1^\circ}$
  & $0.10 e^{-i32.1^\circ}$
  & $1$  
  \\
  \hline
$P_{EW}^{C}-P_{EW}^{E}$
  & $0.04 e^{-i36.7^\circ}$
  & $0.05 e^{-i31.0^\circ}$
  & $1.40 e^{i5.7^\circ}$
  & $0.04 e^{-i79.5^\circ}$
  & $0.05 e^{-i74.9^\circ}$
  & $1.37 e^{i4.6^\circ}$
  \\
  \hline
 $P_{EW}^{A}+P_{EW}^{C}$
  & $0.03 e^{-i44.2^\circ}$
  & $0.05 e^{-i38.5^\circ}$
  & $1.53e^{i5.7^\circ}$
  & $0.03 e^{-i89.8^\circ}$
  & $0.05 e^{-i83.6^\circ}$
  & $1.55 e^{i6.1^\circ}$
  \\
  \hline
$\bar T+2\bar A$
  & $2.66 e^{-i63.9^\circ}$ 
  & $2.43 e^{-i56.2^\circ}$ 
  & $0.92 e^{i7.7^\circ}$
  & $2.66 e^{i73.8^\circ}$ 
  & $2.43 e^{i81.5^\circ}$ 
  & $0.92 e^{i7.7^\circ}$
  \\
 \hline
$\bar C_1+2\bar E$
 & $0.90 e^{-i134.9^\circ}$ 
 & $1.29 e^{-i129.3^\circ}$ 
 & $1.44 e^{i5.6^\circ}$
 & $0.96 e^{i2.8^\circ}$ 
 & $1.29 e^{i8.4^\circ}$ 
 & $1.44 e^{i5.6^\circ}$
 \\
 \hline
$\bar C_2+2\bar P-\frac{1}{3}\bar P^{C}_{EW,2}$
 & $1.97 e^{-i36.6}$ 
 & $2.22 e^{-i49.1}$ 
 & $1.13e^{-i12.5^\circ}$
 & $2.99 e^{-i29.7}$ 
 & $3.34 e^{-i27.1}$ 
 & $1.12 e^{i2.6^\circ}$
 \\
 \hline
$\bar P_{EW} +2\bar P^{A}_{EW}$
  & $0.09 e^{i 28.4^\circ}$
  & $0.08 e^{i 37.4^\circ}$
  & $0.92 e^{i8.9^\circ}$
  & $0.09 e^{-i 15.8^\circ}$
  & $0.08 e^{-i 6.7^\circ}$
  & $0.92 e^{i9.2^\circ}$
  \\
\hline
$\bar P^{C}_{EW,1}+2\bar P^{E}_{EW}$
 & $0.04 e^{-i58.8^\circ}$ 
 & $0.05 e^{-i46.9^\circ}$
 & $1.34 e^{i11.9^\circ}$
 & $0.03 e^{-i72.1^\circ}$ 
 & $0.04 e^{-i68.8^\circ}$
 & $1.53 e^{i3.3^\circ}$
 \\
\hline
$\tilde C+\tilde E +\tilde P$
 & $1.34 e^{-i65.0^\circ}$ 
 & $1.56 e^{-i2.6^\circ}$ 
 & $1.16 e^{i62.4^\circ}$
 & $1.89 e^{-i32.2^\circ}$ 
 & $1.92 e^{i59.1^\circ}$ 
 & $1.02 e^{i91.3^\circ}$
 \\
$+\frac{3}{2}\widetilde {PA}-\frac{1}{3}\tilde P^{C}_{EW}-\frac{1}{3}\tilde P^{E}_{EW}$
\end{tabular}
\end{ruledtabular}
}
\end{table}

In Tables \ref{tab:TADS0} and \ref{tab:TADS-1} we show the topological amplitudes of $\overline B_q\to PP$ and $B_q\to PP$ decays before rescattering ($A^0$) and after rescattering ($A_{FSI}$) in the unit of $10^{-8}$ GeV.~\footnote{The $A^0$ are obtained by using the rescattering parameters as shown in Table~\ref{tab:parameters1}, but with $\tau$, $\nu$, $\delta$ and $\sigma$ set to zero. Do not confuse it with the annihilation amplitude, where they may share the same notation occasionally.} 
These amplitudes are expressed in certain combinations as noted in Eq. (\ref{eq: combinations}). 
Note that without lost of generality the overall phase ($\delta_{27}$) is set to 0 from now on for simplicity.
The ratios $A_{FSI}/A^0$ are also shown.
These results are obtained using the best fitted solution and Eqs.~(\ref{eq: TAQCDF1}), (\ref{eq: TAQCDF2}), (\ref{eq: TAQCDF3}), (\ref{eq: rcombinations1}), (\ref{eq: rcombinations2}) and (\ref{eq: rcombinations3}). Both $\Delta S=0$ and $\Delta S=-1$ amplitudes are shown.
Note that we do not use them directly in the fitting. 
In fact, they can be obtained only after the best fit result is available. 
Nevertheless they will provide useful information.

From Table~\ref{tab:TADS0}, we see that before rescattering, we have the following order for $\overline B_q\to PP$ amplitudes:
\be
&&|T^0+C^0| 
 > |\bar T^0+2\bar A^0| 
 > |\bar C^0_2+2\bar P^0-\frac{1}{3}\bar P^{C0}_{EW,2}| 
\non\\
&& 
 > |P^0-C^0+\frac{1}{3}P_{EW}^{C0}| 
 >|\tilde C^0+\tilde E^0 +\tilde P^0+\frac{3}{2}\tilde {PA}^0-\frac{1}{3}\tilde P^{C0}_{EW}-\frac{1}{3}\tilde P^{E0}_{EW}| 
 > |C^0-E^0| 
\non\\
&&
 \gtrsim |A^0+C^0| 
 \gtrsim |\bar C^0_1+2\bar E^0| 
 >|PA^0-\frac{4}{9} C^0+\frac{13}{9} E^0-\frac{1}{3} P_{EW}^{C0}|, 
 \non
 \en
while the rest are rather small.
After rescattering, we have:
\be
&&|T+C| 
 > |\bar T+2\bar A| 
 > |P-C+\frac{1}{3}P_{EW}^{C}| 
\non\\
&&
 \gtrsim |\bar C_2+2\bar P-\frac{1}{3}\bar P^{C}_{EW,2}| 
 > |\tilde C+\tilde E +\tilde P+\frac{3}{2}\tilde {PA}-\frac{1}{3}\tilde P^{C}_{EW}-\frac{1}{3}\tilde P^{E}_{EW}| 
 \gtrsim |C-E| 
\non\\
&&
 \gtrsim |A+C| 
 >|\bar C_1+2\bar E| 
 >|PA-\frac{4}{9} C+\frac{13}{9} E-\frac{1}{3} P_{EW}^{C}|, 
\non
\en
where $ |C-E|$, $|A+C|$ and $|\bar C_1+2\bar E|$ are enhanced by $40\sim~44$\%, 
$|P-C+\frac{1}{3}P_{EW}^{C}|$ by 26\%
and
$|\tilde C+\tilde E +\tilde P+\frac{3}{2}\tilde {PA}-\frac{1}{3}\tilde P^{C}_{EW}-\frac{1}{3}\tilde P^{E}_{EW}|$ by 16\%.
Note that the orders of 
$|\bar C_2+2\bar P-\frac{1}{3}\bar P^{C}_{EW,2}|$
and
$ |P-C+\frac{1}{3}P_{EW}^{C}| $
are switched after turning on Res.
Sub-leading tree amplitudes and penguin amplitudes are enhanced.
We will return to this shortly.
Note that except in $\tilde C+\tilde E +\tilde P+\frac{3}{2}\tilde {PA}-\frac{1}{3}\tilde P^{C}_{EW}-\frac{1}{3}\tilde P^{E}_{EW}$ 
Res does not introduce sizable phases to these topological amplitude combinations.

Similarly, from Table~\ref{tab:TADS0}, we see that before rescattering, we have the following order for the conjugated $B_q\to PP$ decay amplitudes:
\be
&&|T^0+C^0| 
 > |\bar C^0_2+2\bar P^0-\frac{1}{3}\bar P^{C0}_{EW,2}| 
 > |\bar T^0+2\bar A^0| 
\non\\
&& 
 >|\tilde C^0+\tilde E^0 +\tilde P^0+\frac{3}{2}\tilde {PA}^0-\frac{1}{3}\tilde P^{C0}_{EW}-\frac{1}{3}\tilde P^{E0}_{EW}| 
 > |C^0-E^0| 
 \gtrsim |A^0+C^0| 
\non\\
&&
 \gtrsim |\bar C^0_1+2\bar E^0| 
 > |P^0-C^0+\frac{1}{3}P_{EW}^{C0}| 
 >|PA^0-\frac{4}{9} C^0+\frac{13}{9} E^0-\frac{1}{3} P_{EW}^{C0}|, 
 \non
 \en
while the rest are rather small.
Note that the above order is different form the one in $\overline B_q\to PP$ decays.
After rescattering, only the first two terms switch order,
where $ |\bar C_2+2\bar P-\frac{1}{3}\bar P^{C}_{EW,2}|$ is enhanced by 12\%, 
$|P-C+\frac{1}{3}P_{EW}^{C}|$ by 17\%
and
$ |C-E|$,
$|A+C|$ 
and
$|\bar C_1+2\bar E|$ by $40\sim44$\%.
Note that Res introduces sizable phases to some of these topological amplitude combinations
and $ |\bar C_2+2\bar P-\frac{1}{3}\bar P^{C}_{EW,2}|$,
$|\tilde C+\tilde E +\tilde P+\frac{3}{2}\tilde {PA}-\frac{1}{3}\tilde P^{C}_{EW}-\frac{1}{3}\tilde P^{E}_{EW}|$ and
$|PA-\frac{4}{9} C+\frac{13}{9} E-\frac{1}{3} P_{EW}^{C}|$
are quite different to those in $\overline B_q\to PP$ decays.

Some comments will be useful.
(i) A large number of combinations of topological amplitudes are sizable. 
(ii) After rescattering one sees that the phases introduced to $\bar B\to PP$ and $B\to PP$ amplitudes are quite different.
(iii) The above facts imply that the effect of Res on direct CP violations can be complicate and rich.
(iv) The enhancement of rescattering on some of the $\Delta S=0$ topological amplitudes can be up to $55\%$, 
such as on $P^A_{EW}+P^C_{EW}$, 
but their sizes are still small even after the enhancement. 
Nevertheless this may have impact on some suppressed modes.

It is useful to see the above enhancements in details. 
It is clear from Eq.~(\ref{eq: rcombinations1}) that the effects of Res on $T+C$ and $P_{EW}+P^C_{EW}$ 
are just adding the common phase $\delta_{27}$ to them.
The effects on other combinations of topological amplitudes are more interesting.
In $\overline B_q\to PP$ decays, considering only the dominant contributions in Eq. (\ref{eq: rcombinations1}), we have
\be
C-E
&\simeq&
     (1+ir'_0+ir'_a)(C^{0}-E^{0})
     +i(r'_e-r'_a)(T^{0}+C^{0}),
\non\\
A+C
&\simeq&
     (1+ir'_0+ir'_a)(A^{0}+C^{0})
    +\frac{2}{3}i(r'_e-r'_a)(T^{0}+C^{0})
    +\frac{1}{3}i(\bar r'_e+2\bar r'_a)(\bar T^{0}+2\bar A^{0}).
\non\\
\en
We can estimation the above values by taking the central values of $(1+ir'_0+ir'_a)$, $i(r'_e-r'_a)$ and  $i(\bar r'_e+2\bar r'_a)$ from  Eq. (\ref{eq:rfit}) and the central values of $C^{0}-E^{0}$,  $A^{0}+C^{0}$, $T^{0}+C^{0}$ and $\bar T^{0}+2\bar A^{0}$ from Table ~\ref{tab:TADS0}, obtaing
\be
\frac{C-E}{C^{0}-E^{0}}\simeq 1.4\,e^{-i 7^\circ}, %
\quad
\frac{A+C}{A^0+C^0}\simeq 1.4 \,e^{-i 4^\circ},
\en
which are close the values of $1.40\,e^{i0.7^\circ}$ 
and $1.42\,e^{i1.7^\circ}$ 
shown in Table~\ref{tab:TADS0}. 
Even using a crude estimation by taking $(1+ir'_0+ir'_a)\simeq 1$ and $i(r'_e-r'_a)\simeq i(\bar r'_e+2\bar r'_a)\simeq -i \lambda$, one still get
$1.5\, e^{-i 19^\circ}$ and $1.5\, e^{-i 16^\circ}$, which are not too far off.
It is clear that the effect of Res in $C-E$ mainly comes from the exchange and annihilation rescatterings fed from the $T^0+C^0$ amplitude,
while those in $A+C$ comes from the exchange and annihilation rescatterings fed from both $T^0+C^0$ and $\bar T^0+2\bar A^0$ amplitudes.

Similarly 
from Eq. (\ref{eq: rcombinations3}), we have
\be
&&(\tilde C+\tilde E+\tilde P+\frac{3}{2}\widetilde {PA}-\frac{1}{3}\tilde P^{C}_{EW} -\frac{1}{3}\tilde P^{E}_{EW})
\non\\
&&\simeq\left[1+i\left(\check r'_0+\frac{4\check r'_a+2\check r'_e+3\check r'_t}{6}\right)\right]
\bigg(\tilde C^{0}+\tilde E^{0}+\tilde P^{0}
+\frac{3}{2}\widetilde {PA}^{0}
-\frac{1}{3}\tilde P^{C0}_{EW} -\frac{1}{3}\tilde P^{E0}_{EW}\bigg)
\non\\
&&\quad
+i\left(\hat r'_t +\frac{4\hat r'_a +2\hat r'_e}{3}\right) 
\bigg[\frac{3}{2}(T^{0}+C^{0})+\frac{14}{3}(C^{0}-E^{0})+4 \bigg(P^{0}-C^{0}+\frac{1}{3}P_{EW}^{C0}\bigg)
\non\\
&&\qquad
+6 \bigg(PA^{0}-\frac{4}{9} C^{0}+\frac{13}{9} E^{0}-\frac{1}{3} P_{EW}^{C0}\bigg)\bigg],
\en
and we find that
the $T^{0}+C^{0}$ and $C^{0}-E^{0}$ terms give (sizable) destructive contributions, while
$P^{0}-C^{0}+\frac{1}{3}P_{EW}^{C0}$ and $PA^{0}-\frac{4}{9} C^{0}+\frac{13}{9} E^{0}-\frac{1}{3} P_{EW}^{C0}$ 
terms give (sizable) constructive contributions via the same Res parameter $i\hat r'_t +i(4\hat r'_a +2\hat r'_e)/3$.
The final result of the $16\%$ enhancement in 
$|\tilde C+\tilde E+\tilde P+\frac{3}{2}\widetilde {PA}-\frac{1}{3}\tilde P^{C}_{EW} -\frac{1}{3}\tilde P^{E}_{EW}|$ 
is the complicate interplay of these contributions.

We now turn to the Res effect on the penguin amplitudes.
From [see  Eq. (\ref{eq: rcombinations1})]
\be
P-C+\frac{1}{3}P^{C}_{EW}
&\simeq &\bigg[(1+ir'_0+ir'_a)-\frac{2}{3}i(r'_e-r'_a)\bigg] \bigg(P^{0}-C^{0}+\frac{1}{3}P^{C0}_{EW}\bigg)
\non\\
&&\qquad
-i(r'_e-r'_a)
(T^{0} + C^{0})
+\frac{1}{3}i(\bar r'_e+2\bar r'_a)(\bar C^{0}_2+2\bar P^{0}-\frac{1}{3}\bar P^{C0}_{EW,2}),
\en
we obtain for $\overline B\to PP$ decay:
\be
\frac{P-C+\frac{1}{3}P^{C}_{EW}}{P^{0}-C^{0}+\frac{1}{3}P^{C0}_{EW}}\simeq 1.3\, e^{i 10^\circ},
\en
which is close to the value $1.26\, e^{i 10.0^\circ}$ shown in Table~\ref{tab:TADS0}. 
where the main contribution is from the $r'_e-r'_a$ rescattering term fed from $T^{0} + C^{0}$.

\begin{table}[t!]
\caption{Same as Fig.~\ref{tab:TADS0}, but for $\Delta S=-1$ transition decay amplitudes.
 \label{tab:TADS-1}
}
{\footnotesize
\begin{ruledtabular}
\begin{tabular}{ccccccccc}
 & $A^0(\overline B)$
 & $A_{FSI}(\overline B)$ 
 & $A_{FSI}/A^0(\overline B)$
 & $A^0(B)$
 & $A_{FSI}(B)$ 
 & $A_{FSI}/A^0(B)$
 \\ 
  \hline
$T^{\prime}+C^{\prime}$
  &$ 0.75e^{-i 79.8^\circ}$ 
  &$ 0.75e^{-i 79.8^\circ}$ 
  & $1$
  &$ 0.75e^{i 57.9^\circ}$ 
  &$ 0.75e^{i 57.9^\circ}$
  & $1$
  \\
    \hline
$C^{\prime}-E^{\prime}$
  & $0.26e^{-i119.5^\circ}$ 
  & $0.36e^{-i118.8^\circ}$ 
  & $1.40e^{i0.7^\circ}$
  & $0.26e^{i18.2^\circ}$
  & $0.36e^{i19.0^\circ}$
  & $1.40 e^{i0.7^\circ}$
  \\
  \hline
$A^{\prime}+C^{\prime}$
  & $0.25e^{-i122.4^\circ}$ 
  & $0.35e^{-i120.7^\circ}$ 
  & $1.42e^{i1.7^\circ}$  
  & $0.25e^{i15.3^\circ}$
  & $0.35e^{i17.0^\circ}$
  & $1.42e^{i1.7^\circ}$
  \\
  \hline
$P^{\prime}-C^{\prime}+\frac{1}{3}P_{EW}^{\prime C}$  
  & $4.36 e^{i164.2^\circ}$ 
  & $4.00 e^{i174.0^\circ}$ 
  & $0.92 e^{i9.8^\circ}$
  & $4.64 e^{i170.3^\circ}$
  & $4.48 e^{-i178.6^\circ}$
  & $0.97 e^{i11.1^\circ}$
  \\
  \hline
$PA^{\prime}-\frac{4}{9} C^{\prime}$
  & $0.42 e^{i3.5^\circ}$ 
  & $0.99 e^{i73.2^\circ}$ 
  & $2.37 e^{i69.7^\circ}$
  & $0.29 e^{-i31.2^\circ}$
  & $0.74 e^{i81.8^\circ}$
  & $2.58 e^{i113.0^\circ}$
  \\
 $+\frac{13}{9} E^{\prime}-\frac{1}{3} P_{EW}^{\prime C}$
  &   
  \\
  \hline
$P^{\prime}_{EW}+P_{EW}^{\prime C}$
  & $0.46e^{i168.9^\circ}$ 
  & $0.46e^{i168.9^\circ}$ 
  & $1$
  & $0.46e^{i171.0^\circ}$
  & $0.46e^{i171.0^\circ}$
  & $1$
  \\
  \hline
$P_{EW}^{\prime C}-P_{EW}^{\prime E}$
  & $0.18e^{i120.8^\circ}$
  & $0.26e^{i124.8^\circ}$
  & $1.40e^{i3.9^\circ}$
  & $0.18e^{i122.9^\circ}$
  & $0.26e^{i126.8^\circ}$
  & $1.40e^{i3.9^\circ}$
  \\
  \hline
 $P_{EW}^{\prime A}+P_{EW}^{\prime C}$
  & $0.14 e^{i113.6^\circ}$
  & $0.22 e^{i118.9^\circ}$
  & $1.53 e^{i5.3^\circ}$
  & $0.14 e^{i115.6^\circ}$
  & $0.22 e^{i121.0^\circ}$
  & $1.53 e^{i5.4^\circ}$
  \\
  \hline
$\bar T^{\prime}+2\bar A^{\prime}$
  & $0.61 e^{-i63.9^\circ}$ 
  & $0.56 e^{-i56.2^\circ}$ 
  & $0.92e^{i7.7^\circ}$
  & $0.61 e^{i73.8^\circ}$
  & $0.56 e^{i81.5^\circ}$
  & $0.92e^{i7.7^\circ}$
  \\
 \hline
$\bar C^{\prime}_1+2\bar E^{\prime}$
 & $0.21 e^{-i134.9^\circ}$
 & $0.30 e^{-i129.3^\circ}$
 & $1.44 e^{i5.6^\circ}$
 & $0.21 e^{i2.8^\circ}$
 & $0.30 e^{i8.4^\circ}$
 & $1.44 e^{i5.6^\circ}$
 \\
 \hline
$\bar C^{\prime}_2+2\bar P^{\prime}-\frac{1}{3}\bar P^{\prime C}_{EW,2}$
 & $10.62 e^{i152.6}$ 
 & $11.13 e^{i149.9}$ 
 & $1.05 e^{-i2.7^\circ}$
 & $10.38 e^{i152.4}$
 & $10.85 e^{i148.8}$
 & $1.05 e^{-i3.6^\circ}$
 \\
 \hline
$\bar P^{\prime}_{EW} +2\bar P^{\prime A}_{EW}$
  & $0.42 e^{-i 174.8^\circ}$ 
  & $0.38 e^{-i 165.9^\circ}$ 
  & $0.92 e^{i8.8^\circ}$
  & $0.42 e^{-i 172.6^\circ}$
  & $0.38 e^{-i 163.8^\circ}$
  & $0.92 e^{i8.9^\circ}$
  \\
\hline
$\bar P^{\prime C}_{EW,1}+2\bar P^{\prime E}_{EW}$
 & $0.08 e^{i61.0^\circ}$ 
 & $0.12 e^{i93.9^\circ}$
 & $1.48 e^{i33.0^\circ}$
 & $0.08 e^{i63.0^\circ}$ 
 & $0.12 e^{i93.0^\circ}$
 & $1.48 e^{i33.0^\circ}$
 \\
\hline
$\tilde C^{\prime}+\tilde E^{\prime} +\tilde P^{\prime}$
 & $6.04 e^{i140.0^\circ}$ 
 & $5.01 e^{-i141.5^\circ}$ 
 & $0.83 e^{i78.5^\circ}$
 & $5.89 e^{i138.1^\circ}$
 & $4.88 e^{-i146.1^\circ}$
 & $0.83e^{i75.8^\circ}$
 \\
$+\frac{3}{2}\widetilde {PA}^{\prime}-\frac{1}{3}\tilde P^{\prime C}_{EW}-\frac{1}{3}\tilde P^{\prime E}_{EW}$
\end{tabular}
\end{ruledtabular}
}
\end{table}

We now turn to $\Delta S=-1$ processes. The results are shown in Table~\ref{tab:TADS-1}. 
We see from the table that before rescattering, 
we have the following order for $\overline B_q\to PP$ amplitudes:
\be
&&|\bar C^{\prime 0}_2+2\bar P^{\prime 0}-\frac{1}{3}\bar P^{\prime C0}_{EW,2}| 
>|\tilde C^{\prime 0}+\tilde E^{\prime 0} +\tilde P^{\prime 0}+\frac{3}{2}\tilde {PA}^{\prime 0}-\frac{1}{3}\tilde P^{\prime C0}_{EW}-\frac{1}{3}\tilde P^{\prime E0}_{EW}| 
\non\\
&&
> |P^{\prime 0}-C^0+\frac{1}{3}P_{\prime EW}^{C0}| 
\gg |T^{\prime 0}+C^{\prime 0}| 
 > |\bar T^{\prime 0}+2\bar A^{\prime 0}| 
 > |P^{\prime 0}_{EW}+P^{\prime C0}_{EW}| 
 \non\\
&&\gtrsim |PA^{\prime 0}-\frac{4}{9} C^{\prime 0}+\frac{13}{9} E^{\prime 0}-\frac{1}{3} P_{EW}^{\prime C0}| 
 \gtrsim |\bar P^{\prime 0}_{EW}+2\bar P^{\prime A0}_{EW}| 
 > |C^{\prime 0}-E^{\prime 0}| 
 \gtrsim |A^{\prime 0}+C^{\prime 0}|, 
\non
\en
while the rest are rather small.
Note that as expected penguin amplitudes dominate over trees. 
In fact, even the electroweak penguin amplitudes, which were neglected in the $\Delta S=0$ case, cannot be neglected now. 
After rescattering, the above orders are rearranged into:
\be
&&
|\bar C^\prime_2+2\bar P^\prime-\frac{1}{3}\bar P^{\prime C}_{EW,2}| 
 >|\tilde C^\prime+\tilde E^\prime +\tilde P^\prime+\frac{3}{2}\tilde {PA}^\prime-\frac{1}{3}\tilde P^{\prime C}_{EW}-\frac{1}{3}\tilde P^{\prime E}_{EW}| 
 > |P^\prime-C^\prime+\frac{1}{3}P_{EW}^{\prime C}|  
\non\\
&&
 \gg |PA^\prime-\frac{4}{9} C^\prime+\frac{13}{9} E^\prime-\frac{1}{3} P_{EW}^{\prime C}| 
 >|T^\prime+C^\prime| 
 > |\bar T^\prime+2\bar A^\prime| 
\non\\
&&
 > |P^{\prime}_{EW}+P^{\prime C}_{EW}| 
 > |\bar P^{\prime }_{EW}+2\bar P^{\prime A}_{EW}| 
 \gtrsim |C^\prime-E^\prime| 
 \gtrsim |A^\prime+C^\prime|. 
\non
\en
We see that the combinations with sub-leading tree amplitudes, $C'-E'$ and $A'+C'$, are enhanced, 
while the one with the penguin term, $P'-C'+P^{\prime C}_{EW}/3$, is slightly reduced.
Note that $|PA^\prime-\frac{4}{9} C^\prime+\frac{13}{9} E^\prime-\frac{1}{3} P_{EW}^{\prime C}|$ is enhanced by a factor of 2,
but $|\tilde C^\prime+\tilde E^\prime +\tilde P^\prime+\frac{3}{2}\tilde {PA}^\prime-\frac{1}{3}\tilde P^{\prime C}_{EW}-\frac{1}{3}\tilde P^{\prime E}_{EW}|$ is reduced by about 20\%.
Similar pattern occurs in the conjugated $B_q\to PP$ decays.

The effect of rescattering on $A'+C'$ is similar to the one in $A+C$. 
It is enhanced from the exchange and annihilation rescatterings fed from both $T^{\prime 0}+C^{\prime 0}$ 
and $\bar T^{\prime 0}+2\bar A^{\prime 0}$ amplitudes. 
We also note that the effect of rescattering on $P^{\prime C}_{EW}-P^{\prime E}_{EW}$ is similar to the one in $C^{\prime}-E^{\prime}$,
but with tree amplitudes replaced by electroweak penguins. 
Hence $P^{\prime C}_{EW}-P^{\prime E}_{EW}$ is affected most from $P^{\prime 0}_{EW}+P^{\prime C0}_{EW}$ and the effect is an enhancement in size.

It is useful to see the enhancement and reduction in $|PA^\prime-\frac{4}{9} C^\prime+\frac{13}{9} E^\prime-\frac{1}{3} P_{EW}^{\prime C}|$
and $|\tilde C^\prime+\tilde E^\prime +\tilde P^\prime+\frac{3}{2}\tilde {PA}^\prime-\frac{1}{3}\tilde P^{\prime C}_{EW}-\frac{1}{3}\tilde P^{\prime E}_{EW}|$, respectively, in more detail. 
In $\overline B_q\to PP$ decays, 
keeping only the $(PA^{\prime 0}-\frac{4}{9} C^{\prime 0}+\frac{13}{9} E^{\prime 0}-\frac{1}{3} P_{EW}^{\prime C})$ and
the $(P^{\prime 0} - C^{\prime 0} + \frac{1}{3} P_{EW}^{\prime C0})$ terms in the corresponding formula shown in Eq. (\ref{eq: rcombinations1}), 
we obtain 
\be
\frac{PA^{\prime}-\frac{4}{9} C^{\prime}+\frac{13}{9} E^{\prime}-\frac{1}{3} P_{EW}^{\prime C}}
{PA^{\prime 0}-\frac{4}{9} C^{\prime 0}+\frac{13}{9} E^{\prime 0}-\frac{1}{3} P_{EW}^{\prime C}}
\simeq 2.6\, e^{i 52^\circ},
\en
which is close to the value $2.37\,e^{i69.7^\circ}$ shown in Table~\ref{tab:TADS-1}.
Similarly using the corresponding formula in Eq. (\ref{eq: rcombinations1}) and keep only the $(\tilde C^{\prime 0}+\tilde E^{\prime 0}+\tilde P^{\prime 0}
+\frac{3}{2}\widetilde {PA}^{(\prime)0}
-\frac{1}{3}\tilde P^{\prime C0}_{EW} -\frac{1}{3}\tilde P^{\prime E0}_{EW})$ and 
the $(P^{\prime 0} - C^{\prime 0} + \frac{1}{3} P_{EW}^{\prime C0})$ terms
we obtain
\be
\frac
{\tilde C^{\prime}+\tilde E^{\prime}+\tilde P^{\prime}+\frac{3}{2}\widetilde {PA}^{\prime}-\frac{1}{3}\tilde P^{\prime C}_{EW} -\frac{1}{3}\tilde P^{\prime E}_{EW}}
{\tilde C^{\prime 0}+\tilde E^{\prime 0}+\tilde P^{\prime 0}+\frac{3}{2}\widetilde {PA}^{(\prime)0}
-\frac{1}{3}\tilde P^{\prime C0}_{EW} -\frac{1}{3}\tilde P^{\prime E0}_{EW}}
\simeq 0.8\,e^{i 76^\circ},
\en
which is close to the value $0.83\,e^{i 78.5^\circ}$ shown in Table~\ref{tab:TADS-1}.
In both cases the most important contributions are from the $(P^{\prime 0} - C^{\prime 0} + \frac{1}{3} P_{EW}^{\prime C0})$ term.

\begin{table}[t!]
\caption{Some topological amplitudes of $\Delta S=0,-1$, $\bar B_q\to PP$ and $B_q\to PP$ decays before rescattering ($A^0$) and after rescattering ($A_{FSI}$) in the unit of $10^{-9}$ GeV. 
These results are obtained using the best fitted solution and Eqs.~(\ref{eq: TAQCDF1}), (\ref{eq: TAQCDF2}), (\ref{eq: TAQCDF3}) and
 (\ref{eq: delta TA1}). 
We use an additional assumption, $r'_e=0$ as suggested from U(3) symmetry on TA [see Eq. (\ref{eq: re=0})].  
Without lost of generality the overall phase ($\delta_{27}$) for $A_{FSI}$ is set to 0. 
Results in combinations of $\bar A$ and $\tilde A$ can be found in Tables \ref{tab:TADS0} and \ref{tab:TADS-1}.
 \label{tab:TADS0-1}
}
{\footnotesize
\begin{ruledtabular}
\begin{tabular}{ccccccccc}
 & $A^0(\overline B)$
 & $A_{FSI}(\overline B)$ 
 & $A_{FSI}/A^0(\overline B)$
 & $A^0(B)$
 & $A_{FSI}(B)$ 
 & $A_{FSI}/A^0(B)$
  \\ 
  \hline
$T$
  &$ 25.84e^{-i 63.5^\circ}$ 
  &$ 25.84e^{-i 63.5^\circ}$ 
  & $1$
  &$ 25.84e^{i 74.2^\circ}$ 
  &$ 25.84e^{i 74.2^\circ}$ 
  & $1$
  \\
    \hline
$C$
  & $10.45 e^{-i123.9^\circ}$ 
  & $10.45 e^{-i123.9^\circ}$ 
  & $1$
  & $10.45e^{i13.8^\circ}$ 
  & $10.45e^{i13.8^\circ}$ 
  & $1$  
  \\
  \hline
$E$  
  & $1.19e^{i102.6^\circ}$ 
  & $5.46e^{i71.0^\circ}$ 
  & $4.61e^{-i31.5^\circ}$
  & $1.19e^{-i119.7^\circ}$ 
  & $5.46e^{-i151.2^\circ}$ 
  & $4.61e^{-i31.5^\circ}$  
  \\
  \hline
  $A$  
  & $0.38e^{-i77.4^\circ}$ 
  & $4.78e^{-i113.8^\circ}$ 
  & $12.67e^{-i36.4^\circ}$
  & $0.38e^{i60.3^\circ}$ 
  & $4.78e^{i23.9^\circ}$ 
  & $12.67e^{-i36.4^\circ}$  
  \\
  \hline
$P$
  & $8.89 e^{i8.6^\circ}$ 
  & $12.26e^{i34.3^\circ}$ 
  & $1.38e^{i25.7^\circ}$
  & $9.94 e^{-i31.6^\circ}$ 
  & $6.43e^{-i47.6^\circ}$ 
  & $0.65e^{-i16.0^\circ}$
  \\
  \hline
$PA$
  & $0.76 e^{-i166.4^\circ}$ 
  & $7.95 e^{-i116.2^\circ}$ 
  & $10.50e^{i50.2^\circ}$
  & $0.76 e^{i149.3^\circ}$ 
  & $5.19 e^{-i1.6^\circ}$  
  & $6.86e^{-i150.9^\circ}$
  \\
  \hline
$P_{EW}$
  & $0.86e^{i29.0^\circ}$ 
  & $0.86e^{i29.0^\circ}$ 
  & $1$
 & $0.86e^{-i15.3^\circ}$ 
  & $0.86e^{-i15.3^\circ}$ 
  & $1$  
  \\
  \hline
$P_{EW}^{C}$
  & $0.29e^{-i46.8^\circ}$ 
  & $0.29e^{-i46.8^\circ}$ 
  & $1$
  & $0.29e^{-i89.1^\circ}$ 
  & $0.29e^{-i89.1^\circ}$ 
  & $1$
  \\
  \hline
   $P_{EW}^{E}$
  & $0.11 e^{i170.5^\circ}$ 
  & $0.27 e^{i166.0^\circ}$ 
  & $2.42e^{-i4.4^\circ}$
  & $0.11 e^{i126.2^\circ}$ 
  & $0.26 e^{i121.2^\circ}$ 
  & $2.31e^{-i5.0^\circ}$
  \\
    \hline
 $P_{EW}^{A}$
  & $0.02 e^{i13.6^\circ}$ 
  & $0.18 e^{-i24.7^\circ}$ 
  & $11.21e^{-i38.3^\circ}$
  & $0.02 e^{-i30.7^\circ}$ 
  & $0.18 e^{-i68.8^\circ}$ 
  & $11.08e^{-i.38.1^\circ}$
  \\
   \hline
$T^{\prime}$
  &$ 5.98e^{-i 63.5^\circ}$ 
  &$ 5.98e^{-i 63.5^\circ}$ 
  & $1$
  &$ 5.98e^{i 74.2^\circ}$ 
  &$5.98e^{i 74.2^\circ}$ 
  & $1$
  \\
    \hline
$C^{\prime}$
  & $2.41e^{-i123.9^\circ}$ 
  & $2.41e^{-i123.9^\circ}$ 
  & $1$
  & $2.41e^{i13.8^\circ}$ 
  & $2.41e^{i13.8^\circ}$ 
  & $1$
  \\
  \hline
$E^{\prime}$
  & $0.27e^{i102.6^\circ}$ 
  & $1.26e^{i71.0^\circ}$ 
  & $4.61e^{-i31.5^\circ}$  
  & $0.27e^{-i119.7^\circ}$ 
  & $1.26e^{-i151.2^\circ}$ 
  & $4.61e^{-i31.5^\circ}$
  \\
  \hline
  $A^{\prime}$
  & $0.09e^{-i77.4^\circ}$ 
  & $1.10e^{-i113.8^\circ}$ 
  & $12.67e^{-i36.4^\circ}$  
  & $0.09e^{i60.3^\circ}$ 
  & $1.10e^{i23.9^\circ}$ 
  & $12.67e^{-i36.4^\circ}$
  \\
  \hline
$P^{\prime}$  
  & $44.12 e^{i167.6^\circ}$ 
  & $40.99e^{i177.5^\circ}$ 
  & $0.93e^{i9.9^\circ}$
  & $43.90 e^{i169.6^\circ}$ 
  & $42.29e^{-i178.7^\circ}$ 
  & $0.96e^{i11.7^\circ}$
  \\
  \hline
$PA^{\prime}$
  & $3.54 e^{-i9.6^\circ}$ 
  & $7.43 e^{i78.3^\circ}$
  & $2.10e^{i87.9^\circ}$
  & $3.54 e^{-i7.5^\circ}$ 
  & $9.56 e^{i68.4^\circ}$
  & $2.70e^{i75.9^\circ}$
  \\
  \hline
$P^{\prime}_{EW}$
  & $4.01e^{-i174.2^\circ}$ 
  & $4.01e^{-i174.2^\circ}$ 
  & $1$
  & $4.01e^{-i172.1^\circ}$ 
  & $4.01e^{-i172.1^\circ}$ 
  & $1$
  \\
  \hline
$P_{EW}^{\prime C}$
  & $1.38e^{i111.0^\circ}$ 
  & $1.38e^{i111.0^\circ}$ 
  & $1$
  & $1.38e^{i113.0^\circ}$ 
  & $1.38e^{i113.0^\circ}$ 
  & $1$
  \\
    \hline
 $P_{EW}^{\prime E}$
  & $0.53 e^{-i32.7^\circ}$ 
  & $1.27 e^{-i40.3^\circ}$ 
  & $2.40e^{-i7.6^\circ}$
  & $0.53 e^{-i30.6^\circ}$ 
  & $1.27 e^{-i38.2^\circ}$ 
  & $2.41e^{-i7.6^\circ}$
   \\
  \hline
 $P_{EW}^{\prime A}$
  & $0.07 e^{i170.4^\circ}$ 
  & $0.83 e^{i132.2^\circ}$ 
  & $11.14 e^{-i38.2^\circ}$
  & $0.07 e^{i172.5^\circ}$ 
  & $0.83 e^{i134.3^\circ}$ 
  & $11.15 e^{-i38.2^\circ}$
\end{tabular}
\end{ruledtabular}
}
\end{table}

\subsubsection{Rescattering effects on some Individual Topological Amplitudes}

The results in Tables~\ref{tab:TADS0} and \ref{tab:TADS-1} are all we can have, if no further assumption is made. 
It is, however, desirable to reveal the effect of Res on some individual topological amplitudes instead of their combinations.
To explore the effect one needs the information of various $r'_i$ instead of their combinations shown in Eq.~(\ref{eq:rfit}). 
For example, the Res effect on exchange amplitude is given by [see Eq.~(\ref{eq: delta TA1})]
\be
\delta E^{(\prime)}
&=&ir'_0 E^{(\prime)0}+ ir'_a T^{(\prime)0}-\frac{1}{3} i (r'_e+2r'_a) C^{(\prime)0}+\frac{1}{3}i(-2 r'_e+5 r'_a) E^{(\prime)0}
\non\\
         &&           +\frac{1}{3}i(\bar r'_e+2\bar r'_a)(\bar C^{(\prime)0}_1+2\bar E^{(\prime)0}).
\en
It is clear that we need the information of $r'_0$, $r'_a$, $r'_e$ and so on to obtain $\delta E^{(\prime)}$. 
From the fit we only have information on some combinations of these rescattering parameters, 
such as $1+i (r'_0+ r'_a)$, $i(r'_e-r'_a)$ and so on [see Eq. (\ref{eq:rfit})], but not on individual ones.
To study the effect of Res on individual topological amplitudes, we make an additional assumption:
\be
r'_e=0,
\en
which is suggested by the U(3) symmetry on TA~[see Eq.~(\ref{eq: re=0})].
Note that we only assume $r'_e=0$ and do not impose any condition on $\bar r'_e$, $\hat r'_e$ and $\check r'_e$.
Hence we are not using the full U(3) symmetry, but rather consider the case of a suppressed $r'_e$.
Using the above assumption and the results in Eq. (\ref{eq:rfit}) one can now extract the effect of Res on some individual TAs of interest.
The results are shown in Table~\ref{tab:TADS0-1}. One should keep in mind of the assumption made. 
Note that the above assumption will affect our interpretation of the effect of Res on individual topological amplitudes, 
but not on the interpretation of the effect of Res on the combinations of topological amplitudes as discussed previously. 
In other words, the above assumption will affect the results stated in Table~\ref{tab:TADS0-1}, but not on those in Tables~\ref{tab:TADS0} and \ref{tab:TADS-1}. 

From Table~\ref{tab:TADS0-1} we see that, before Res, for $\overline B_q\to PP$ and $B_q\to PP$ decays, we have
\be
&&|T^0|
>|C^0|
>|P^0|
\gg|E^0|
>|P_{EW}^0|
>|PA^0|
>|A^0|
\gtrsim |P_{EW}^{0C}|
>|P_{EW}^{0E}|
\gg|P_{EW}^{0A}|,
\non\\
&&|P^{\prime 0}|
\gg |T^{\prime 0}|
>|P_{EW}^{\prime 0}|
>|PA^{\prime 0}|
\gtrsim |C^{\prime 0}|
>|P_{EW}^{\prime C0}|
>|P_{EW}^{\prime E0}|
>|E^{\prime 0}|
\gg |A^{\prime 0}|
\gtrsim |P_{EW}^{\prime A0}|,
\en
while after Res, we have
\be
&&|T|
>|P|
>|C|
>|PA|
>|E|
>|A|
\gg|P_{EW}|
> |P_{EW}^{C}|
>|P_{EW}^{E}|
>P_{EW}^{A}|,
\non\\
&&|P^{\prime }|
\gg |PA^{\prime }|
>|T^{\prime }|
>|P_{EW}^{\prime }|
> |C^{\prime }|
>|P_{EW}^{\prime C}|
> |P_{EW}^{\prime E}|
\gtrsim|E^{\prime }|
>|A^{\prime }|
> |P_{EW}^{\prime A}|,
\en
for $\overline B_q\to PP$ decays, and
\be
&&|T|
>|C|
>|P|
>|E|
>|PA|
>|A|
\gg|P_{EW}|
> |P_{EW}^{C}|
>|P_{EW}^{E}|
>|P_{EW}^{A}|,
\non\\
&&|P^{\prime}|
\gg |PA^{\prime}| 
>|T^{\prime}| 
> |P^{\prime}_{EW}|
> |C^{\prime}| 
> |P_{EW}^{\prime C}| 
> |P_{EW}^{\prime E}| 
\gtrsim |E^{\prime}| 
>  |A^{\prime}| 
> |P_{EW}^{\prime A}|,
\en
for $B_q\to PP$ decays.
Note that the positions of  $|P|$ and $|PA|$ in the above orders are different in $\overline B_q\to PP$ and $B_q\to PP$ decays. We will come to that later.

We see from Table~\ref{tab:TADS0-1} that $|E|$, $|E'|$, $|A|$, $|A'|$, $|PA|$, $|PA'|$, $|P^{A,E}_{EW}|$ and $|P^{\prime A,E}_{EW}|$ are enhanced significantly with factors ranging from $2\sim 11$,
while $|P|$ is enhanced by $35\%$ in $\overline B_q\to PP$ decay, 
but is suppressed by $35\%$ in $B_q\to PP$ decay and $|P'|$ are suppressed by $6\%$ and $3\%$ in $\overline B_q\to PP$ and $B_q\to PP$ decays, respectively.
Note that in particular $|A|$ and $|A'|$ are enhanced by a factor of $11.5$. It is useful to look into the enhancement.
From Eq.~(\ref{eq: delta TA1}), we have
\be
A^{(\prime)}&=&(1+i r'_0-\frac{2}{3} i r'_e +\frac{5}{3} ir'_a  )A^{(\prime)0}
-\frac{1}{3}i(r'_e+2r'_a) T^{(\prime)0}+ir'_a C^{(\prime)0}
\non\\
&&+\frac{1}{3}i(\bar r'_e+2\bar r'_a)(\bar T^{(\prime)0}+2\bar A^{(\prime)0}).
\en
Now make use of $r'_e=0$ and Eq. (\ref{eq:rfit}), we obtain
\be
\frac{A^{(\prime)}}{A^{(\prime)0}}=0.99 e^{i 20.0^\circ}+9.48 e^{-i64.4^\circ}+5.75 e^{i 55.2^\circ}+4.45^{-i 64.8^\circ}
=12.67 e^{-i 36.4},
\en
where the terms in the right hand side of the first equality are from $A^{(\prime)0}$, $T^{(\prime)0}$, $C^{(\prime)0}$, $\bar T^{(\prime)0}+2\bar A^{(\prime)0}$ contributions, respectively.
We see that the $T^{(\prime)0}$, $C^{(\prime)0}$, $\bar T^{(\prime)0}+2\bar A^{(\prime)0}$ terms give sizable contributions to $A^{(\prime)}$, 
via $r'_a$, $r'_a$ and $\bar r_e+2 \bar r'_a$ rescatterings, respectively,
and enhance its size significantly. 
Similarly we have
\be
\frac{E^{(\prime)}}{E^{(\prime)0}}=0.99 e^{i 20.0^\circ}+4.53 e^{-i64.4^\circ}+1.22 e^{i 55.2^\circ}+0.48^{i 44.2^\circ}=4.61 e^{-i 31.5},
\label{eq: E/E0}
\en
where the terms in the right hand side of the first equality are from $A^{(\prime)0}$, $T^{(\prime)0}$, $C^{(\prime)0}$, $\bar C^{(\prime)0}_1+2\bar E^{(\prime)0}$ contributions, respectively. The dominate contribution is from $T^{(\prime)0}$ via annihilation rescattering $r'_a$.

\begin{table}[t!]
\caption{Same as Table~\ref{tab:TADS0-1}, but for $u$-penguins and $c$-penguins.
 \label{tab:ucpenguinsDS0-1}
}
{\footnotesize
\begin{ruledtabular}
\begin{tabular}{ccccccccc}
 & $A^0(\overline B)$
 & $A_{FSI}(\overline B)$ 
 & $A_{FSI}/A^0(\overline B)$
 & $A^0(B)$
 & $A_{FSI}(B)$ 
 & $A_{FSI}/A^0(B)$
  \\ 
  \hline
$P^u$
  & $3.51 e^{i108.3^\circ}$ 
  & $6.98 e^{i84.7^\circ}$ 
  & $1.99e^{-i23.6^\circ}$
  & $3.51 e^{-i113.9^\circ}$ 
  & $6.98 e^{-i137.6^\circ}$ 
  & $1.99e^{-i23.6^\circ}$
  \\
  \hline
  $P^c$
  & $10.09 e^{-i11.4^\circ}$ 
  & $9.49 e^{-i0.2^\circ}$ 
  & $0.94e^{i11.2^\circ}$ 
  & $10.09 e^{-i11.4^\circ}$ 
  & $9.49 e^{-i0.2^\circ}$ 
  & $0.94e^{i11.2^\circ}$ 
  \\
  \hline
$PA^u$
  & $0.31 e^{-i77.4^\circ}$ 
  & $5.98 e^{-i120.1^\circ}$ 
  & $19.55e^{-i42.7^\circ}$
  & $0.31 e^{i60.3^\circ}$ 
  & $5.98 e^{i17.6^\circ}$ 
  & $19.55e^{-i42.7^\circ}$
  \\
  \hline
$PA^c$
  & $0.81 e^{i171.4^\circ}$ 
  & $2.02 e^{-i104.6^\circ}$ 
  & $2.49 e^{i84.0^\circ}$
  & $0.81 e^{i171.4^\circ}$ 
  & $2.02 e^{-i104.6^\circ}$ 
  & $2.49 e^{i84.0^\circ}$
  \\
  \hline
$P^{u\prime}$  
  & $0.81 e^{i108.3^\circ}$ 
  & $1.61 e^{i84.7^\circ}$
  & $1.99 e^{-i23.6^\circ}$
  & $0.81 e^{-i113.9^\circ}$ 
  & $1.61 e^{-i137.6^\circ}$
  & $1.99 e^{-i23.6^\circ}$
  \\
  \hline
  $P^{c\prime}$  
  & $43.71 e^{i168.6^\circ}$
  & $41.10e^{i179.8^\circ}$
  & $0.94e^{i11.2^\circ}$
  & $43.71 e^{i168.6^\circ}$
  & $41.10e^{i179.8^\circ}$
  & $0.94e^{i11.2^\circ}$
  \\
  \hline
$PA^{u\prime}$
  & $0.07e^{-i77.4^\circ}$ 
  & $1.38 e^{-i120.1^\circ}$
  & $19.55e^{-i42.7^\circ}$
  & $0.07e^{i60.3^\circ}$ 
  & $1.38 e^{i17.6^\circ}$
  & $19.55e^{-i42.7^\circ}$
  \\
  \hline
  $PA^{c\prime}$
  & $3.51 e^{-i8.6^\circ}$ 
  & $8.75 e^{i75.4^\circ}$ 
  & $2.49 e^{i84.0^\circ}$
  & $3.51 e^{-i8.6^\circ}$ 
  & $8.75 e^{i75.4^\circ}$ 
  & $2.49 e^{i84.0^\circ}$
\end{tabular}
\end{ruledtabular}
}
\end{table}

As noted previously $P^{(\prime)}$ and $PA^{(\prime)}$ receive different Res contributions in $\overline B_q\to PP$ and $B_q\to PP$ decays.
It is interesting to investigate the effects of Res on these penguin amplitudes in details. 
First, we decompose $P^{(\prime)}$ into the so-called $u$-penguin ($P^{(\prime)u}$) and $c$-penguin ($P^{(\prime)c}$) 
as $P^{(\prime)}=P^{(\prime)u}+P^{(\prime)c}$ according to the different CKM factors.
Now from Eq.~(\ref{eq: delta TA1}), we have
\be
P^{(\prime)u}&=&
\left[1+ir'_0 +\frac{1}{3}i(-2 r'_e +5 r'_a)\right] P^{(\prime)u0}
+i r'_a T^{(\prime)0}
-\frac{1}{3}i( r'_e +2 r'_a) C^{(\prime)0}
\non\\
&&-\frac{1}{3} i r'_a P^{(\prime)u0}_{EW}
+\frac{1}{9}i(r'_e +2 r'_a) P_{EW}^{(\prime)uC0}
+ \frac{1}{3}i(\bar r'_e+2\bar r'_a)(\bar C^{(\prime)0}_2+2\bar P^{(\prime)u0}-\frac{1}{3}\bar P^{(\prime)uC0}_{EW,2}),
\non\\
P^{(\prime)c}&=&\left[1+ir'_0+\frac{1}{3}i(- 2r'_e +5 r'_a)\right] P^{(\prime)c0}
\non\\
&&-\frac{1}{3} i r'_a P^{(\prime)c0}_{EW}
+\frac{1}{9}i(r'_e +2 r'_a) P_{EW}^{(\prime)cC0}
+ \frac{1}{3}i(\bar r'_e+2\bar r'_a)(2\bar P^{(\prime)c0}-\frac{1}{3}\bar P^{(\prime)cC0}_{EW,2}).
\en
Using these formulas and the best fit parameters, we obtain
\be
\frac{P^{(\prime)u}}{P^{(\prime)u0}}
=0.99 e^{i 20.0^\circ}+1.53 e^{-i70.2^\circ}+0.41 e^{i 49.4^\circ}+0.01^{-i 56.1^\circ}+0.10^{-i 2.3^\circ}
=1.99 e^{-i 23.6^\circ},
\label{eq: PuPu0}
\en
where the terms in the right hand side of the first equality are from $P^{(\prime)u0}$, $T^{(\prime)0}$, $C^{(\prime)0}$, $P^{(\prime)u0, (\prime)uC0}_{EW}$
and
$\bar C^{(\prime)0}_2+2\bar P^{(\prime)u0}-\frac{1}{3}\bar P^{(\prime)uC0}_{EW,2}$, respectively,
and
\be
\frac{P^{(\prime)c}}{P^{(\prime)c0}}=0.99 e^{i 20.0^\circ}+0.01^{-i 46.8^\circ}+0.15^{-i 94.6^\circ}
=0.94 e^{i 11.2^\circ},
\en
where the terms in the right hand side of the first equality are from $P^{(\prime)c0}$, $P^{c(\prime)0, c(\prime)C0}_{EW}$
and
$2\bar P^{(\prime)c0}-\frac{1}{3}\bar P^{(\prime)cC0}_{EW,2}$, respectively.
It is clear that $T^{(\prime)0}$ and $C^{(\prime)0}$ only contribute to $P^{(\prime)u}$ (via the annihilation rescattering $r'_a$) 
and $|P^{(\prime)u}|$ is enhanced by about a factor of 2. 
On the other hand $P^{(\prime)c}$ is only slightly affected by rescattering and is still close to the original $P^{(\prime)c0}$.
The results are shown in Table~\ref{tab:ucpenguinsDS0-1}.

It is useful to note that the ratio of $u$-penguin and $c$-penguin in $\Delta S=0$ process before rescattering is expected to proportional to the CKM factors giving
\be
\bigg|\frac{P^u}{P^c}\bigg|\simeq \bigg|\frac{V_{ub} V^*_{ud}}{V_{cb} V^*_{cd}}\bigg|\simeq 0.38.
\en
The estimation is close to the ratio $|P^{u0}/P^{c0}|=3.51/10.09\simeq0.35$ using $P^{u0}$ and $P^{c0}$ shown in Table~\ref{tab:ucpenguinsDS0-1}.
The CKM ratio implies that $u$-penguin and the $c$-penguin are not as hierarchical as in the $\Delta S=-1$ case.
Furthermore, when rescattering is turned on, the $u$-penguin and $c$-penguin receive different contributions as only $P^u$ can receive contribution fed from $T^0$,
see Eq.~(\ref{eq: PuPu0}),
and, consequently, the above ratio is enhanced to $6.98/9.49\simeq 0.74$ (see Table~\ref{tab:ucpenguinsDS0-1}). 
These will affect the CP asymmetries of $\Delta S=0$ modes to be discussed later.

We now turn to $PA^{(\prime)}$. 
Similarly we decompose $PA^{(\prime)}$ into  $PA^{(\prime)u}+PA^{(\prime)c}$ and from Eq.~(\ref{eq: delta TA1}) we have
\be
PA^{(\prime)u}&=& 
\frac{1}{3}(3+3ir'_0
- ir'_e +16 ir'_a +12 ir'_t) PA^{(\prime)u0}
+ i r'_t T^{(\prime)0}
+\frac{1}{9}(2 i r'_e +4 i r'_a -3 i r'_t) C^{(\prime)0}
\non\\
&&
+\frac{2}{9}(i r'_e+11 ir'_a+12  ir'_t) E^{(\prime)0}
+\frac{2}{9}( ir'_e+11 ir'_a +12 ir'_t ) P^{(\prime)u0}
\non\\
&&
+\bigg(-\frac{1}{3} i r'_t P^{(\prime)u0}_{EW}
+\frac{1}{27}i(-2  r'_e -4 r'_a +3 r'_t) P^{(\prime)uC0}_{EW}
-\frac{2}{27} i (r'_e +11 r'_a +12 r'_t )P_{EW}^{(\prime)uE0}\bigg)
\non\\
&&-\frac{2}{9}i(\bar r'_e+2\bar r'_a)
\bigg(\bar C^{(\prime)0}_1+2\bar E^{(\prime)0}+\bar C^{(\prime)0}_2+2\bar P^{(\prime)u0}-\frac{1}{3}\bar P^{(\prime)uC0}_{EW,2}
-\frac{1}{3}\bar P^{(\prime)uC0}_{EW,1} -\frac{2}{3}\bar P^{(\prime)uE0}_{EW}\bigg)
\non\\
&&+\frac{1}{3}i(\hat r'_t +\frac{4\hat r'_a +2\hat r'_e}{3})
\bigg(\tilde C^{(\prime)0}+\tilde E^{(\prime)0}+\tilde P^{(\prime)u0} +\frac{3}{2}\widetilde {PA}^{(\prime)u0}-\frac{1}{3}\tilde P^{(\prime)uC0}_{EW} -\frac{1}{3}\tilde P^{(\prime)uE0}_{EW}\bigg),
\non\\
PA^{(\prime)c}&=& 
\frac{1}{3}(3+3ir'_0
- ir'_e +16 ir'_a +12 ir'_t) PA^{(\prime)c0}
+\frac{2}{9}( ir'_e+11 ir'_a +12 ir'_t ) P^{(\prime)c0}
\non\\
&&+\bigg(-\frac{1}{3} i r'_t P^{(\prime)c0}_{EW}
+\frac{1}{27}i(-2  r'_e -4 r'_a +3 r'_t) P^{(\prime)cC0}_{EW}
-\frac{2}{27} i (r'_e +11 r'_a +12 r'_t )P_{EW}^{(\prime)cE0}
\bigg)
\non\\
&&-\frac{2}{9}i(\bar r'_e+2\bar r'_a)
\bigg(2\bar P^{(\prime)c0}-\frac{1}{3}\bar P^{(\prime)cC0}_{EW,2}
-\frac{1}{3}\bar P^{(\prime)cC0}_{EW,1} -\frac{2}{3}\bar P^{(\prime)cE0}_{EW})\bigg)
\non\\
&&+\frac{1}{3}i(\hat r'_t +\frac{4\hat r'_a +2\hat r'_e}{3})
\bigg(\tilde P^{(\prime)c0} +\frac{3}{2}\widetilde {PA}^{(\prime)c0}-\frac{1}{3}\tilde P^{(\prime)cC0}_{EW}
-\frac{1}{3}\tilde P^{(\prime)cE0}_{EW}\bigg),
\en
Using these formulas and the best fit parameters, we obtain
\be
\frac{PA^{(\prime)u}}{PA^{(\prime)u0}}
&=&0.94 e^{i 2.2^\circ}
+22.39 e^{-i67.4^\circ}
+6.17 e^{i 53.7^\circ}
+0.78^{i 91.1^\circ}
\non\\
&&
+2.31^{i 96.8^\circ}
+0.09^{-i 47.6^\circ}
+1.91^{i 28.5^\circ}
+1.52^{-i 99.7^\circ}
\non\\
&=&19.55 e^{-i 42.7^\circ},
\label{eq: PAu/PAu0}
\en
where the terms in the right hand side of the first equality are from 
$PA^{(\prime)u0}$, 
$T^{(\prime)0}$, 
$C^{(\prime)0}$, 
$E^{(\prime)0}$, 
$P^{(\prime)u0}$, 
$P^{(\prime)u0, (\prime)uC0,(\prime)uE0}_{EW}$,
$\bar C^{(\prime)0}_1+2\bar E^{(\prime)0}+\bar C^{(\prime)0}_2+2\bar P^{(\prime)u0}-\frac{1}{3}\bar P^{(\prime)uC0}_{EW,2}
-\frac{1}{3}\bar P^{(\prime)uC0}_{EW,1} -\frac{2}{3}\bar P^{(\prime)uE0}_{EW}$
and 
$\tilde C^{(\prime)0}+\tilde E^{(\prime)0}+\tilde P^{(\prime)u0} +\frac{3}{2}\widetilde {PA}^{(\prime)u0}-\frac{1}{3}\tilde P^{(\prime)uC0}_{EW} -\frac{1}{3}\tilde P^{(\prime)uE0}_{EW}$
contributions, respectively.
Note that $|PA^{(\prime)u}|$ is enhanced by a factor of 18,
and the main contributions are from $T^{(\prime)0}$, $C^{(\prime)0}$ and $P^{(\prime)u0}$ terms via the total annihilation rescattering $r'_t$, the annihilation $r'_a$ and total annihilation $r'_t$ rescatterings, respectively.
In particular, the enhancement from $T^{(\prime)0}$ via $r'_t$ is the most prominent one.

Similarly we have
\be
\frac{PA^{(\prime)c}}{PA^{(\prime)c0}}
=0.94 e^{i 2.2^\circ}
+2.51 e^{i 88.2^\circ}
+0.09^{-i 47.0^\circ}
+1.26^{-i 97.3^\circ}
+1.41 e^{i 117.7^\circ}
=2.49 e^{i 84.0^\circ},
\label{eq: PAc/PAc0}
\en
where the terms in the right hand side of the first equality are from $PA^{(\prime)c0}$, 
$P^{(\prime)c0}$, 
$P^{(\prime)c0, (\prime)cC0,(\prime)cE0}_{EW}$,
$2\bar P^{(\prime)c0}-\frac{1}{3}\bar P^{(\prime)cC0}_{EW,2}
-\frac{1}{3}\bar P^{(\prime)cC0}_{EW,1} -\frac{2}{3}\bar P^{(\prime)cE0}_{EW}$
and 
$\tilde P^{(\prime)c0} +\frac{3}{2}\widetilde {PA}^{(\prime)c0}-\frac{1}{3}\tilde P^{(\prime)cC0}_{EW} -\frac{1}{3}\tilde P^{(\prime)cE0}_{EW}$
contributions, respectively. Note that $|PA^{(\prime)c}|$ is enhanced by a factor of $2.5$, while the main contribution is from the $P^{(\prime)c0}$ term 
via the annihilation $r'_a$ and total annihilation $r'_t$ rescatterings. The effect of rescattering in $PA^{(\prime)c}$ is not as prominent as in the $PA^{(\prime)u}$ case.

We see that in the presence of rescattering, the resulting $|PA^u|$ is even greater than $|PA^c|$, while $PA^{(\prime)u}$ can no longer be neglected (see Table~\ref{tab:ucpenguinsDS0-1}).
The above observations can shed light on the results in the following discussions.

\subsection{Numerical results for decay rates and CP asymmetries}
 
In this part we will present the numerical results on 
rates in $\overline B {}^0$ and $B^-$ decays, 
direct CP violations in $\overline B {}^0$ and $B^-$ Decays,
rates and direct CP asymmetries in $\ov B{}^0_s$ decays,
and 
time-dependent CP violations in $\ov B{}^0$ and $\ov B{}^0_s$ decays.

\subsubsection{Rates in $\overline B {}^0$ and $B^-$ Decays}

In Table~\ref{tab:table-br}, we show the CP-average rates of $\ov
B{}^0,B^-\to PP$ decays. In the table, Fac and Res denote
the factorization (without rescattering) and the rescattering results, respectively. 
To see the effect of rescattering, we also show the results from the rescattering solution, but with all rescattering phases turn off, i.e. with rescattering turn off,
in the parentheses.
In the table the contributions from various modes to $\chi^2_{\rm min}$ in the best fitted solutions
are also shown.

\begin{table}[t!]
\caption{ \label{tab:table-br} 
\small Branching ratios of various $\overline B{}_{u,d}\to PP$ modes in units of $10^{-6}$.
Fac and Res denote factorization and rescattering
results, respectively. 
Experimental results are taken from \cite{HFAG, PDG}. 
Contributions to $\chi^2_{\rm min}$ from the best fitted solutions 
are also shown. The values in parenthesis are the results from the rescattering solution, but with all rescattering rescattering phases turn off.}
\begin{ruledtabular}
\begin{tabular}{lccccc}
 Mode
      &Exp
      &Fac
      &Res
      & $\chi^{2\,\rm (Fac)}_{\rm min}$ 
      & $\chi^{2\,\rm (Res)}_{\rm min}$ 
      \\
\hline
 $\ov B{}^0\to K^-\pi^+$
        & $19.57^{+0.53}_{-0.52}$
        & $19.3_{-0.3}^{+0.3}$ 
        & $20.7_{-0.3}^{+0.3}$ (23.1)
        & 0.2
        & 4.7 (44.7)
        \\
 $\ov B {}^0\to \ov K {}^0\pi^0$
        & $9.93\pm0.49$
        & $8.5\pm0.1$ 
        & $9.6_{-0.1}^{+0.2}$ (10.7)
        & 8.2
        & 0.4 (2.4)
        \\
 $\overline B {}^0\to \ov K {}^0\eta$
        & $1.23_{-0.24}^{+0.27}$
        & $1.3\pm0.1$ 
        & $1.6\pm0.1$ (1.6) 
        & 0.0
        & 1.6 (1.8)
        \\
 $\overline B {}^0\to \ov K {}^0\eta'$
        & $66.1\pm3.1$
        & $70.0\pm1.2$ 
        & $68.3_{-1.4}^{+2.6}$ (64.6)
        & 1.6
        & 0.5 (0.2)
        \\
        \hline
 $B^-\to \ov K{}^0\pi^-$
        & $23.79\pm0.75$
        & $21.1\pm0.3$ 
        & $22.5\pm0.3$ (25.4)
        & 11.8
        & 3.1 (4.5)
        \\
 $B^-\to K^-\pi^0$
        & $ 12.94_{-0.51}^{+0.52} $
        & $12.1\pm0.1$ 
        & $12.3\pm0.2$ (13.8)
        & 2.7
        & 1.7 (2.7)
        \\
 $B^-\to K^-\eta$
        & $2.36_{-0.21}^{+0.22}$
        & $1.7\pm0.1$
        & $2.1_{-0.2}^{+0.1}$ (2.1)
        & 8.2
        & 1.5 (1.1)
        \\
 $B^-\to K^-\eta'$
        & $71.1\pm 2.6$
        & $74.7\pm1.3$
        & $71.4_{-1.5}^{+2.9}$ (66.5)
        & 1.9
        & 0.0 (3.1)
        \\
        \hline
 $B^-\to \pi^-\pi^0$
        & $5.48^{+0.35}_{-0.34}$
        & $4.7\pm0.1$
        & $4.9_{-0.1}^{+0.2}$ (4.9)
        & 5.7
        & 2.9 (2.9)
        \\
 $B^-\to K^0 K^-$
        & $1.32\pm0.14$
        & $1.43\pm 0.03$
        & $1.31\pm0.03$ (1.5)
        & 0.6
        & 0.0 (1.8)
        \\
 $B^-\to \pi^-\eta$
        & $4.02\pm0.27$
        & $3.4\pm0.1$
        & $4.2\pm0.1$ (4.3)
        & 4.1
        & 0.3 (0.8)
        \\
 $B^-\to \pi^-\eta'$
        & $2.7^{+0.5}_{-0.4}$
        & $2.9\pm0.1$
        & $3.5\pm0.1$ (3.3)
        & 0.1
        & 3.3 (1.5)
        \\
\hline
 $\ov B {}^0\to \pi^+\pi^-$
        & $5.10\pm 0.19$
        & $6.2\pm0.1$
        & $5.3\pm0.1$ (6.0)
        &  36.1
        &   0.7 (23.7)
        \\
 $\ov B {}^0\to \pi^0 \pi^0$
        & $1.59\pm0.26$\footnotemark[1]
        & $0.98_{-0.03}^{+0.05}$
        & $1.09_{-0.05}^{+0.06}$ (0.82)
        & 5.6
        & 3.7 (9.7)
        \\
 $\ov B {}^0\to \eta\eta$
        & $0.76\pm0.29$
        & $0.28\pm0.01$
        & $0.41_{-0.06}^{+0.04}$ (0.11)
        & 2.8
        & 1.5 (5.1)
        \\
 $\ov B {}^0\to \eta \eta'$
        & $0.5\pm0.4(<1.2)$
        & $0.32_{-0.01}^{+0.02}$
        & $0.30_{-0.04}^{+0.05}$ (0.26)
        & 0.2
        & 0.2 (0.4)
        \\
 $\ov B {}^0\to \eta'\eta'$
        & $0.6\pm0.6(<1.7)$
        & $0.24\pm0.01$
        & $0.40_{-0.12}^{+0.15}$ (0.08)
        & 0.4
        & 0.1 (0.7)
        \\
 $\ov B {}^0\to K^+ K^-$
        & $0.084\pm0.024$
        & $0.065\pm0.002$
        & $0.100_{-0.007}^{+0.012}$ (0.03)
        & 0.6
        & 0.5 (4.3)
        \\
 $\ov B {}^0\to K^0\ov K^0$
        & $1.21\pm0.16$
        & $1.67\pm0.03$
        & $1.19\pm0.03$ (1.21)
        & 8.4
        & 0.0 (0.0)
        \\
 $\ov B {}^0\to \pi^0 \eta$
        & $0.41\pm0.17$
        & $0.37\pm0.01$
        & $0.36_{-0.00}^{+0.02}$ (0.41)
        & 0.1
        & 0.1 (0.0)
        \\
 $\ov B {}^0\to \pi^0\eta'$
        & $1.2\pm0.6$\footnotemark[2]
        & $0.52\pm0.02$
        & $0.60\pm0.02$ (0.47)
        & 1.3
        & 1.0 (1.5)
        \\
 \end{tabular}
 \footnotetext[1]{An $S$ factor of 1.4 is included in the uncertainty.}
 \footnotetext[2]{Taken from PDG with an $S$ factor of 1.7 included in the uncertainty.}
 \end{ruledtabular}
\end{table}

From the table, we see that, except for rates in $\ov B{}^0\to K^-\pi^+$, $\ov K{}^0\eta$ and $B^-\to
\pi^-\eta^{\prime}$ 
decays, the $\chi^2$ in Res for the other modes are lower than the Fac ones. 
In particular, the $\chi^2$ in the $\ov B{}^0\to \ov K{}^0\pi^0$, $\pi^+\pi^-$, $K^0\ov K{}^0$ and $B^-\to\ov K{}^0\pi^-$, 
$K^-\eta$, $\pi^-\pi^0$, $\pi^-\eta$ rates are improved significantly, as Fac encounters difficulties to fit some of these rates well. 
In fact, in Fac the $\chi^2$ in $\ov B{}^0\to\pi^+\pi^-$ is as large as $36.1$, while it is reduced to $0.7$ in Res.
We see that in each group the $\chi^2$ is improved in the presence of Res.
The total $\chi^2$ from these 21($=4+4+4+9$) modes reduced from 
$100.7(=10.1 + 24.7 + 10.6 + 55.3)$ to 
$27.7(=7.2 + 6.3 + 6.4 + 7.8)$ (the breakdown can be found in Table~\ref{tab:chisquare} as well).
Overall speaking rescattering significantly improves the fit in this sector, especially in the last group, and can reproduce all the measured $\ov B_{u,d}\to PP$ rates reasonably well.

Note that both Fac and Res can successfully reproduce the newly measured $\ov B{}^0\to \pi^0\eta$ and $ K^+K^-$ rates~\cite{Pal:2015ewa,Aaij:2016elb}.
On the other hand,
both Fac and Res results on the $\ov B{}^0\to\pi^0\pi^0$ rate have tension with the data, 
while Res is somewhat better as its $\chi^2$ (=3.7) is smaller than the one (5.6) in Fac. 
It should be note that the uncertainty in the present data is still large
and it will be interesting to see the updated measurement.  
Both Fac and Res fits on the $B^-\to \pi^-\pi^0$ rates are smaller than the experimental result. 
The $\chi^2$ from Fac on this mode is $5.7$, while the Res fit improves it to $2.9$ with a slightly large rate, 
but both results are in tension with data.

We will investigate how rescattering improves the fit in $\ov B {}^0_{d}\to \pi^+\pi^-, \pi^0\pi^0, K^+K^-$ and $B^-\to\ov K{}^-\pi^0$ rates.
For simplicity we will concentrate on the dominant contributions to the decay amplitudes in the following discussion.
By neglecting the electroweak penguin contributions, the $\ov B{}^0\to\pi^+\pi^-$ amplitude in Eq.~(\ref{eq: TAgroup4}) can be expressed as
\be
A_{\ov B {}^0_{d}\to \pi^+\pi^-}
&\simeq&T+P+E+PA.
\label{eq: pipi TA}
\en
Using the results in Sec.~\ref{subsec: FSITA}, we see that before rescattering and after rescattering, we have (in unit of $10^{-8}$ GeV)
\be
(A_{\ov B {}^0_{d}\to \pi^+\pi^-})^0
&\simeq& 2.58e^{-i 63.5^\circ}+0.89 e^{i8.6^\circ}+0.19 e^{i102.6^\circ}+0.08 e^{-i166.4^\circ}
\non\\
&\simeq& 2.98 e^{-i 47.0^\circ}+0.14 e^{i 135.4^\circ}\simeq 2.84 e^{-i 47.1^\circ},
\non\\
(A_{B {}^0_{d}\to \pi^+\pi^-})^0
&\simeq& 2.58e^{i 74.2^\circ}+0.99 e^{-i31.6^\circ}+0.12e^{-i119.7^\circ}+0.08 e^{i149.3^\circ}
\non\\
&\simeq& 2.50 e^{i 51.8^\circ}+0.14 e^{-i 152.6^\circ}\simeq 2.38 e^{i 53.2^\circ},
\non\\
(A_{\ov B {}^0_{d}\to \pi^+\pi^-})_{FSI}
&\simeq& 2.58e^{-i 63.5^\circ}+1.23e^{i34.3^\circ}+0.55e^{i71.0^\circ}+0.79 e^{-i116.2^\circ}
\non\\
&\simeq& 2.71 e^{-i 136.8^\circ}+0.26 e^{-i 131.4^\circ}\simeq 2.70 e^{-i 42.4^\circ},
\non\\
(A_{B {}^0_{d}\to \pi^+\pi^-})_{FSI}
&\simeq& 2.58e^{i 74.2^\circ}+0.64 e^{-i47.6^\circ}+0.55e^{-i152.2^\circ}+0.52 e^{-i1.6^\circ}
\non\\
&\simeq& 2.31e^{i 60.6^\circ}+0.28e^{-i81.7^\circ}\simeq 2.10 e^{i 55.9^\circ},
\label{eq: pipiTA}
\en
respectively,
where expressions with four terms are given in the order of $T$, $P$, $E$ and $PA$ and those in two terms are
with the first two terms ($T+P$) and the last two terms ($E+PA$) summed separately.
Before we proceed we may compare the above estimation to our full numerical results, 
where we have $(A_{\ov B {}^0_{d}\to \pi^+\pi^-})^0$, $(A_{B {}^0_{d}\to \pi^+\pi^-})^0$, $(A_{\ov B {}^0_{d}\to \pi^+\pi^-})_{FSI}$ and $(A_{B {}^0_{d}\to \pi^+\pi^-})_{FSI}$ given by 
$2.86 e^{-i 47.1^\circ}$, 
$2.36 e^{i 52.8^\circ}$, 
$2.71 e^{-i40.8^\circ}$ and 
$2.16 e^{i57.2^\circ}$ (in unit of $10^{-8}$ GeV), respectively, which are close to the above estimation.

Note that $T+P$ are dominant contributions, while $E+PA$ are sub-leading contributions, and these two groups interfere destructively.
In the presence of rescattering, the sizes of the dominant parts, $T+P$, are reduced, 
while the sizes of the destructive and sub-leading parts, $E+PA$, are enhanced,
resulting more effective destructive interferences.
From the estimation we see that 
the $\ov B {}^0_{d}\to \pi^+\pi^-$ rate is reduced by about $15\%$ bringing $\B(\ov B{}^0\to\pi^+\pi^-)\simeq 6\times 10^{-6}$ down to $\sim 5\times 10^{-6}$, 
which agrees well with the data [$(5.1\pm 0.19)\times 10^{-6}$] shown in Table~\ref{tab:table-br} 
and, consequently, the quality of the fit is improved significantly.

Similarly for $\overline B{}^0\to\pi^0\pi^0$ decays, we have
\be
\sqrt 2 A_{\ov B {}^0_{d}\to \pi^0 \pi^0}
 &\simeq&-C+P+E+ PA, 
\en
which is close to the above $\overline B{}^0\to\pi^+\pi^-$ amplitudes, but with $T$ replaced by $-C$.
Before rescattering and after rescattering, we have (in unit of $10^{-8}$ GeV)
\be
\sqrt2 (A_{\ov B {}^0_{d}\to \pi^0\pi^0})^0
&\simeq& 1.05e^{i 56.1^\circ}+0.89 e^{i8.6^\circ}+0.19 e^{i102.6^\circ}+0.08 e^{-i166.4^\circ}
\non\\
&\simeq& 1.82e^{i37.9^\circ}+0.08 e^{-i166.4^\circ}\simeq 1.75 e^{i 38.9^\circ},
\non\\
\sqrt 2 (A_{B {}^0_{d}\to \pi^0 \pi^0})^0
&\simeq& 1.05e^{-i 166.2^\circ}+0.99 e^{-i31.6^\circ}+0.12e^{-i119.7^\circ}+0.08 e^{i149.3^\circ}
\non\\
&\simeq&
0.90e^{-i104.5^\circ}+0.08 e^{i149.3^\circ}\simeq 0.88 e^{-i 109.2^\circ},
\non\\
\sqrt 2 (A_{\ov B {}^0_{d}\to \pi^0\pi^0})_{FSI}
&\simeq&1.05e^{i 56.1^\circ}+1.23e^{i34.3^\circ}+0.55e^{i71.0^\circ}+0.79 e^{-i116.2^\circ}
\non\\
&\simeq&  2.73e^{i49.5^\circ}+0.79 e^{-i116.2^\circ}\simeq 1.97 e^{i 43.8^\circ},
\non\\
\sqrt 2 (A_{B {}^0_{d}\to \pi^0\pi^0})_{FSI}
&\simeq&1.05e^{-i 166.2^\circ}+0.64 e^{-i47.6^\circ}+0.55e^{-i152.2^\circ}+0.52 e^{-i1.6^\circ}
\non\\
&\simeq&1.45 e^{-i137.0^\circ}+0.52 e^{-i1.6^\circ}\simeq 1.14 e^{-i 118.3^\circ},
\en
respectively,
where terms are given in the order of $-C$, $P$, $E$ and $PA$ 
and the expressions with the first three terms ($-C+P+E$) combined are also shown.
The above estimation is close to the values in the full numerical results with $\sqrt2 (A_{\ov B {}^0_{d}\to \pi^0\pi^0})^0$, $\sqrt 2 (A_{B {}^0_{d}\to \pi^0 \pi^0})^0$, $\sqrt 2 (A_{\ov B {}^0_{d}\to \pi^0\pi^0})_{FSI}$ and $\sqrt 2 (A_{B {}^0_{d}\to \pi^0\pi^0})_{FSI}$
given by 
$1.67 e^{i39.6^\circ}$, 
$0.89 e^{-i 114.8^\circ}$, 
$1.96 e^{i45.0^\circ}$ and 
$1.08 e^{-i 126.0^\circ}$ in the unit of $10^{-8}$ GeV, 
respectively.

In the above estimation the first three terms and the last term interfere destructively.
With Res $P$ and $E$ are enhanced giving a larger $-C+P+E$, 
while the enhanced $PA$ cannot be neglected anymore, 
producing a slightly larger decay amplitude and resulting a 35\% enhancement in rate, 
which brings the rate up from $\B(\ov B{}^0\to\pi^0\pi^0)\simeq 0.8\times 10^{-6}$ to $\sim 1.1\times 10^{-6}$ as shown in Table~\ref{tab:table-br}.
As noted previously the rate is still smaller than the central value of the data, which however accompanies with large uncertainty. 

For the newly observed $\ov B {}^0_{d}\to K^+ K^-$ mode, we note that as shown in Table~\ref{tab:table-br} rescattering enhances the rate by $0.100/0.03=3.33$ times.
It will be useful to see the enhancement in details. 
From Tables~\ref{tab:TADS0} and \ref{tab:ucpenguinsDS0-1} and Eq.~(\ref{eq: TAgroup4}),
\be
A_{\ov B {}^0_{d}\to K^+ K^-}&=&E+PA^u+PA^c+\frac{1}{3} P_{EW}^A,
\en
we have (in unit of $10^{-8}$ GeV)
\be
(A_{\ov B {}^0_{d}\to K^+ K^-})^0
&\simeq& 0.119 e^{i102.6^\circ}+0.031 e^{-i 77.4^\circ}+0.081 e^{i171.4^\circ}+0.001 e^{i 13.6^\circ}
\non\\
&\simeq& 0.139 e^{i 135.2^\circ},
\non\\
(A_{B {}^0_{d}\to K^+ K^-})^0   
&\simeq& 0.119 e^{-i119.7^\circ}+0.031 e^{i 60.3^\circ}+0.081 e^{i171.4^\circ}+0.001 e^{-i 30.7^\circ}
\non\\
&\simeq& 0.139 e^{-i 152.4^\circ},
\non\\
(A_{\ov B {}^0_{d}\to K^+ K^-} )_{FSI}  
&\simeq& 0.546 e^{i 71.0^\circ}+0.598 e^{-i120.1^\circ}+0.202 e^{-i 104.6^\circ}+0.006 e^{-i24.7^\circ}
\non\\
&\simeq& 0.261 e^{-i 130.1^\circ}
\non\\
(A_{B {}^0_{d}\to K^+ K^-} )_{FSI}       
&\simeq&0.546 e^{-i151.2^\circ}+0.552 e^{i17.1^\circ}+0.202 e^{-i 105.4^\circ}+0.006 e^{-i68.8^\circ}
\non\\
&\simeq& 0.286 e^{-i 81.4^\circ},
\label{eq: B0KK}
\en
for the decay amplitudes before and after rescattering,
where terms are given in the order of $E$, $PA^u$, $PA^c$ and $P_{EW}^A/3$.
Compare the above estimation to the values in our full numerical result, which have
$0.200 e^{i 135.0^\circ}$, 
$0.200 e^{-i 152.6^\circ}$, 
$0.332 e^{-i 139.8^\circ}$, and 
$0.350 e^{-i 87.2^\circ}$ for
$(A_{\ov B {}^0_{d}\to K^+ K^-})^0$, $(A_{B {}^0_{d}\to K^+ K^-})^0 $, $(A_{\ov B {}^0_{d}\to K^+ K^-} )_{FSI}$ 
and $(A_{B {}^0_{d}\to K^+ K^-} )_{FSI}$ in unit of $10^{-8}$ GeV, respectively.
The discrepancy is mainly from SU(3) breaking effects, which are not included in the above equation.
In fact, by scaling the numbers in Eq. (\ref{eq: B0KK}) by $(f_K/f_\pi)^2$, the sizes become 
0.199, 0.199, 0.373 and 0.409, 
which agree better to the above values now. 

From the above equation, we see that $E$, $PA^u$, $PA^c$ and $P^A_{EW}$ are all enhanced.
Note that $E$ interferes destructively with $PA^u$ and $PA^c$ in $A_{\ov B {}^0_{d}\to K^+ K^-}$, 
while $PA^u$ interferes destructively with $E$ and $PA^c$ in $A_{B {}^0_{d}\to K^+ K^-}$.  
The result is an enhancement of $3.8$ in the averaged rate,
which is close to our numerical result ($0.100/0.03=3.33$) as shown in Table~\ref{tab:table-br}.
We will return to this mode again in the discussion of direct CP asymmetry.

Finally we turn to the $B^-\to\overline K^0\pi^-$ decay. From Eq.~(\ref{eq: TAgroup2}) we have
\be
A_{B^-\to \ov K^0\pi^-}
&=&A'+P^{\prime u}+P^{\prime c}+\frac{1}{3}(-P_{EW}^{\prime\,C}+2P_{EW}^{\prime\, E}),
\en
which gives before and after rescattering (in unit of $10^{-8}$ GeV)
\be
(A_{B^-\to \ov K^0\pi^-})^0
&\simeq&
0.01 e^{-i77.4^\circ}
+0.08 e^{i 108.3^\circ}
+4.37 e^{i 168.6^\circ}
+0.08 e^{-i53.3^\circ}
\simeq
4.35 e^{i168.4^\circ},
\non\\
(A_{B^+\to  K^0\pi^+})^0
&\simeq&
0.01 e^{i60.3^\circ}
+0.08 e^{-i 113.9^\circ}
+4.37 e^{i 168.6^\circ}
+0.08 e^{-i51.2^\circ}
\simeq4.33 e^{i170.2^\circ},
\non\\
(A_{B^-\to \ov K^0\pi^-})_{FSI}
&\simeq&
0.11 e^{-i113.8^\circ}
+0.16 e^{i 84.7^\circ}
+4.11 e^{i 179.8^\circ}
+0.13 e^{-i50.3^\circ}
\simeq 4.06 e^{-i 179.7^\circ},
\non\\
(A_{B^+\to K^0\pi^+})_{FSI}
&\simeq&
0.11 e^{i23.9^\circ}
+0.16 e^{-i 137.6^\circ}
+4.11 e^{i 179.8^\circ}
+0.13 e^{-i48.2^\circ}
\simeq4.05 e^{-i 178.0^\circ},
\non\\
\label{eq: K0pi estimation}
\en
respectively,
where terms are given in the order of $A'$, $P^{\prime u}$, $P^{\prime c}$ and $(-P_{EW}^{\prime\,C}+2P_{EW}^{\prime\, E})/3$.
Note that in our numerical result, we have
$5.17 e^{i 167.2^\circ}$, 
$5.29 e^{i171.2^\circ}$, 
$4.86 e^{i 179.6^\circ}$ and 
$4.98 e^{i 175.7^\circ}$, for
$(A_{B^-\to \ov K^0\pi^-})^0$, $(A_{B^+\to  K^0\pi^+})^0$, $(A_{B^-\to \ov K^0\pi^-})_{FSI}$ and $(A_{B^+\to K^0\pi^+})_{FSI}$ in unit of $10^{-8}$ GeV, respectively.
By scaling the values in the Eq.~(\ref{eq: K0pi estimation})  by $f_K/f_\pi$, the sizes become 
$5.20$, $5.17$, $4.85$ and $4.83$, respectively, 
which are close to the numerical results. 
In the full numerical result either in the presence of rescattering or without it, the sizes of $A_{B^+\to  K^0\pi^+}$ is slightly greater than $A_{B^-\to \ov K^0\pi^-}$, 
but it is the other way around in the estimation.
In fact, in the numerical result, we have 
$P^{\prime u}=0.10 e^{i 107.8^\circ}$ and $P^{\prime c}=5.19 e^{i167.5^\circ}$ in $(A_{B^-\to \ov K^0\pi^-})^0$ and
$P^{\prime u}=0.10 e^{-i 114.4^\circ}$ and $P^{\prime c}=5.33 e^{i169.6^\circ}$ in $(A_{B^+\to K^0\pi^+})^0$. 
The latter $|P^{\prime c}|$ in $(A_{B^+\to K^0\pi^+})^0$ is greater than the one in $(A_{B^-\to \ov K^0\pi^-})^0$. The difference can be traced to the non-vanishing first Gegenbauer moment of the kaon wave function ($\alpha_1^{\bar K}=-\alpha_1^K=0.2)$, which will change sign in changing from $\ov K$ to $K$.
This will affect the direct CP asymmetry and such a feature is absent in the above estimation.

From Eq.~(\ref{eq: K0pi estimation}) we see that $A'+\frac{1}{3}(-P_{EW}^{\prime\,C}+2P_{EW}^{\prime\, E})$ interferes destructively to 
the dominating $P^{\prime c}$ term.
Since the sizes of $A'$ and $\frac{1}{3}(-P_{EW}^{\prime\,C}+2P_{EW}^{\prime\, E})$ are enhanced, 
while the size of $P^{\prime c}$ is slightly reduced, 
the size of the total amplitude is reduced under the rescattering resulting a reduction of $13\%$ in the averaged rate, 
which brings the rate from $\B(B^-\to\ov K{}^0\pi^+)\simeq 25\times 10^{-6}$ to
$\sim 22\times 10^{-6}$, which is closer to the data [$(23.79\pm0.75)\times 10^{-6}$] as shown in Table~\ref{tab:table-br}.


\subsubsection{Direct CP Violations in $\overline B {}^0$
and $B^-$ Decays}

\begin{table}[t!]
\caption{ \label{tab:table-acp} Same as Table~\ref{tab:table-br},
except for the direct CP asymmetries $\A$ (in units of percent)
in various $\overline B{}_{u,d}\to PP$ modes. }
\begin{ruledtabular}
\begin{tabular}{lccccc}
 Mode
      &Exp
      &Fac
      &Res
      &$\chi^{2\,\rm(Fac)}_{\rm min}$
      &$\chi^{2\,\rm(Res)}_{\rm min}$
      \\
\hline
 $\ov B{}^0\to K^-\pi^+$
        & $-8.2\pm0.6$
        & $-8.0\pm0.1$
        & $-8.2\pm0.3$ $(-9.5)$
        & 0.1
        & 0.0 (4.8)
        \\
 $\ov B {}^0\to \ov K {}^0\pi^0$
        & $-1\pm13$\footnotemark[1] 
        & $-15.2\pm0.6$
        & $-14.3\pm1.0$ $(-8.5)$
        & 1.2
        & 1.0 (0.3)
        \\
 $\overline B {}^0\to \ov K {}^0\eta$
        & --
        & $-29.3_{-1.9}^{+1.3}$
        & $-27.7_{-2.2}^{+1.4}$ $(-17.5)$
        & --
        & --
        \\
 $\overline B {}^0\to \ov K {}^0\eta'$
        & $5\pm4$
        & $7.8\pm0.2$
        & $6.1_{-0.9}^{+1.3}$ $(6.3)$
        & 0.5
        & 0.1 (0.1)
        \\
        \hline
 $B^-\to \ov K{}^0\pi^-$
        & $-1.7\pm1.6$ 
        & $-3.5\pm0.1$
        & $-2.4_{-0.4}^{+0.6}$ $(-2.3)$
        & 1.2
        & 0.2 (0.1)
        \\
 $B^-\to K^-\pi^0$
        & $4.0\pm 2.1 $
        & $4.0\pm0.4$
        & $4.9_{-1.1}^{+0.8}$ $(-1.9)$
        & 0.0
        & 0.2 (7.8)
        \\
 $B^-\to K^-\eta$
        & $-37\pm 8$
        & $-42.0_{-3.7}^{+2.5}$
        & $-33.9\pm2.6$ $(-10.8)$
        & 0.4
        & 0.1 (10.7)
        \\
 $B^-\to K^-\eta'$
        & $1.3\pm 1.7 $
        & $4.5_{-0.1}^{+0.2}$
        & $1.8_{-0.7}^{+1.6}$ $(2.7)$
        & 3.6
        & 0.1 (0.7)
        \\
        \hline
 $B^-\to \pi^-\pi^0$
        & $2.6\pm3.9$
        & $-0.11\pm0.00$
        & $-0.09\pm0.01$ $(-0.09)$
        & 0.5
        & 0.5 (0.5)
        \\
 $B^-\to K^0 K^-$
        & $-8.7\pm10$
        & $-5.7\pm0.1$
        & $-4.8_{-5.3}^{+3.8}$ $(-8.8)$
        & 0.1
        & 0.2 (0.0)
        \\
 $B^-\to \pi^-\eta$
        & $-14\pm5$
        & $-11.9_{-0.7}^{+0.8}$
        & $-10.3_{-1.6}^{+1.7}$ $(0.8)$
        & 0.2
        & 0.5 (8.7)
        \\
 $B^-\to \pi^-\eta'$
        & $6\pm15$
        & $37.8_{-1.3}^{+0.8}$
        & $43.6_{-2.4}^{+2.0}$ $(34.6)$
        & 4.5
        & 6.3 (3.6)
        \\
\hline
 $\ov B {}^0\to \pi^+\pi^-$
        & $31\pm5$
        & $14.0\pm0.4$
        & $22.5_{-1.0}^{+0.9}$ $(19.1)$
        & 11.5
        & 2.9 (5.7)
        \\
 $\ov B {}^0\to \pi^0 \pi^0$
        & $34\pm22$
        & $79.1_{-1.5}^{+1.2}$
        & $53.7_{-7.1}^{+3.3}$ $(55.9)$
        & 4.2
        & 0.8 (1.0)
        \\
 $\ov B {}^0\to \eta\eta$
        & --
        & $-64.5_{-1.4}^{+1.5}$
        & $-31.1_{-5.5}^{+7.2}$ $(-73.5)$
        & --
        & --
        \\
 $\ov B {}^0\to \eta \eta'$
        & --
        & $-35.6\pm1.1$
        & $-29.8_{-8.0}^{+9.4}$ $(-52.1)$
        & --
        & --
        \\
 $\ov B {}^0\to \eta'\eta'$
        & --
        & $-20.0\pm0.4$
        & $-7.6_{-19.8}^{+19.2}$ $(-12.9)$
        & --
        & --
        \\
 $\ov B {}^0\to K^+ K^-$
        & --
        & $0$
        & $-5.2_{-5.0}^{+5.2}$ (0)
        & --
        & --
        \\
 $\ov B {}^0\to K^0\ov K^0$
        & $-6\pm36$\footnotemark[2] 
        & $-8.4\pm 0.1$
        & $-41.8_{-3.9}^{+2.6}$ $(-10.0)$
        & 0.0
        & 1.0 (0.0)
        \\
 $\ov B {}^0\to \pi^0 \eta$
        & --
        & $-45.6_{-1.7}^{+1.8}$
        & $-40.9_{-3.6}^{+4.6}$ $(-36.3)$
        & --
        & --
        \\
 $\ov B {}^0\to \pi^0\eta'$
        & --
        & $-30.4_{-0.5}^{+0.9}$
        & $-8.8\pm1.4$ $(-8.8)$
        & --
        & --
        \\
 \end{tabular}
  \footnotetext[1]{An $S$ factor of 1.4 is included in the uncertainty.}
  \footnotetext[2]{An $S$ factor of 1.4 is included in the uncertainty.}
\end{ruledtabular}
\end{table}

Results for direct CP asymmetries ($\A$) in $\ov B{}_{u,d}\to PP$
decays are summarized in Table~\ref{tab:table-acp}. 
The Fac and Res fits give similar results in the first group of data, namely the direct CP asymmetries in
$\ov B{}^0\to K^-\pi^+$, $\ov K{}^0\pi^0$ and $\ov K{}^0\eta'$ decays.
Both can explain the so-call $K\pi$ CP puzzle by producing positive $\A(B^-\to K^-\pi^0)$ and negative $\A(\overline B{}^0\to K^-\pi^+)$, 
but the Res give a slightly larger $\A(B^-\to K^-\pi^0)$.
Fac fits better than Res in the $B^-\to \pi^-\eta'$ and $B{}^0\to K^0\ov K{}^0$ modes, 
while Res fits better than Fac in the $B^-\to \ov K{}^0\pi^-$, $K^-\eta'$, $\ov B{}^0\to\pi^+\pi^-$ and $\pi^0\pi^0$ modes.
In particular, the $\chi^2$ in $\A$ of $\ov B{}^0\to\pi^+\pi^-$ is reduced significantly from 11.5 (Fac) to 2.9 (Res).  
Overall speaking the fit in Res in this sector (see also Table II) is better than Fac, as the corresponding $\chi^2$ are 
13.9($=1.1+0.6+7.5+4.7$) and 29.2($=1.8+5.2+6.5+15.7$), respectively.

It is interesting to see how rescattering solve the so-call $K\pi$ CP puzzle, where  
experimental data gives $\Delta\A\equiv\A(K^-\pi^+)-\A(K^-\pi^0)=(12.2\pm2.2)\%$, in details.
The $\ov B{}^0\to K^-\pi^+$ and $B^-\to K^- \pi^0$ decay amplitudes can be expressed as
\be
 A_{\ov B {}^0_{d}\to K^-\pi^+}
 &=&T'+P'+\frac{1}{3}(2 P_{EW}^{\prime\,C}-P_{EW}^{\prime\,E}),
 \non\\
\sqrt 2A_{B^-\to K^- \pi^0}
&=&T'+C'+A'+P'+ P'_{EW}+\frac{2}{3} P_{EW}^{\prime\,C}+\frac{2}{3} P_{EW}^{\prime\, E}.
\en
It is useful to note that these two amplitudes are related by the following relation:
\be
\sqrt 2A_{B^-\to K^- \pi^0}
&=&A_{\ov B {}^0_{d}\to K^-\pi^+}
+C'+A'
+P'_{EW}
+P^{\prime E}_{EW}.
\en
Using the values in Table~\ref{tab:TADS-1} and the above equation, we have (in unit of $10^{-8}$ GeV and in the corresponding order of the above equation) before and after Res
\be
\sqrt 2(A_{B^-\to K^- \pi^0})^0
&\simeq& 
4.12e^{i173.0^\circ}
+0.24 e^{-i 123.9^\circ}
+0.01 e^{-i77.4^\circ}
+0.40 e^{-i 174.2^\circ}
+0.05 e^{-i32.7^\circ}
\non\\
&\simeq& 4.58 e^{i177.2^\circ},
\non\\
\sqrt 2(A_{B^+\to K^+ \pi^0})^0
&\simeq& 
4.45e^{i160.8^\circ}
+0.24 e^{i 13.8^\circ}
+0.01 e^{i60.3^\circ}
+0.40 e^{-i 172.1^\circ}
+0.05 e^{-i30.6^\circ}
\non\\
&\simeq& 4.55 e^{i161.5^\circ},
\non\\
\sqrt 2(A_{B^-\to K^- \pi^0})_{FSI}
&\simeq&
 3.90e^{-i 176.4^\circ}
 +0.24 e^{-i 123.9^\circ}
 +0.11 e^{-i 113.8^\circ}
 +0.40 e^{-i 174.2^\circ}
 +0.13 e^{-i 40.3^\circ}
 \non\\
 &\simeq& 4.43 e^{-i171.3^\circ},
\non\\
\sqrt 2(A_{B^+\to K^+ \pi^0})_{FSI}
&\simeq& 
4.18e^{i171.9^\circ}
+0.24 e^{i 13.8^\circ}
+0.11 e^{i 23.9^\circ}
+0.40 e^{-i 172.1^\circ}
+0.13 e^{-i38.2^\circ}
\non\\
&\simeq& 4.14 e^{i172.2^\circ},
\label{eq: Kpi0ACP}
\en
respectively.
In our full numerical results, for $\ov B{}^0\to K^-\pi^+$ decay, 
we have 
$4.91 e^{i 172.0^\circ}$, 
$5.40 e^{i 161.8^\circ}$, 
$4.68 e^{-i 176.8^\circ}$ and 
$5.08 e^{i 174.0^\circ}$ for
$(A_{\ov B {}^0_{d}\to K^-\pi^+})^0$,  $(A_{B {}^0_{d}\to K^+\pi^-})^0$, $(A_{\ov B {}^0_{d}\to K^-\pi^+})_{FSI}$ and 
$(A_{ B {}^0_{d}\to K^+\pi^-})_{FSI}$ in unit of $10^{-8}$ GeV, respectively, which are close to the scaled (by $f_K/f_\pi$) estimations,
$4.93 e^{i 177.2^\circ}$, 
$5.32 e^{i 161.5^\circ}$, 
$4.66 e^{-i 171.3^\circ}$ and 
$4.99 e^{i 172.2^\circ}$, from Eq.~(\ref{eq: Kpi0ACP}).
For $B^-\to K^-\pi^0$ decays, we have 
$5.40 e^{i 176.4^\circ}$, 
$5.50 e^{i 162.7^\circ}$, 
$5.26 e^{-i 171.6^\circ}$ and 
$5.01 e^{i 174.8^\circ}$ for
$\sqrt2(A_{B^-\to K^-\pi^0})^0$,  $\sqrt2(A_{B^-\to K^-\pi^0})^0$, $\sqrt2(A_{B^-\to K^-\pi^0})_{FSI}$ and 
$\sqrt2(A_{B^-\to K^-\pi^0})_{FSI}$ in unit of $10^{-8}$ GeV, respectively, which are close to the scaled (by $f_K/f_\pi$) estimations,
$5.48 e^{i 177.2^\circ}$, 
$5.44 e^{i 161.5^\circ}$, 
$5.29 e^{-i 171.3^\circ}$ and 
$4.94 e^{i 172.2^\circ}$, from Eq.~(\ref{eq: Kpi0ACP}).

From Eq.~(\ref{eq: Kpi0ACP}) we see that the asymmetries are 
$\A(\ov B {}^0_{d}\to K^-\pi^+)\simeq -7.7\%$, $\A(B^-\to K^- \pi^0)\simeq 0.6\%$ and $\Delta \A\simeq 8.3\%$ before Res, 
which are not too far from the values $-9.5\%$, $-1.9\%$ and $7.6\%$ shown in Table~\ref{tab:table-acp}, and
$\A(\ov B {}^0_{d}\to K^-\pi^+)\simeq -6.8\%$, $\A(B^-\to K^- \pi^0)\simeq 6.8\%$ and $\Delta\A\simeq 13.6\%$ after Res, 
which are close to the values $-8.2\%$, $4.9\%$ and $13.0\%$ shown in Table~\ref{tab:table-acp}. 
As noted in the discussion of the $B^-\to \ov K{}^0\pi^-$ rate in the last sub-section,
the first Gegenbauer moment of the kaon wave function is the main source of the discrepancies between the estimations and the full numerical results.

As shown in Eq.~(\ref{eq: Kpi0ACP}), it is interesting that before rescattering the $C'$ and $P'_{EW}$ terms are the sources of deviation of 
$\A(B^-\to K^- \pi^0)$ from $\A(\ov B {}^0_{d}\to K^-\pi^+)$,
while with the presence of Res, the sizes of $A'$ and $P^{\prime E}_{EW}$ are enhanced 
and hence further enlarges the deviation of $\A(\ov B {}^0_{d}\to K^-\pi^+)$ and $\A(B^-\to K^- \pi^0)$ producing a larger $\Delta \A$.
Note that comparing to the discussion in $\overline B{}^0\to\pi^+\pi^-$ and $\pi^0\pi^0$ decay rates [see discussion after Eq.~(\ref{eq: pipi TA})], 
we see that the correlation of the effects of Res on these two sectors is not prominent. 
Indeed, in the $\pi^0\pi^0$ mode the most affected TAs under rescattering are $P$, $E$ and $PA$,
while at here $A'$ and $P^{\prime E}_{EW}$ are the most affected and relevant ones.

We now turn to $\A(\ov B{}^0\to\pi^+\pi^-)$.
From previous discussion [see Eq. (\ref{eq: pipiTA})], 
we find that before Res 
$A(\overline B{}^0\to\pi^+\pi^-)\simeq2.84 e^{-i 47.1^\circ}\times 10^{-8}$ GeV and 
$A(B{}^0\to\pi^+\pi^-)\simeq2.38 e^{i 53.2^\circ}\times 10^{-8}$ GeV, 
giving $\A\simeq 18\%$,
while in the presence of Res, the sizes of the dominant parts, $T+P$, are reduced, 
but the sizes of the destructive but the sub-leading parts, $E+PA$, are enhanced,
resulting richer interferences, giving 
$A(\overline B{}^0\to\pi^+\pi^-)\simeq2.70 e^{-i 42.4^\circ}\times 10^{-8}$ GeV and 
$A(B{}^0\to\pi^+\pi^-)\simeq2.10 e^{i 55.9^\circ}\times 10^{-8}$ GeV,
and, consequently,
producing an enhanced $\A\simeq 24.7\%$, which is closer to the data, $(31\pm 5)\%$.

Note that the results of Fac and Res in $\A(\ov B{}^0\to K^0\ov K{}^0)$ are different, 
while with large uncertainty the present data, $\A(\ov B{}^0\to K^0\ov K{}^0)=(-6\pm36)\%$, allows both.  
Note that the uncertainty in the data is enlarged by an $S$ factor of 1.4, as Belle and BaBar give very different results in $\A(\ov B{}^0\to K_s K_s)$, namely,
Belle gives $\A(\ov B{}^0\to K_s K_s)=-0.38\pm0.38\pm0.5$~\cite{Nakahama:2007dg}, while BaBar gives $0.40\pm0.41\pm0.06$~\cite{Aubert:2006gm}.
The result of Res, 
$\A(\ov B {}^0\to K^0\ov K^0)=-0.418_{-0.039}^{+0.026}$,
prefers the Belle result.
One should be reminded that Res can reproduce the $\ov B{}^0\to K^0\ov K{}^0$ CP-averaged rate much better than Fac (see Table~\ref{tab:table-br}).
We need more data to clarify the situation and to verify these predictions.

It will be useful to see the effect of Res on the $\ov B {}^0_{d}\to K^0 \ov K {}^0$ direct CP asymmetry.
From Eq. (\ref{eq: TAgroup4}), we can approximate the $\ov B {}^0_{d}\to K^0 \ov K {}^0$ amplitude as
\be
A_{\ov B {}^0_{d}\to K^0 \ov K {}^0}
 &\simeq&P+PA\simeq P^u+PA^u+P^c+PA^c.
 \en
From Table~\ref{tab:ucpenguinsDS0-1}, before Res and after FSI, we have (in unit of $10^{-8}$ GeV)
\be
(A_{\ov B {}^0_{d}\to K^0 \ov K {}^0})^0&\simeq& 
0.35 e^{i 108.3^\circ}+0.03 e^{-i77.4^\circ}+1.01 e^{-i11.4^\circ}+0.08 e^{i 171.4^\circ}
\simeq 0.81 e^{i 8.1^\circ},
\non\\
(A_{B {}^0_{d}\to K^0 \ov K {}^0})^0
&\simeq& 0.35 e^{-i 113.9^\circ}+0.03 e^{i60.3^\circ}+1.01 e^{-i11.4^\circ}+0.08 e^{i 171.4^\circ}
\simeq 0.92 e^{-i 31.6^\circ},
\non\\
(A_{\ov B {}^0_{d}\to K^0 \ov K {}^0})_{FSI}
&\simeq& 0.70 e^{i 84.7^\circ}+0.60 e^{-i120.1^\circ}+0.95 e^{-i0.2^\circ}+0.20 e^{-i104.6^\circ}
\simeq 0.66 e^{-i 1.9^\circ},
\non\\
(A_{B {}^0_{d}\to K^0 \ov K {}^0})_{FSI}
&\simeq& 0.70 e^{-i 137.6^\circ}+0.60 e^{i17.6^\circ}+0.95 e^{-i0.2^\circ}+0.20 e^{-i104.6^\circ}
\simeq1.07 e^{-i 27.2^\circ},
\non\\
\label{eq: AK0K0 Res}
\en
respectively,
where the values of $P^u$, $PA^u$, $P^c$ and $PA^c$ are shown in the corresponding order.
In our full numerical result, we have 
$1.12 e^{i 8.6^\circ}$, 
$1.24 e^{-i 33.3^\circ}$, 
$0.90 e^{i 1.6^\circ}$ and 
$1.40 e^{-i 27.8^\circ}$ for 
$(A_{\ov B {}^0_{d}\to K^0 \ov K {}^0})^0$, $(A_{B {}^0_{d}\to K^0 \ov K {}^0})^0$,
$(A_{\ov B {}^0_{d}\to K^0 \ov K {}^0})_{FSI}$ and $(A_{B {}^0_{d}\to K^0 \ov K {}^0})_{FSI}$ in unit of $10^{-8}$ GeV, 
respectively, which are close to the scaled [by $(f_K/f_\pi)^2$] estimations,
$1.16 e^{i 8.1^\circ}$,  
$1.31 e^{-i 31.6^\circ}$, 
$ 0.95 e^{-i 1.9^\circ}$ and 
$1.53 e^{-i 27.2^\circ}$,
from Eq.~(\ref{eq: AK0K0 Res}).

In Eq. (\ref{eq: AK0K0 Res}), we see that both $P^u$ and the $PA^u$ terms are enhanced under Res (mainly through rescattering from $T^0$)
and produce richer inference pattern contributing to the direct CP asymmetry.
The $\ov B {}^0_{d}\to K^0 \ov K {}^0$ amplitude is reduced, 
while the amplitude of the conjugated decay mode, $B {}^0_{d}\to K^0 \ov K {}^0$, is enhanced under Res,
producing an enlarged direct CP asymmetry,
which is changed from $-12\%$ to $-45\%$ and hence close to the Belle result.

As shown in Table~\ref{tab:table-acp}, we see that before Res the direct CP asymmetry of $\ov B{}^0\to K^+K^-$ is vanishing. 
Indeed, as one can infer from Eq. (\ref{eq: B0KK}) that the rates of $\ov B{}^0\to K^+K^-$ and $B{}^0\to K^+K^-$ are the same before Res.
This can be understood in the following. 
In QCDF, $E$, $PA$ and $P_{EW}^A$ can be expressed in terms of the so-called $A^i_1$ and $A^i_2$ terms, 
and these $A^i_1$ and $A^i_2$ terms are identical 
when the asymptotic distribution amplitudes are used (as in the present case)~\cite{Beneke:2003zv}.
Since we have $A_{\ov B{}^0\to K^+K^-}=E+PA+P_{EW}^A/3$ 
and these three topological amplitudes all have a common strong phase resulting a vanishing direct CP asymmetry.
Note that in the presence of Res, $E$ and $PA^u$ are enhanced mostly from $T^0$ [see Eqs. (\ref{eq: E/E0}) and (\ref{eq: PAu/PAu0})], 
while $PA^c$ from $P^c$ [see Eq. (\ref{eq: PAc/PAc0})], consequently,
the strong phases of these terms are no longer degenerate.
In fact, from Eq. (\ref{eq: B0KK}) one can infer that the direct CP asymmetry is estimated to be $-18\%$, 
which can be compared to the value of $(-7.7_{-6.2}^{+6.0})\%$ obtained in the full numerical result as shown in Table~\ref{tab:table-acp}.

For prediction, we see that except $\ov B {}^0\to K^+ K^-$, the sizes of the predicted direct CP asymmetries from Res are smaller than those in Fac.

\subsubsection{Rates and Direct CP asymmetries in $\ov B{}^0_s$
Decays}

\begin{table}
\caption{ \label{tab:table-Bs} Same as Table~\ref{tab:table-br},
except for the branching ratios (upper table) in the unit of
$10^{-6}$ and direct CP asymmetries (lower table) in the unit of
percent for various $\overline B_s\to PP$ modes.}
\begin{ruledtabular}
\begin{tabular}{lccccc}
 Mode
      &Exp
      &Fac
      &Res
      &$\chi^{2\,\rm(Fac)}_{\rm min}$
      &$\chi^{2\,\rm(Res)}_{\rm min}$
      \\
\hline
 $\B(\ov B{}_s^0\to K^+\pi^-)$
        & $5.5\pm0.5$
        & $5.5\pm0.1$
        & $5.5_{-1.8}^{+0.4}$ (6.3)
        & 0.0
        & 0.0 (2.6)
        \\
 $\B(\ov B {}_s^0\to K^0\pi^0)$
        & --
        & $0.59\pm0.01$
        & $1.02_{-0.13}^{+3.64}$ (0.68)
        & --
        & --
        \\
 $\B(\ov B {}_s^0\to K^0\eta)$
        & --
        & $0.18_{-0.00}^{+0.01}$
        & $0.48_{-0.06}^{+1.87}$ (0.22)
        & --
        & --
        \\
 $\B(\ov B {}_s^0\to K^0\eta')$
        & --
        & $1.76\pm0.03$
        & $2.02_{-0.19}^{+4.30}$ (1.75)
        & --
        & --
        \\
        \hline
 $\B(\ov B {}_s^0\to \pi^+\pi^-)$
        & $0.671\pm 0.083$
        & $0.30\pm0.01$
        & $0.67_{-0.06}^{+0.49}$ (0.14)
        & 20.2
        & 0.0 (41.1)
        \\
 $\B(\ov B {}_s^0\to \pi^0 \pi^0)$
        & --
        & $0.15\pm0.00$
        & $0.33_{-0.03}^{+0.25}$ (0.07)
        & --
        & --
        \\
 $\B(\ov B {}_s^0\to \eta\eta)$
        & --
        & $24.7_{-0.4}^{+0.3}$
        & $19.6_{-6.5}^{+0.6}$ (20.4)
        & --
        & --
        \\
 $\B(\ov B {}_s^0\to \eta \eta')$
        & --
        & $67.2_{-1.4}^{+0.9}$
        & $75.1_{-3.5}^{+67.4}$ (68.7)
        & --
        & --
        \\
 $\B(\ov B{}_s^0\to \eta'\eta')$
        & $33.1\pm7.1$
        & $60.5_{-1.1}^{+0.8}$
        & $34.9_{-4.7}^{+16.0}$ (46.6)
        & 16.0
        & 0.0 (3.6)
        \\
 $\B(\ov B{}_s^0\to K^+ K^-)$
        & $24.8\pm1.7$
        & $32.7_{-0.6}^{+0.5}$
        & $24.6_{-0.6}^{+2.7}$ (24.5)
        & 21.3
        & 0.0 (0.0)
        \\
 $\B(\ov B{}_s^0\to K^0\ov K^0)$
        & $19.6\pm9.5$
        & $34.3_{-0.6}^{+0.5}$
        & $24.6_{-1.0}^{+0.7}$ (25.6)
        & 2.4
        & 0.3 (0.4)
        \\
 $\B(\ov B{}_s^0\to \pi^0 \eta)$
        & --
        & $0.07\pm0.00$
        & $0.07_{-0.00}^{+0.09}$ (0.06)
        & --
        & --
        \\
 $\B(\ov B{}_s^0\to \pi^0\eta')$
        & --
        & $0.09_{-0.00}^{+0.00}$
        & $0.11_{-0.01}^{+0.10}$ (0.10)
        & --
        & --
        \\
        \hline
        \hline
 $\A(\ov B{}_s^0\to K^+\pi^-)$
        & $26\pm4$
        & $17.4_{-0.5}^{+0.4}$
        & $24.8_{-1.0}^{+22.1}$ (28.2)
        & 4.6
        & 0.1 (0.3)
        \\
 $\A(\ov B{}_s^0\to K^0\pi^0)$
        & --
        & $66.8_{-1.6}^{+1.5}$
        & $74.9_{-50.8}^{+4.8}$ (53.7)
        & --
        & --
        \\
 $\A(\ov B{}_s^0\to K^0\eta)$
        & --
        & $88.1_{-1.2}^{+0.9}$
        & $81.2_{-54.8}^{+6.9}$ (78.2)
        & --
        & --
        \\
 $\A(\ov B{}_s^0\to K^0\eta')$
        & --
        & $-38.7_{-0.5}^{+0.9}$
        & $-38.6_{-2.2}^{+13.0}$ $(-34.4)$
        & --
        & --
        \\
        \hline
 $\A(\ov B{}_s^0\to \pi^+\pi^-)$
        & --
        & $0$
        & $1.7_{-2.5}^{+0.5}$ (0)
        & --
        & --
        \\
 $\A(\ov B{}_s^0\to \pi^0 \pi^0)$
       & --
        & $0$
        & $1.7_{-2.5}^{+0.5}$ (0)
        & --
        & --
        \\
 $\A(\ov B{}_s^0\to \eta\eta)$
        & --
        & $-2.4\pm0.1$
        & $-3.7_{-8.2}^{+0.6}$ $(-2.8)$
        & --
        & --
        \\
 $\A(\ov B{}_s^0\to \eta \eta')$
        & --
        & $-0.01\pm0.01$
        & $0.95_{-0.19}^{+0.39}$ $(-0.01)$
        & --
        & --
        \\
 $\A(\ov B{}_s^0\to \eta'\eta')$
        & --
        & $2.0\pm0.0$
        & $-1.2_{-4.7}^{+1.0}$ $(1.9)$
        & --
        & --
        \\
 $\A(\ov B{}_s^0\to K^+ K^-)$
        & $-14\pm 11$
        & $-5.8\pm0.0$
        & $-10.5_{-0.4}^{+1.1}$ $(-9.9)$
        & 0.6
        & 0.1 (0.1)
        \\
 $\A(\ov B{}_s^0\to K^0\ov K^0)$
        & --
        & $-0.9\pm0.0$
        & $0.9_{-0.3}^{+2.2}$ $(-0.6)$
        & --
        & --
        \\
 $\A(\ov B{}_s^0\to \pi^0 \eta)$
        & --
        & $46.0_{-1.2}^{+1.5}$
        & $92.9_{-15.4}^{+2.9}$ $(69.9)$
        & --
        & --
        \\
 $\A(\ov B{}_s^0\to \pi^0\eta')$
        & --
        & $64.3_{-1.1}^{+1.4}$
        & $77.7_{-6.9}^{+8.5}$ (54.0)
        & --
        & --
        \\
 \end{tabular}
\end{ruledtabular}
\end{table}

We show the CP-averaged rates and direct CP violations of $\ov B{}^0_s\to PP$ decays in Table~\ref{tab:table-Bs}. 
There are five measured $\ov B{}_s$ decay rates, 
namely $K^+\pi^-$, $\pi^+\pi^-$, $\eta'\eta'$, $K^+K^-$ and $K^0\ov K{}^0$ decay rates. 
Among them $\ov B{}_s\to\pi^+\pi^-$ and $\eta'\eta'$ decays
are newly observed by LHCb~\cite{Aaij:2015qga,Aaij:2016elb}.
From the table we see that both Fac and Res can fit the $\ov B{}_s\to K^+\pi^-$ rate well,
but Fac is having difficulties in fitting all other four modes: in particular the $\chi^2$ of $\pi^+\pi^-$, $\eta'\eta'$ and $K^+K^-$ are as large as 
20.2, 16.0 and 21.3, respectively, while Res can fit all $B_s$ decay modes very well and brings down these $\chi^2$ efficiently, giving $0.0$, $0.0$ and $0.0$, respectively.
Note that the rates of the two newly measured modes ($\pi^+\pi^-$ and $\eta'\eta'$) can be easily reproduced in the Res fit, but not in the Fac fit.
For other modes, we see from the table that Res predicts larger rates in $\ov B{}^0_s\to K^0\pi^0$, $K^0\eta$, $\pi^0\pi^0$ decays, 
but gives similar predictions on $K^0\eta'$, $\eta\eta$, $\eta\eta'$, $\pi^0\eta$ and $\pi^0\eta'$ rates.

The $\ov B{}^0_s\to\pi^+\pi^-$ rate in the factorization calculation is too small compared to data. 
As shown in Table~\ref{tab:table-Bs}, through Res the rate can be enhanced significantly.  
It is useful to see the enhancement of the $\pi^+\pi^-$ rate more closely. 
From Eq.~(\ref{eq: TABsgroup4}),
\be
 A_{\ov B {}^0_{s}\to \pi^+\pi^-}
 &=&E'+PA^{\prime u}+PA^{\prime c}+\frac{1}{3}P_{EW}^{\prime \,A},
\en
and the values in Tables~\ref{tab:TADS-1} and \ref{tab:ucpenguinsDS0-1}, before and after Res, we have (in unit of $10^{-9}$ GeV) 
\be
(A_{\ov B {}^0_{s}\to \pi^+\pi^-})^0
&\simeq& 0.27 e^{i102.6^\circ}+0.07 e^{-i77.4^\circ}+3.51 e^{-i 8.6^\circ}+0.02 e^{i 170.4}\simeq 3.42 e^{-i5.4^\circ},
\non\\
(A_{\ov B {}^0_{s}\to \pi^+\pi^-})^0
&\simeq& 0.27 e^{-i120.1^\circ}+0.07 e^{i60.3^\circ}+3.51 e^{-i 8.6^\circ}+0.02 e^{i 172.5}\simeq 3.42 e^{-i11.7^\circ},
\non\\
(A_{\ov B {}^0_{s}\to \pi^+\pi^-})_{FSI}
&\simeq& 1.26 e^{i71.0^\circ}+1.38 e^{-i120.1^\circ}+8.75 e^{i 75.4^\circ}+0.28 e^{i132.2^\circ}\simeq 8.88 e^{i78.7^\circ},
\non\\
(A_{\ov B {}^0_{s}\to \pi^+\pi^-})_{FSI}
&\simeq& 1.26 e^{-i151.2^\circ}+1.38 e^{i17.6^\circ}+8.75 e^{i 75.4^\circ}+0.28 e^{i134.3^\circ}\simeq 8.76 e^{i75.3^\circ},
\label{eq: Bspipi}
\en
respectively, where terms are given in the order of $E'$, $PA^{\prime u}$, $PA^{\prime c}$ and $P_{EW}^{\prime \,A}/3$.
In our full numerical result, we have 
$4.17 e^{-i 5.3^\circ}$, 
$4.17 e^{-i11.7^\circ}$, 
$9.19 e^{i 66.7^\circ}$ and 
$9.04 e^{i 64.8^\circ}$ for 
$(A_{\ov B {}^0_{s}\to \pi^+\pi^-})^0$, $(A_{\ov B {}^0_{s}\to \pi^+\pi^-})^0$, $(A_{\ov B {}^0_{s}\to \pi^+\pi^-})_{FSI}$ and $(A_{\ov B {}^0_{s}\to \pi^+\pi^-})_{FSI}$ in unit of $10^{-9}$ GeV,
respectively, which are close to the scaled (by $f_{B_s}/f_B\simeq f_K/f_\pi$) estimations, 
$4.15 e^{-i5.4^\circ}$, $4.15 e^{-i11.8^\circ}$, $10.74 e^{i78.7^\circ}$ and $10.64 e^{i75.3^\circ}$, from Eq. (\ref{eq: Bspipi}).

From Eq. (\ref{eq: Bspipi}), we see that
the sizes of the amplitudes of the $\ov B{}^0_s$ and the conjugated $B^0_s$ decays 
are enhanced by factors of $2.58$ and $2.56$, respectively,
where the enhancements are mainly from the enhancement in $PA^{\prime c}$. 
Consequently, the CP-averaged rate is enhanced by a factor of $6.6$, 
while $\A$ is changed from $0$ to $0.9\%$ as $E'$ and $PA^{\prime u}$ are also enhanced.
Note that the above estimation of rate enhancement is somewhat larger than the one in our full numerical result ($0.67/0.14=4.79$) in Table~\ref{tab:table-Bs},
but the direct CP asymmetry is close the value ($1.9\%$) shown in the table. 
The reason of the vanishing $\A$ before Res is similar to those in the $\ov B{}^0\to K^+K^-$ decay as discussed previously.
Hence, in the presence of Res, $E'$ and $PA^{\prime u}$ are enhanced mostly from $T^{\prime 0}$ [see Eqs. (\ref{eq: E/E0}) and (\ref{eq: PAu/PAu0})], 
while $PA^{\prime c}$ from $P^{\prime c}$ [see Eq. (\ref{eq: PAc/PAc0})], which help to enhance the $\overline B{}^0_s\to \pi^+\pi^-$ rate and bring in non-vanishing direct CP asymmetry. 

We now compare our results to the data in direct CP asymmetries.
There are two reported measurements in direct CP asymmetries of  $\ov B{}_s$ modes: 
$\A(\ov B{}_s^0\to K^+\pi^-)$ and $\A(\ov B_s\to K^+K^-)$.
A better measurement is reported in the $K^+\pi^-$ mode with a much reduced uncertainty.
From the table we see that Res gives a better fit to this data than Fac with $\chi^{2(\rm Fac)}=4.6$ and $\chi^{2(\rm Res)}=0.1$.
On the other hand both Fac and Res can fit $\A(\ov B_s\to K^+K^-)$ well, 
as the uncertainty in data is still large to accommodate both results, but Res has a smaller $\chi^2$.

For predictions on direct CP asymmetries, 
we note that the signs of $\A(\ov B{}_s\to\eta'\eta')$ and $\A(\ov B{}_s\to K^0\ov K{}^0)$ are opposite in Fac and Res;
Res predicts non-vanishing $\A(\ov B{}_s\to\pi^+\pi^-,\pi^0\pi^0)$ and larger $\A(\ov B{}_s\to\pi^0\eta)$, 
while predictions of Fac and Res on other modes are similar. These predictions can be checked in near future.

\subsubsection{Time-dependent CP violations in $\ov B{}^0$ and $\ov
B{}^0_s$ Decays}

\begin{table}
\caption{ \label{tab:table-S} Results on the time-dependent CP
asymmetry $\sin2\beta_{\rm eff}$ (for the first three modes) and $S$ of various $\overline B_{d,s}\to PP$ modes. }
\begin{ruledtabular}
\begin{tabular}{lcrrcc}
 Mode
      &Exp
      &Fac~$\qquad$
      &Res~$\qquad$
      &$\chi^{2\,\rm (Fac)}_{\rm min}$
      &$\chi^{2\,\rm(Res)}_{\rm min}$
      \\
\hline
 $\ov B {}^0\to \ov K{}^0\pi^0$
        & $0.57\pm0.17$
        & $0.798\pm0.002$
        & $0.806^{+0.010}_{-0.003}$ (0.793)
        & 1.8
        & 1.9 (1.7)
        \\
 $\ov B{}^0\to \ov K{}^0\eta$
        & --
        & $0.672_{-0.015}^{+0.009}$
        & $0.728_{-0.018}^{+0.030}$ (0.757)
        &
        &
        \\
 $\ov B{}^0\to \ov K{}^0\eta'$
        & $0.63\pm0.06$
        & $0.689^{+0.001}_{-0.002}$
        & $0.683_{-0.008}^{+0.006}$ (0.693)
        & 1.0
        & 0.8 (1.1)
        \\
        \hline
 $\ov B{}^0\to \pi^+\pi^-$
        & $-0.66\pm 0.06$
        & $-0.477_{-0.041}^{+0.039}$
        & $-0.598\pm0.040$ $(-0.578)$
        & 9.3
        & 1.1 (1.9)
        \\
 $\ov B{}^0\to \pi^0 \pi^0$
        & --
        & $0.602\pm0.023$
        & $0.675_{-0.049}^{+0.055}$ (0.778)
        &
        &
        \\
 $\ov B{}^0\to \eta\eta$
        & --
        & $-0.741_{-0.015}^{+0.014}$
        & $-0.663_{-0.033}^{+0.031}$ $(-0.669)$
        &
        &
        \\
 $\ov B{}^0\to \eta \eta'$
        & --
        & $-0.847_{-0.014}^{+0.013}$
        & $-0.953_{-0.021}^{+0.028}$ $(-0.795)$
        \\
 $\ov B{}^0\to \eta'\eta'$
        & --
        & $-0.922_{-0.004}^{+0.003}$
        & $-0.753_{-0.089}^{+0.067}$ $(-0.962)$
        \\
 $\ov B{}^0\to K^+ K^-$
        & --
        & $-0.835_{-0.017}^{+0.016}$
        & $-0.992_{-0.007}^{+0.017}$ $(-0.895)$
        \\
 $\ov B{}^0\to K_S\ov K_S$
        & $-0.38^{+0.69}_{-0.77}\pm0.09$
        & $-0.016\pm0.002$
        & $-0.231_{-0.042}^{+0.048}$ $(-0.037)$
        & 0.2
        & 0.0 (0.2)
        \\
        & $-1.28^{+0.80}_{-0.73}{}^{+0.11}_{-0.16}$
        \\
 $\ov B{}^0\to \pi^0 \eta$
        & --
        & $0.215_{-0.006}^{+0.005}$
        & $-0.473_{-0.068}^{+0.043}$ $(-0.494)$
        \\
 $\ov B{}^0\to \pi^0\eta'$
        & --
        & $-0.002_{-0.012}^{+0.010}$
        & $-0.414_{-0.025}^{+0.035}$ $(-0.440)$
        &
        &
        \\
         \hline
 $\ov B {}_s^0\to \pi^+\pi^-$
        & --
        & $0.152\pm0.001$
        & $0.071_{-0.009}^{+0.011}$ (0.149)
        &
        &
        \\
 $\ov B {}_s^0\to \pi^0 \pi^0$
        & --
        & $0.152\pm0.001$
        & $0.071_{-0.009}^{+0.011}$ (0.149)
        \\
 $\ov B{}_s^0\to \eta\eta$
        & --
        & $-0.005\pm0.000$
        & $-0.035_{-0.067}^{+0.004}$ $(-0.027)$
        \\
 $\ov B{}_s^0\to \eta \eta'$
        & --
        & $-0.004\pm0.000$
        & $0.005_{-0.001}^{+0.007}$ $(0.006)$
        \\
 $\ov B{}_s^0\to \eta'\eta'$
        & --
        & $0.021\pm0.000$
        & $0.046_{-0.003}^{+0.006}$ (0.025)
        \\
 $\ov B{}_s^0\to K^+ K^-$
        & $0.30\pm0.13$
        & $0.200\pm0.002$
        & $0.149_{-0.066}^{+0.005}$ (0.176)
        & 0.6
        & 1.4 (1.0)
        \\
 $\ov B{}_s^0\to K^0\ov K^0$
        & --
        & $-0.022_{-0.000}^{+0.001}$
        & $-0.019_{-0.017}^{+0.004}$ $(-0.027)$
        \\
 $\ov B{}_s^0\to \pi^0 \eta$
        & --
        & $-0.059_{-0.004}^{+0.009}$
        & $0.100_{-0.475}^{+0.050}$ (0.308)
        \\
 $\ov B {}_s^0\to \pi^0\eta'$
        & --
        & $0.232_{-0.008}^{+0.013}$
        & $-0.016_{-0.319}^{+0.065}$ $(0.053)$
        \\
        \hline
 $\ov B{}_s^0\to K_S\pi^0$
        & --
        & $-0.738_{-0.020}^{+0.017}$
        & $-0.311_{-0.092}^{+0.541}$ $(-0.784)$
        \\
 $\ov B{}_s^0\to K_S\eta$
        & --
        & $-0.296_{-0.037}^{+0.041}$
        & $0.274_{-0.076}^{+0.369}$ $(-0.273)$
        \\
 $\ov B{}_s^0\to K_S\eta'$
        & --
        & $-0.395_{-0.004}^{+0.011}$
        & $-0.049_{-0.052}^{+0.367}$ $(-0.276)$
        \\
\end{tabular}
\end{ruledtabular}
\end{table}

Results on time-dependent CP-asymmetries $S$ are given in
Table~\ref{tab:table-S}. 
%
%
We fit to data on mixing induced CP asymmetries. 
There are reported experimental results of mixing induced CP asymmetries in the following 5 modes:
$\ov B {}^0\to \ov K{}^0\pi^0$, $\ov B{}^0\to \ov K{}^0\eta'$, $\ov B{}^0\to \pi^+\pi^-$,
$\ov B{}^0\to K_S\ov K_S$ and $\ov B{}_s^0\to K^+ K^-$.
Since the measurements are subtle, the experimental progress in this sector is slower than those in rates and direct CP asymmetries. 
Currently, the $\ov B{}^0\to \ov K{}^0\pi^0$ mode was updated up to 2010;
the $\ov B{}^0\to \ov K{}^0\eta'$ mode was updated up to 2014;
the $\ov B{}^0\to \pi^+\pi^-$ mode was updated up to 2013,
the $\ov B{}^0\to K_S\ov K{}_S$ mode was updated up to 2007
and
the $\ov B{}_s\to K^+K^-$ mode was included in these measurement in 2013~\cite{updates,Aaij:2013tna,Aubert:2006gm,Nakahama:2007dg}. 
New data are eagerly awaited.
Note that for the
$\ov B{}^0\to K^0\ov K{}^0$ mode, the mixing induced CP asymmetry
obtained by Belle
($-0.38^{+0.69}_{-0.77}\pm0.09$~\cite{Nakahama:2007dg})
and BaBar
($-1.28^{+0.80+0.11}_{-0.73-0.16}$~\cite{Aubert:2006gm}) are different. 
As the central value of the latter exceeds the physical range,
we only include the former one in our fit.

From Table~\ref{tab:table-S} we see that fit in Res for the $\ov B{}^0\to\pi^+\pi^-$ mode is much better than the one in Fac, where the 
$\chi^2$ are 1.1 and 9.3 for the former and the latter, respectively.
On the contrary, the fit in Fac is better than Res in the $\ov B{}_s\to K^+K^-$ mode, where the $\chi^2$ are 0.6 and 1.4 for the former and the latter, respectively.
Note that the uncertainty in the data of the $\ov B{}_s\to K^+K^-$ mode is much larger than the one in the $\ov B{}^0\to\pi^+\pi^-$ mode. It will be interesting to see the updated data on the $\ov B{}_s\to K^+K^-$ mode.
Overall speaking the quality of fit to mixing induced CP asymmetries is improved ($\chi^2$ reduced from $12.9$ to $5.2$, see also Table II) in the presence of Res.

It is useful to look into the mixing induced asymmetry in the $\ov B {}^0_{d}\to K^0 \ov K {}^0$ mode.
Recall in Eq. (\ref{eq: AK0K0 Res}) that, before and after Res, we have (in unit of $10^{-8}$ GeV, without SU(3) breaking correction)
\be
(A_{\ov B {}^0_{d}\to K^0 \ov K {}^0})^0&\simeq& 0.81 e^{i 8.1^\circ},
\quad
(A_{B {}^0_{d}\to K^0 \ov K {}^0})^0\simeq 0.92 e^{-i 31.6^\circ},
\non\\
(A_{\ov B {}^0_{d}\to K^0 \ov K {}^0})_{FSI}&\simeq&0.66 e^{-i 1.9^\circ},
\quad
(A_{B {}^0_{d}\to K^0 \ov K {}^0})_{FSI}\simeq1.07 e^{-i 27.2^\circ},
\en
respectively.
Using the well known formula:
\be
S=\frac{2 {\rm Im} \lambda_A}{1+|\lambda_A|^2}
\en
with
\be
\lambda_A\equiv \frac{q}{p} \frac{A_{\ov B {}^0_{d}\to K^0 \ov K {}^0}}{A_{B {}^0_{d}\to K^0 \ov K {}^0}}
=e^{-i2\beta}\frac{A_{\ov B {}^0_{d}\to K^0 \ov K {}^0}}{A_{B {}^0_{d}\to K^0 \ov K {}^0}},
\en
we obtain
$S\simeq -0.08$ and $-0.29$ without and with Res, respectively, which are close to the values reported in Table~\ref{tab:table-S}.
As explained previously, although $\ov B {}^0_{d}\to K^0 \ov K {}^0$ is a pure penguin mode,
its $S$ is not necessary close to $-\sin2\beta$, 
as the $u$-penguin contribution is not negligible ($|P^{0 u}/P^{0 c}|\simeq 0.35$, see Table~\ref{tab:ucpenguinsDS0-1}). 
When Res is turned on, the $u$-penguin and $c$-penguin receive different contributions, where it is clear that trees can only contribute to the former giving $|P^u/P^c|\simeq 0.74$ (see Table~\ref{tab:ucpenguinsDS0-1}),
and, consequently, the value of $S$ can be changed drastically.

We now compare the predictions of Fac and Res on mixing induced CP asymmetries. We note that they have different predictions on the mixing induced CP asymmetries of $\ov B{}^0\to\eta\eta$, $\eta\eta'$, $\pi^0\eta$, $\pi^0\eta'$, $\ov B{}_s\to \pi^0\eta$, $\pi^0\eta'$, $K_s\pi^0$, $K_s\eta$ and $K_S\eta'$ modes.
In particular, the signs of central values of the asymmetries of $\ov B{}^0\to\pi^0\eta$, $\pi^0\eta'$, $\ov B{}_s\to \pi^0\eta$, $\pi^0\eta'$, and $K_s\eta$ are opposite.

\section{conclusion}

Various new measurements in charmless $B_{u,d,s}\to PP$ modes
are reported by Belle and LHCb.
These include the rates of
$B^0\to\pi^0\pi^0$, $\eta\pi^0$, $B_s\to\eta'\eta'$,  $B^0\to K^+K^-$ and $B^0_s\to\pi^+\pi^-$ decays. 
Some of these modes are highly suppressed and are among the rarest $B$ decays.
Direct CP asymmetries on various modes
are constantly updated. 
It is well known that direct CP asymmetries and rates of suppressed modes are sensitive to final state interaction. 
As new measurements are reported and more data will be collected,
it is interesting and timely to studied the rescattering on $B_{u,d,s}\to PP$ decays.
We perform a $\chi^2$ analysis with all available data on
CP-averaged rates and CP asymmetries in $\overline B{}_{u,d,s}\to PP$
decays.
Our numerical results are compared to data and those from factorization approach.
The quality of the fit is improved significantly in the presence of Res, 
especially in the decay rates in the $\overline B{}^0$ $\Delta S=0$ sector and in rates and direct CP asymmetries in the $\overline B{}^0_s$ decay modes.
Indeed, the $\chi^2$ in the $\ov B{}^0\to \ov K{}^0\pi^0$, $\pi^+\pi^-$, $K^0\ov K{}^0$, $B^-\to\ov K{}^0\pi^-$, 
$K^-\eta$, $\pi^-\pi^0$, $\pi^-\eta$ and $\ov B{}_s^0\to \pi^+\pi^-$, $\eta'\eta'$ and $K^+K^-$ rates, 
and in $\ov B{}^0\to\pi^+\pi^-$ and $\ov B{}_s^0\to K^+\pi^-$ direct CP asymmetries are improved significantly.
Res also fit. better to the semileptonic data on $|V_{ub}| F^{B\pi}(0)$ [see Eq.~(\ref{eq: SL})].

The relations on topological amplitudes and rescattering are explored 
and they help to provide a better understanding of the effects of rescattering.
As suggested by U(3) symmetry on topological amplitudes and FSI, a vanishing exchange rescattering scenario is considered.
The exchange, annihilation, $u$-penguin, $u$-penguin annihilation and some electroweak penguin amplitudes are enhanced significantly
via annihilation and total annihilation rescatterings.
In particular, the $u$-penguin annihilation amplitude is sizably enhanced by the tree amplitude via total annihilation rescattering.
These enhancements affect rates and CP asymmetries.
For example, 
the enhanced $PA^u$ changes the $\ov B {}^0_{d}\to K^0 \ov K {}^0$ direct CP asymmetry significantly;
the enhanced $P$, $E$ and $PA$ produce (through complicate interference) a slightly larger $\ov B{}^0\to\pi^0\pi^0$ decay amplitude and resulting a 35\% enhancement in rate; 
$A'$ and $P^{\prime E}_{EW}$ are enhanced 
and enlarges the deviation of $\A(\ov B {}^0_{d}\to K^-\pi^+)$ and $\A(B^-\to K^- \pi^0)$ producing a larger $\Delta \A$;
the $\ov B{}^0_s\to\pi^+\pi^-$ rate is sizably enhanced through the enhancement in $PA^{\prime c}$;
the $|P^u/P^c|$ ratio is enhanced from $0.35$ to $0.74$ and can change mixing induced CP asymmetries drastically.

For the comparison of the predictions of Fac and Res, we observed the following points. 
(i) Belle and BaBar give very different results in $\A(\ov B{}^0\to K_s K_s)$ mode, namely
Belle gives $\A(\ov B{}^0\to K_s K_s)=-0.38\pm0.38\pm0.5$~\cite{Nakahama:2007dg}, while BaBar gives $0.40\pm0.41\pm0.06$~\cite{Aubert:2006gm}.
The result of Res prefers the Belle result, while Fac prefers a negative but less sizable direct CP asymmetry.
(ii) Except $\ov B {}^0\to K^+ K^-$, 
the sizes of the predicted direct CP asymmetries of $B^-, \ov B{}^0\to PP$ modes from Res are smaller than those in Fac. 
(iii) For $B_s$ decay rates,  Res predicts larger rates in $\ov B{}^0_s\to K^0\pi^0$, $K^0\eta$, $\pi^0\pi^0$ decays, 
but gives similar predictions on $K^0\eta'$, $\eta\eta$, $\eta\eta'$, $\pi^0\eta$ and $\pi^0\eta'$ rates.
(iv) For predictions on direct CP asymmetries, 
we note that the signs of $\A(\ov B{}_s\to\eta'\eta')$ and $\A(\ov B{}_s\to K^0\ov K{}^0)$ are opposite in Fac and Res;
Res predicts non-vanishing $\A(\ov B{}_s\to\pi^+\pi^-,\pi^0\pi^0)$ and larger $\A(\ov B{}_s\to\pi^0\eta)$, 
while predictions of Fac and Res on other modes are similar.
(v) Finally, Fac and Res have different predictions on the mixing induced CP asymmetries of $\ov B{}^0\to\eta\eta$, $\eta\eta'$, $\pi^0\eta$, $\pi^0\eta'$, $\ov B{}_s\to \pi^0\eta$, $\pi^0\eta'$, $K_s\pi^0$, $K_s\eta$ and $K_S\eta'$ modes.
In particular, the signs of central values of the asymmetries of $\ov B{}^0\to\pi^0\eta$, $\pi^0\eta'$, $\ov B{}_s\to \pi^0\eta$, $\pi^0\eta'$, and $K_s\eta$ modes are opposite. 
These predictions can be checked in the future.

\begin{acknowledgments}

This work is
supported in part by Ministry of Science and Technology of R.O.C. under Grant Nos. 103-2112-M-033-002-MY3 and 106-2112-M-033-004-MY3.

\end{acknowledgments}

\appendix

\section{FSI Formulas}


The weak Hamiltonian for charmless $\overline B_q$ decays can be written as $H_{\rm W}=(G_f/\sqrt2) \sum_{r,l}V_{qb}V^*_{rq} c^r_l {\cal O}^r_l$, where $c_l^r$ are Wilson coefficients, ${\cal O}_l^r$ are Wilson operators and $V_{qb,rq}$ are the relevant CKM matrix elements, see for example~\cite{Buras}.
From the time invariant of the Wilson operator $ {\cal O}^r_l$, we obtain
$
(\langle i;{\rm out}| {\cal O}^r_l|\overline B\rangle)^*
=\left(\langle i;{\rm out}|\right)^*
U^\dagger_T U_T ({\cal O}^r_l)^* U^\dagger_T U_T |B\rangle^*,
$
where $U_T$ is the time-reversal transformation operator.
Using the time-reversal invariant of the operators, $ {\cal O}^r_l=U_T ({\cal O}^r_l)^* U^\dagger_T$ and the appropriate phase convention of states,  
$U_T|{\rm out\,(in)}\rangle^*=|{\rm in\,(out)}\rangle$, we have
\be
(\langle i;{\rm out}| {\cal O}^r_l|\overline B\rangle)^*
&=&
\langle i;{\rm in}|{\cal O}^r_l|B\rangle
=\sum_j \langle i;{\rm in}|\,j;{\rm out}\rangle
          \langle j;{\rm out}|{\cal O}_i|\overline B\rangle,
\non
\end{eqnarray}
where we have inserted a complete set in the last step.
Therefore using the time-reversal invariant property of the operator ${\cal O}^r_l$ in a $\ov B\to PP$ decay, 
one obtains
\be
(\langle i;{\rm out}| {\cal O}^r_l|\overline B\rangle)^*
&=&\sum_j {\cal S}^*_{ji} \langle j;{\rm out}|{\cal O}^r_l|\overline B\rangle,
\label{eq:timerev}
\end{eqnarray}
where ${\cal S}_{ij}\equiv\langle i;{\rm out}|\,j;{\rm in}\rangle$ is
the strong interaction $S$-matrix element, $j$ denotes all possible states.
Eq. (\ref{eq:timerev}) is the master formula of FSI.

One can easily verify that the solution of the above equation is given by
 \be
 A_i({\cal O}^r_l)=\sum_{k=1}^N\Sc^{1/2}_{ik} A^0_k({\cal O}^r_l),
 \label{eq:master0}
 \en
where we have $A_i({\cal O}^r_l)\equiv \langle i;{\rm out}|{\cal O}^r_l|\overline B\rangle $ and $A^0({\cal O}^r_l)$ are real amplitudes.
Putting back the coefficients, we obtain the master formula Eq. (\ref{eq:master}),
and we can now state clearly $A_i\equiv \langle i;{\rm out}|H_{\rm W}|\overline B\rangle=
 (G_f/\sqrt2) \sum_{r,l}V_{qb}V^*_{rq} c^r_lA_i({\cal O}^r_l)$ and $A^0_i\equiv
 (G_f/\sqrt2) \sum_{r,l}V_{qb}V^*_{rq} c^r_lA^0_i({\cal O}^r_l)$.

Without loss of generality, we can re-express the $S$-matrix in
Eq.~(\ref{eq:master}) as
 \be
 \Sc_{ik}=\sum_{j=1}^n(\Sc_1)_{ij} (\Sc_2)_{jk},
 \label{eq:S1S2}
 \en
where $\Sc_1$ is a non-singular $n\times n$ matrix with $n$ the
total number of charmless $PP$ states and $\Sc_2$ is defined
through the above equation, i.e. $\Sc_2\equiv \Sc^{-1}_{1} \Sc$.
As mentioned previously (in the introduction)
the factorization amplitudes contain a large portion of rescattering effects as encoded in ${\cal S}_2$,
while some
residual rescattering among a small group of states is still allowed and needs to be explored:
 \be
 \Sc_1=\Sc_{res},\quad
 A^{ fac}_j=\sum_{k=1}^N(\Sc^{1/2}_2)_{jk}A^0_k,
 \label{eq:res}
 \en
with $N$ the total number of states entering Eq.~(\ref{eq:master}),
$A^{\rm fac}_j$ the factorization amplitude and $\Sc_{res}$ the rescattering matrix to govern rescattering among $PP$ states.

We collect the rescattering formulas used in this work.
We have
\begin{eqnarray}
\left(
\begin{array}{l}
 A_{\ov B {}^0_{d}\to K^-\pi^+}\\
 A_{\ov B {}^0_{d}\to \ov K {}^0 \pi^0}\\
 A_{\ov B {}^0_{d}\to \ov K {}^0 \eta_8}\\
 A_{\ov B {}^0_{d}\to \ov K {}^0 \eta_1}
\end{array}
\right)
 &=& \Sc_{res,1}^{1/2}
 \left(
\begin{array}{l}
 A^{fac}_{\ov B {}^0_{d}\to K^-\pi^+}\\
 A^{fac}_{\ov B {}^0_{d}\to \ov K {}^0 \pi^0}\\
 A^{fac}_{\ov B {}^0_{d}\to \ov K {}^0 \eta_8}\\
 A^{fac}_{\ov B {}^0_{d}\to \ov K {}^0 \eta_1}
\end{array}
\right),
 \label{eq:FSIB0Kpi}
\end{eqnarray}
for group-1 modes,
\begin{eqnarray}
\left(
\begin{array}{l}
 A_{B^-\to \ov K^0\pi^-}\\
 A_{B^-\to K^- \pi^0}\\
 A_{B^-\to K^- \eta_8}\\
 A_{B^-\to K^- \eta_1}
\end{array}
\right)
 &=& \Sc_{res,2}^{1/2}
 \left(
\begin{array}{l}
 A^{fac}_{B^-\to \ov K^0\pi^-}\\
 A^{fac}_{B^-\to K^- \pi^0}\\
 A^{fac}_{B^-\to K^- \eta_8}\\
 A^{fac}_{B^-\to K^- \eta_1}
\end{array}
\right),
 \label{eq:FSIBKpi0}
\end{eqnarray}
for group-2 modes,
\begin{eqnarray}
\left(
\begin{array}{l}
 A_{B^-\to \pi^-\pi^0}\\
 A_{B^-\to K^0 K^-}\\
 A_{B^-\to \pi^- \eta_8}\\
 A_{B^-\to \pi^- \eta_1}
\end{array}
\right)
 &=& \Sc_{res,3}^{1/2}
 \left(
\begin{array}{l}
 A^{fac}_{B^-\to \pi^-\pi^0}\\
 A^{fac}_{B^-\to K^0 K^-}\\
 A^{fac}_{B^-\to \pi^- \eta_8}\\
 A^{fac}_{B^-\to \pi^- \eta_1}
\end{array}
\right),
 \label{eq:FSIBpipi0}
\end{eqnarray}
for group-3 modes and
\begin{eqnarray}
\left(
\begin{array}{l}
 A_{\ov B {}^0_{d}\to \pi^+\pi^-}\\
 A_{\ov B {}^0_{d}\to \pi^0 \pi^0}\\
 A_{\ov B {}^0_{d}\to \eta_8 \eta_8}\\
 A_{\ov B {}^0_{d}\to \eta_8 \eta_1}\\
 A_{\ov B {}^0_{d}\to \eta_1 \eta_1}\\
 A_{\ov B {}^0_{d}\to K^+ K^-}\\
 A_{\ov B {}^0_{d}\to K^0 \ov K {}^0}\\
 A_{\ov B {}^0_{d}\to \pi^0 \eta_8}\\
 A_{\ov B {}^0_{d}\to \pi^0 \eta_1}
\end{array}
\right)
 &=& \Sc_{res,4}^{1/2}
 \left(
\begin{array}{l}
 A^{fac}_{\ov B {}^0_{d}\to \pi^+\pi^-}\\
 A^{fac}_{\ov B {}^0_{d}\to \pi^0 \pi^0}\\
 A^{fac}_{\ov B {}^0_{d}\to \eta_8 \eta_8}\\
 A^{fac}_{\ov B {}^0_{d}\to \eta_8 \eta_1}\\
 A^{fac}_{\ov B {}^0_{d}\to \eta_1 \eta_1}\\
 A^{fac}_{\ov B {}^0_{d}\to K^+ K^-}\\
 A^{fac}_{\ov B {}^0_{d}\to K^0 \ov K {}^0}\\
 A^{fac}_{\ov B {}^0_{d}\to \pi^0 \eta_8}\\
 A^{fac}_{\ov B {}^0_{d}\to \pi^0 \eta_1}
\end{array}
\right),
 \label{eq:FSIB0pipi}
\end{eqnarray}
for group-4 modes,
%
where we define ${\cal S}^{1/2}_{res,i}=(1+i{\cal T}_i)^{1/2}\equiv 1+i{\cal T}_i^{(1/2)}$, before incorporating SU(3) breaking effect, with
\begin{eqnarray}
 {\cal T}_1 &=& \left(
\begin{array}{cccc}
r_0+r_a
      &\frac{-r_a+r_e}{\sqrt2}
      &\frac{-r_a+r_e}{\sqrt6}
      &\frac{2\bar r_a+\bar r_e}{\sqrt3}
      \\
\frac{-r_a+r_e}{\sqrt2}
      &r_0+\frac{r_a+r_e}{2}
      &\frac{r_a-r_e}{2\sqrt3}
      &-\frac{2\bar r_a+\bar r_e}{3\sqrt2}
      \\
\frac{-r_a+r_e}{\sqrt6}
      &\frac{r_a-r_e}{2\sqrt3}
      &r_0+\frac{r_a+5r_e}{6}
      &-\frac{2\bar r_a+\bar r_e}{3\sqrt2}
      \\
\frac{2\bar r_a+\bar r_e}{\sqrt3}
      &-\frac{2\bar r_a+\bar r_e}{\sqrt6}
      &-\frac{2\bar r_a+\bar r_e}{3\sqrt2}
      &\tilde r_0+\frac{4\tilde r_a+2\tilde r_e}{3}
\end{array}
\right), \nonumber\\
 {\cal T}_2 &=& \left(
\begin{array}{cccc}
r_0+r_a
      &\frac{r_a-r_e}{\sqrt2}
      &\frac{-r_a+r_e}{\sqrt6}
      &\frac{2\bar r_a+\bar r_e}{\sqrt3}
      \\
\frac{r_a-r_e}{\sqrt2}
      &r_0+\frac{r_a+r_e}{2}
      &\frac{-r_a+r_e}{2\sqrt3}
      &\frac{2\bar r_a+\bar r_e}{3\sqrt2}
      \\
\frac{-r_a+r_e}{\sqrt6}
      &\frac{-r_a+r_e}{2\sqrt3}
      &r_0+\frac{r_a+5r_e}{6}
      &-\frac{2\bar r_a+\bar r_e}{3\sqrt2}
      \\
\frac{2\bar r_a+\bar r_e}{\sqrt3}
      &\frac{2\bar r_a+\bar r_e}{\sqrt6}
      &-\frac{2\bar r_a+\bar r_e}{3\sqrt2}
      &\tilde r_0+\frac{4\tilde r_a+2\tilde r_e}{3}
\end{array}
\right), \nonumber\\
 {\cal T}_3 &=& \left(
\begin{array}{cccc}
r_0+r_a
      &0
      &0
      &0
      \\
0
      &r_0+r_a
      &\sqrt{\frac{2}{3}}(r_a-r_e)
      &\frac{2\bar r_a+\bar r_e}{\sqrt3}
      \\
0
      &\sqrt{\frac{2}{3}}(r_a-r_e)
      &r_0+\frac{2r_a+r_e}{3}
      &\frac{\sqrt2}{3}(2\bar r_a+\bar r_e)
      \\
0
      &\frac{2\bar r_a+\bar r_e}{\sqrt3}
      &\frac{\sqrt2}{3}(2\bar r_a+\bar r_e)
      &\tilde r_0+\frac{4\tilde r_a+2\tilde r_e}{3}
\end{array}
\right),
 \label{eq:T123}
\end{eqnarray}
and
\begin{eqnarray}
&&\hspace{-0.5 cm}\T_4={\rm diag}(r_0,r_0,r_0,\tilde r_0,\check
r_0,r_0,r_0,r_0,\tilde r_0)
\nonumber\\
 &&\hspace{-0.5cm}+ \tiny{
 \left(
\begin{array}{ccccccccc}
2r_a+r_t
       &\frac{2r_a-r_e+r_t}{\sqrt2}
       &\frac{2r_a+r_e+3r_t}{3\sqrt2}
       &\frac{\sqrt2(2\bar r_a+\bar r_e)}{3}
       &\frac{4\hat r_a+2\hat r_e+3\hat r_t}{3\sqrt2}
       &r_a+r_t
       &r_a+r_t
       &0
       &0
       \\
\frac{2r_a-r_e+r_t}{\sqrt2}
       &\frac{2r_a+r_e+r_t}{2}
       &\frac{2r_a+r_e+3r_t}{6}
       &\frac{2\bar r_a+\bar r_e}{3}
       &\frac{4\hat r_a+2\hat r_e+3\hat r_t}{6}
       &\frac{r_a+r_t}{\sqrt2}
       &\frac{r_a+r_t}{\sqrt2}
       &0
       &0
       \\
\frac{2r_a+r_e+3r_t}{3\sqrt2}
       &\frac{2r_a+r_e+3r_t}{6}
       &\frac{2r_a+r_e+r_t}{2}
       &-\frac{2\bar r_a+\bar r_e}{3}
       &\frac{4\hat r_a+2\hat r_e+3\hat r_t}{6}
       &\frac{5r_a-2r_e+3r_t}{3\sqrt2}
       &\frac{5r_a-2r_e+3r_t}{3\sqrt2}
       &0
       &0
       \\
\frac{\sqrt2(2\bar r_a+\bar r_e)}{3}
       &\frac{2\bar r_a+\bar r_e}{3}
       &-\frac{2\bar r_a+\bar r_e}{3}
       &\frac{4\tilde r_a+2\tilde r_e}{3}
       &0
       &-\frac{2\bar r_a+\bar r_e}{3\sqrt2}
       &-\frac{2\bar r_a+\bar r_e}{3\sqrt2}
       &0
       &0
       \\
\frac{4\hat r_a+2\hat r_e+3\hat r_t}{3\sqrt2}
       &\frac{4\hat r_a+2\hat r_e+3\hat r_t}{6}
       &\frac{4\hat r_a+2\hat r_e+3\hat r_t}{6}
       &0
       &\frac{4\check r_a+2\check r_e+3\check r_t}{6}
       &\frac{4\hat r_a+2\hat r_e+3\hat r_t}{3\sqrt2}
       &\frac{4\hat r_a+2\hat r_e+3\hat r_t}{3\sqrt2}
       &0
       &0
       \\
r_a+r_t
       &\frac{r_a+r_t}{\sqrt2}
       &\frac{5r_a-2r_e+3r_t}{3\sqrt2}
       &-\frac{2\bar r_a+\bar r_e}{3\sqrt2}
       &\frac{4\hat r_a+2\hat r_e+3\hat r_t}{3\sqrt2}
       &2 r_a+r_t
       &r_a+r_t
       &\frac{r_a-r_e}{\sqrt3}
       &\frac{2\bar r_a+\bar r_e}{\sqrt6}
       \\
r_a+r_t
       &\frac{r_a+r_t}{\sqrt2}
       &\frac{5r_a-2r_e+3r_t}{3\sqrt2}
       &-\frac{2\bar r_a+\bar r_e}{3\sqrt2}
       &\frac{4\hat r_a+2\hat r_e+3\hat r_t}{3\sqrt2}
       &r_a+r_t
       &2r_a+r_t
       &\frac{-r_a+r_e}{\sqrt3}
       &-\frac{2\bar r_a+\bar r_e}{\sqrt6}
       \\
0
       &0
       &0
       &0
       &0
       &\frac{r_a-r_e}{\sqrt3}
       &\frac{-r_a+r_e}{\sqrt3}
       &\frac{2r_a+r_e}{3}
       &\frac{\sqrt2(2\bar r_a+\bar r_e)}{3}
       \\
0
       &0
       &0
       &0
       &0
       &\frac{2\bar r_a+\bar r_e}{\sqrt6}
       &-\frac{2\bar r_a+\bar r_e}{\sqrt6}
       &\frac{\sqrt2(2\bar r_a+\bar r_e)}{3}
       &\frac{4\tilde r_a+2\tilde r_e}{3}
       \end{array}
 \right)
 }.
 \nonumber\\
\label{eq:T4}
\end{eqnarray}
Note that for identical particle final states,
such as $\pi^0\pi^0$, factors of $1/\sqrt2$ are included in the
amplitudes and the corresponding $\Sc_{res}$ matrix elements. 
The rescattering formulas for $\overline B^{0}_s\to \bar P\bar P$ decays are similar to the $\overline B^{0}_d\to PP$ ones, since strong interaction respect charge conjugation. For example, the rescattering formula for $\overline B^{0}_s\to \bar K^+\bar \pi^-$ is similar to those of $\overline B^{0}_d\to K^-\pi^+$ with trivial replacement on amplitudes.

To include the SU(3) breaking effect, we proceed as outlined in the main text.
First we remove the SU(3) breaking effect in $A^{\rm fac}$ before recattering and put it back after the rescattering. 
For the reasoning one is referred to the main text. 
For convenient we absorb these two action into the rescattering matrices. 
We use ratios of decay constants to model the SU(3) breaking effect. 
For example, in the group-3 modes, in the $\pi^-\pi^0$--$K^0 K^-$--$\pi^- \eta_q$--$\pi^- \eta_s$ basis, we have
\be
\Sc_{res,3}^{1/2}=
 \left(
\begin{array}{cccc}
 1 &0 &0 &0\\
 0 &\big(\frac{f_K}{f_\pi}\big)^2 &0 &0\\
 0 &0 &\frac{f_{\eta_q}}{f_\pi} &0\\
 0 &0 &0 &\frac{f_{\eta_s}}{f_\pi}
\end{array}
\right)
(1+i\T_3^{(1/2)})
 \left(
\begin{array}{cccc}
 1 &0 &0 &0\\
 0 &\big(\frac{f_\pi}{f_K}\big)^2 &0 &0\\
 0 &0 &\frac{f_\pi}{f_{\eta_q}} &0\\
 0 &0 &0 &\frac{f_\pi}{f_{\eta_s}}
\end{array}
\right).
\end{eqnarray}
In the numerical study we follow \cite{Feldmann:1998vh} to use $f_{\eta_q}/f_\pi=1.07$ and $f_{\eta_s}/f_\pi=1.34$.
It is clear that when FSI is turned off the above $\Sc_{res,3}^{1/2}$ is just an identity matrix.
The SU(3) breaking effects are incorporated in other $\Sc_{res,i}^{1/2}$ in a similar fashion.
Note that $\T_i$ in Eqs. (\ref{eq:T123}) and (\ref{eq:T4}) are given in the $\eta_8$--$\eta_1$ basis 
and to incorporate the SU(3) effect, one needs to transform $\T_i$ to the $\eta_q$--$\eta_s$ basis (see below). 

The physical
$\eta,\,\eta^\prime$ mesons are defined through
\begin{equation}
\left(
\begin{array}{c}
\eta\\
      \eta^\prime
\end{array}
\right)= \left(
\begin{array}{cc}
\cos\vartheta &-\sin\vartheta\\
\sin\vartheta &\cos\vartheta
\end{array}
\right) \left(
\begin{array}{c}

\eta_8\\
      \eta_1
\end{array}
\right),
\end{equation}
with the mixing angle $\vartheta\simeq-15.4^\circ$
\cite{Feldmann:1998vh}. For the $\eta^{(\prime)}\eta^{(\prime)}$
states, we have
\begin{equation}
\left(
\begin{array}{c}
\eta\eta\\
      \eta\eta^\prime\\
      \eta'\eta'
\end{array}
\right)= \left(
\begin{array}{ccc}
\cos^2\vartheta
   &-\sqrt2\cos\vartheta \sin\vartheta
   &\sin^2\vartheta\\
\sqrt2\cos\vartheta \sin\vartheta
   &\cos^2\vartheta-\sin^2\vartheta
   &-\sqrt2\cos\vartheta \sin\vartheta\\
\sin^2\vartheta
   &\sqrt2\cos\vartheta \sin\vartheta
   &\cos^2\vartheta
\end{array}
\right) \left(
\begin{array}{c}
\eta_8\eta_8\\
      \eta_8\eta_1\\
      \eta_1\eta_1
\end{array}
\right),
\end{equation}
where the identical particle factor of $1/\sqrt2$ is properly
included in the mixing matrix.
Note that the above formulas can be easily used to transform the $\eta_q$--$\eta_s$ basis into the $\eta_8$--$\eta_1$ basis by replacing the above $\vartheta$ by 
$\tan^{-1}\sqrt2$.

Rescattering
parameters enter $\Sc_{res}$ only through 7 independent
combinations: $1+i(r_0+r_a)$, $i(r_e-r_a)$, $i(r_a+r_t)$, $i(2\bar
r_a+\bar r_e)$, $1+i[\tilde r_0+(4\tilde r_a+2\tilde r_e)/3]$,
$i(4\hat r_a+2\hat r_e+3\hat r_t)$ and $1+i[\check r_0+(4\check
r_a+2\check r_e+3\check r_t)/6]$. 
The solutions to Eqs.~(\ref{eq:SSU3a}) and
(\ref{eq:SSU3b}) are given in Eq. (\ref{eq:solution}).

If the full U(3) symmetry is a good symmetry,
it requires:
 \be
 r_i=\bar r_i=\tilde r_i=\hat r_i=\check r_i,
 \en
for each $i=0,a,e,t$. 
We are constrained to have 
 \be
 r^{(m)}_e r^{(m)}_a=0.
 \en
Consequently, there are two different solutions:
(a)~the annihilation type ($r^{(m)}_a\neq0,\,r^{(m)}_e=0$) with
 \be
 \delta_{27}=\delta_8'=\delta'_1,
 \quad
 \delta_8,
 \quad
 \delta_1,
 \quad
 \tau=-\frac{1}{2} \sin^{-1}\frac{4\sqrt5}{9},
 \quad
 \nu=-\frac{1}{2}\sin^{-1}\frac{4\sqrt2}{9},
 \label{eq:solutionreU3ra}
 \en
and (b)~the exchange type ($r^{(m)}_e\neq0,\,r^{(m)}_a=r^{(m)}_t=0$)
with
 \be
 \delta_{27}=\delta'_8=\delta_1',
 \quad
 \delta_8=\delta_1,
 \quad
 \tau=\frac{1}{2} \sin^{-1}\frac{\sqrt5}{3},
 \quad
 \nu=\frac{1}{2}\sin^{-1}\frac{2\sqrt2}{3}.
 \label{eq:solutionreU3re}
 \en
It is interesting to note that in both solutions of the U(3) case,
a common constraint
 \be
  \delta_{27}=\delta_8'=\delta'_1,
  \label{eq:constrain}
 \en
has to be satisfied.

\section{Derivation of the rescattering effects on topological amplitudes} 

It is straightforward to obtain the rescattering effects on topological amplitudes. 
In analogy to Eq. (\ref{eq:master2}):
$
 A=\Sc_{res}^{1/2}\cdot
 A^{\rm fac}=(1+i\T^{1/2})\cdot A^{\rm fac},
$
we have
$
H_{\rm eff}=(1+i\T^{1/2})\cdot H^0_{\rm eff}=H^0_{\rm eff}+i\T^{1/2}\cdot H^0_{\rm eff},
$
where $H_{\rm eff}$ is given in Eq. (\ref{eq:B2PP Heff}), $\T^{1/2}$ in Eq. (\ref{eq:Tm}), $H^0_{\rm eff}$ is the un-scattered effective Hamiltonian with all $TA$ in $H_{\rm eff}$ replaced by $TA^0$ and the dot in the above equation implies all possible pairing of the $P^{\rm out}P^{\rm out}$ fields in $H_{\rm eff}^0$ to the $P^{\rm in}P^{\rm in}$ fields in $\T^{1/2}$. 
%
It is useful to use $H^{ik}_i=H^k$, $H^{ik}_k=0$, $(H_{EW})^{ik}_k=0$, $(H_{EW})^{ik}_i
=-\frac{1}{3} H^k$, $(\Pi^{\rm in})^a_a=(\Pi^{\rm out})^a_a=0$
and the fact that the paring of creation and annihilation fields gives the following flavor structure:
$
 \la (\Pi^{\rm out})^j_k (\Pi^{\rm in})^a_b\ra\to\delta^j_b\delta^a_k-\frac{1}{3}\delta^j_k\delta^a_b.
$

In bellow we work out the contribution from $T^0$ via the rescattering among $PP$ states for illustration.
We shall concentrate on the flavor structures after the pairings in $(i\T^{1/2}\cdot H^0_{\rm eff})$ and compare them to the operators in $H_{\rm eff}$.~\footnote{There are integrations of momentum and so on, which will not shown explicitly in the following derivation and are absorbed in the definition of $r'_i$. See Ref.~\cite{{Chua:2001br}} for the treatment on this issue.}


\subsection{Pairing $T^0\, \overline B_m H^{ik}_j (\Pi^{\rm out})^j_k (\Pi^{\rm out})^m_i$ and $(i r'_0/2) Tr(\Pi^{\rm in}\Pi^{\rm out})Tr(\Pi^{\rm in}\Pi^{\rm out})$.}
Pairing the $T^0$ term in $H^0_{\rm eff}$ and the $ir'_0$ term from $\T^{1/2}$ gives:
\be
&&
\hspace{-1cm}
  T^0\, \overline B_m H^{ik}_j (\Pi^{\rm out})^j_k (\Pi^{\rm out})^m_i 
 \cdot
  [\frac{ir'_0}{2}(\Pi^{\rm in})^a_b(\Pi^{\rm out})^b_a(\Pi^{\rm in})^c_d(\Pi^{\rm out})^d_c]
\non\\
&&=\frac{ir'_0}{2} T^0\, \overline B_m H^{ik}_j 
  [\la (\Pi^{\rm out})^j_k (\Pi^{\rm in})^a_b\ra
   \la (\Pi^{\rm out})^m_i  (\Pi^{\rm in})^c_d\ra
\non\\
&&\quad
+
\la (\Pi^{\rm out})^j_k  (\Pi^{\rm in})^c_d\ra
\la (\Pi^{\rm out})^m_i (\Pi^{\rm in})^a_b\ra]
(\Pi^{\rm out})^b_a (\Pi^{\rm out})^d_c
\non\\
&&= \frac{ir'_0}{2} T^0\, \overline B_m H^{ik}_j 
[(\delta^j_b \delta^a_k-\frac{1}{3}\delta^j_k \delta^a_b)
(\delta^m_d  \delta^c_i-\frac{1}{3}\delta^m_i  \delta^c_d)]
\non\\
&&\quad
+(\delta^j_d \delta^c_k-\frac{1}{3}\delta^j_k \delta^c_d)
(\delta^m_b\delta^a_i-\frac{1}{3}\delta^m_i\delta^a_b)]
(\Pi^{\rm out})^b_a (\Pi^{\rm out})^d_c
\non\\
&&=\frac{ir'_0}{2} T^0\, \overline B_m H^{ik}_j 
[(\Pi^{\rm out})^j_k (\Pi^{\rm out})^m_i
+(\Pi^{\rm out})^m_i (\Pi^{\rm out})^j_k]
\non\\
&&=ir'_0 T^0\,\overline B_m H^{ik}_j (\Pi^{\rm out})^j_k (\Pi^{\rm out})^m_i.
\en
We note that the last term has the same form of the $T$ operator in $H_{\rm eff}$ and we denote it as $\delta T(T^0)\,\overline B_m H^{ik}_j (\Pi^{\rm out})^j_k (\Pi^{\rm out})^m_i$. 
From the above equation we obtain,
\be
\delta T(T^0)=i r'_0 T^0.
\en


\subsection{Pairing $T^0\, \overline B_m H^{ik}_j (\Pi^{\rm out})^j_k (\Pi^{\rm out})^m_i$ and $i r'_eTr(\Pi^{\rm in}\Pi^{\rm out}\Pi^{\rm in}\Pi^{\rm out})/2$}
Pairing the $T^0$ term in $H^0_{\rm eff}$ and the $ir'_e$ term from $\T^{1/2}$ gives:
\be
&&
\hspace{-1cm}
  T^0\, \overline B_m H^{ik}_j (\Pi^{\rm out})^j_k (\Pi^{\rm out})^m_i 
 \cdot
[\frac{ir'_e}{2} (\Pi^{\rm in})^a_b(\Pi^{\rm out})^b_c(\Pi^{\rm in})^c_d(\Pi^{\rm out})^d_a]
\non\\
&&=ir'_e \frac{1}{2} T^0\, \overline B_m H^{ik}_j 
[\la (\Pi^{\rm out})^j_k (\Pi^{\rm in})^a_b\ra
\la (\Pi^{\rm out})^m_i  (\Pi^{\rm in})^c_d\ra
\non\\
&&\quad
+
\la (\Pi^{\rm out})^j_k  (\Pi^{\rm in})^c_d\ra
\la (\Pi^{\rm out})^m_i (\Pi^{\rm in})^a_b\ra]
(\Pi^{\rm out})^b_c (\Pi^{\rm out})^d_a
\non\\
&&=ir'_e \frac{1}{2} T\, \overline B_m H^{ik}_j 
[(\Pi^{\rm out})^j_i (\Pi^{\rm out})^m_k-\frac{1}{3}\delta^m_i   (\Pi^{\rm out})^j_c (\Pi^{\rm out})^c_k
\non\\
&&\quad+(\Pi^{\rm out})^m_k (\Pi^{\rm out})^j_i-\frac{1}{3} \delta^m_i (\Pi^{\rm out})^a_k (\Pi^{\rm out})^j_a]
\non\\
&&=ir'_e T^0\,\overline B_m H^{ik}_j (\Pi^{\rm out})^j_i (\Pi^{\rm out})^m_k
-\frac{1}{3}ir'_e T^0\,\overline B_i H^{ik}_j (\Pi^{\rm out})^j_l (\Pi^{\rm out})^l_k
\non\\
&&=\delta C(T^0)\,\overline B_m H^{ik}_j (\Pi^{\rm out})^j_i (\Pi^{\rm out})^m_k
+\delta A(T^0)\,\overline B_i H^{ik}_j (\Pi^{\rm out})^j_l (\Pi^{\rm out})^l_k,
\en
which leads to
\be
\delta C(T^0)=ir'_e T^0,
\quad
\delta A(T^0)=-\frac{1}{3}ir'_e T^0.
\en

\subsection{Pairing $T^0\, \overline B_m H^{ik}_j (\Pi^{\rm out})^j_k (\Pi^{\rm out})^m_i$ and $i r'_a Tr(\Pi^{\rm in}\Pi^{\rm in}\Pi^{\rm out}\Pi^{\rm out}),$}
Pairing the $T^0$ term in $H^0_{\rm eff}$ and the $ir'_a$ term from $\T^{1/2}$ gives:
\be
&&
\hspace{-1cm}
 ir'_a  T^0\, \overline B_m H^{ik}_j (\Pi^{\rm out})^j_k (\Pi^{\rm out})^m_i 
[(\Pi^{\rm in})^a_b(\Pi^{\rm in})^b_c(\Pi^{\rm out})^c_d(\Pi^{\rm out})^d_a]
\non\\
&&=ir'_a T^0\, \overline B_m H^{ik}_j 
[\la (\Pi^{\rm out})^j_k (\Pi^{\rm in})^a_b\ra
\la (\Pi^{\rm out})^m_i (\Pi^{\rm in})^b_c\ra
\non\\
&&\quad
+
\la (\Pi^{\rm out})^j_k (\Pi^{\rm in})^b_c\ra
\la (\Pi^{\rm out})^m_i (\Pi^{\rm in})^a_b\ra]
(\Pi^{\rm out})^c_d(\Pi^{\rm out})^d_a
\non\\
&&=ir'_a  T^0\,\overline B_m H^k (\Pi^{\rm out})^m_i (\Pi^{\rm out})^i_k
      +ir'_a  T^0\,\overline B_k H^{ik}_j (\Pi^{\rm out})^j_l (\Pi^{\rm out})^l_i
      \non\\
&&\quad
      -\frac{2}{3}ir'_a T^0\,\overline B_i H^{ik}_j (\Pi^{\rm out})^j_l (\Pi^{\rm out})^l_k
\non\\
&&=\delta P(T^0)\,\overline B_m H^k (\Pi^{\rm out})^m_i (\Pi^{\rm out})^i_k
      +\delta E(T^0)\,\overline B_k H^{ik}_j (\Pi^{\rm out})^j_l (\Pi^{\rm out})^l_i
      \non\\
&&\quad
      +\delta A(T^0)\,\overline B_i H^{ik}_j (\Pi^{\rm out})^j_l (\Pi^{\rm out})^l_k,
\en
which leads to
\be
\delta P(T^0)=ir'_a  T^0,
\quad
\delta E(T^0)=ir'_a  T^0
\quad
\delta A(T^0)=-\frac{2}{3}ir'_a T^0.
\en
\subsection{Pairing $T^0\, \overline B_m H^{ik}_j (\Pi^{\rm out})^j_k (\Pi^{\rm out})^m_i$ and $ir'_t Tr(\Pi^{\rm in}\Pi^{\rm in})Tr(\Pi^{\rm out}\Pi^{\rm out})/4,
$}
Pairing the $T^0$ term in $H^0_{\rm eff}$ and the $ir'_t$ term from $\T^{1/2}$ gives:
\be
&&
\hspace{-1cm}
 ir'_t\frac{1}{4} T^0\, \overline B_m H^{ik}_j (\Pi^{\rm out})^j_k (\Pi^{\rm out})^m_i 
[(\Pi^{\rm in})^a_b(\Pi^{\rm in})^b_a (\Pi^{\rm out})^c_d(\Pi^{\rm out})^d_c]
\non\\
&&=ir'_t \frac{1}{4} T^0\, \overline B_m H^{ik}_j 
[\la (\Pi^{\rm out})^j_k (\Pi^{\rm in})^a_b\ra
\la (\Pi^{\rm out})^m_i (\Pi^{\rm in})^b_a\ra
\non\\
&&
\quad+
\la (\Pi^{\rm out})^j_k (\Pi^{\rm in})^b_a\ra
\la (\Pi^{\rm out})^m_i (\Pi^{\rm in})^a_b\ra]
(\Pi^{\rm out})^c_d(\Pi^{\rm out})^d_c
\non\\
&&= \frac{1}{2}i r'_t T^0\,\overline B_k H^k (\Pi^{\rm out})^l_m(\Pi^{\rm out})^m_l
\non\\
&&= \frac{1}{2}\delta PA(T^0)\,\overline B_k H^k (\Pi^{\rm out})^l_m(\Pi^{\rm out})^m_l,
\en
which leads to
\be
\delta PA(T^0)=i r'_t T^0.
\en

\subsection{Pairing $T^0\, \overline B_m H^{ik}_j (\Pi^{\rm out})^j_k (\Pi^{\rm out})^m_i$ and 
$i(\bar r'_e+2\bar r'_a)$
$Tr(\Pi^{\rm in}\Pi^{\rm out}\Pi^{\rm in})\eta^{\rm out}_1/\sqrt3$}
Pairing the $T^0$ term in $H^0_{\rm eff}$ and the $i(\bar r'_e+2\bar r'_a)$ term from $\T^{1/2}$ gives:
\be
&&
\hspace{-1cm}
 i(\bar r'_e+2\bar r'_a) T^0\, \overline B_m H^{ik}_j (\Pi^{\rm out})^j_k (\Pi^{\rm out})^m_i 
[(\Pi^{\rm in})^a_b(\Pi^{\rm in})^b_c(\Pi^{\rm out})^c_a\eta^{\rm out}_1]
\non\\
&&=i(\bar r'_e+2\bar r'_a) T^0\, \overline B_m H^{ik}_j 
[\la (\Pi^{\rm out})^j_k (\Pi^{\rm in})^a_b\ra
\la (\Pi^{\rm out})^m_i (\Pi^{\rm in})^b_c\ra
\non\\
&&\quad
+
\la (\Pi^{\rm out})^j_k (\Pi^{\rm in})^b_c\ra
\la (\Pi^{\rm out})^m_i (\Pi^{\rm in})^a_b\ra]
(\Pi^{\rm out})^c_a\eta^{\rm out}_1/\sqrt3
\non\\
&&=i(\bar r'_e+2\bar r'_a) T\, \overline B_m H^{ik}_j 
[\delta^j_i(\Pi^{\rm out})^m_k\eta^{\rm out}_1/\sqrt3+
\delta^m_k(\Pi^{\rm out})^j_i\eta^{\rm out}_1/\sqrt3
\non\\
&&\quad
-\frac{1}{3}\delta^m_i  (\Pi^{\rm out})^j_k\eta^{\rm out}_1/\sqrt3
-\frac{1}{3}\delta^m_i (\Pi^{\rm out})^j_k\eta^{\rm out}_1/\sqrt3)]
\non\\
&&=i(\bar r'_e+2\bar r'_a) T^0\,\overline B_m H^k (\Pi^{\rm out})^m_k\eta^{\rm out}_1/\sqrt3
      +i(\bar r'_e+2\bar r'_a) T^0\,\overline B_k H^{ik}_j (\Pi^{\rm out})^j_i\eta^{\rm out}_1/\sqrt3
      \non\\
&&\quad
      -\frac{2}{3}i(\bar r'_e+2\bar r'_a)T^0\,\overline B_i H^{ik}_j (\Pi^{\rm out})^j_k\eta^{\rm out}_1/\sqrt3
\non\\
&&=\delta (\bar C_2+\bar P_1+\bar P_2-\frac{1}{3}\bar P^C_{EW,2})(T^0)\,\overline B_m H^k (\Pi^{\rm out})^m_k\eta^{\rm out}_1/\sqrt3
\non\\
&&\quad
      +\delta (\bar C_1+\bar E_1+\bar E_2)(T^0)\,\overline B_k H^{ik}_j (\Pi^{\rm out})^j_i\eta^{\rm out}_1/\sqrt3
      \non\\
&&\quad
      +\delta (\bar T+\bar A_1+\bar A_2)(T^0)\,\overline B_i H^{ik}_j (\Pi^{\rm out})^j_k\eta^{\rm out}_1/\sqrt3,
\en
which is similar to the pairing of $T^0$ $ir'_a$ and leads to
\be
\delta (\bar C_2+\bar P_1+\bar P_2-\frac{1}{3}\bar P^C_{EW,2})(T^0)&=&i(\bar r'_e+2\bar r'_a) T^0
\non\\
\delta (\bar C_1+\bar E_1+\bar E_2)(T^0)&=&i(\bar r'_e+2\bar r'_a) T^0
\non\\
\delta (\bar T+\bar A_1+\bar A_2)(T^0)&=&-\frac{2}{3}i(\bar r'_e+2\bar r'_a) T^0.
\en


\subsection{Pairing $T^0\, \overline B_m H^{ik}_j (\Pi^{\rm out})^j_k (\Pi^{\rm out})^m_i$ and 
$i(\tilde r'_0 +\frac{4\tilde r'_a +2\tilde r'_e}{3})$
$Tr(\Pi^{\rm in}\Pi^{\rm out})\eta_1^{\rm in}\eta_1^{\rm out}$}
Pairing the $T^0$ term in $H^0_{\rm eff}$ and the $i(\tilde r'_0 +\frac{4\tilde r'_a +2\tilde r'_e}{3})$ term from $\T^{1/2}$ gives vanishing result.

\subsection{Pairing $T^0\, \overline B_m H^{ik}_j (\Pi^{\rm out})^j_k (\Pi^{\rm out})^m_i$ and 
$i(\hat r'_t +\frac{4\hat r'_a +2\hat r'_e}{3})
\eta^{\rm out}_1\eta^{\rm out}_1Tr(\Pi^{\rm in}\Pi^{\rm in})/4$}
Pairing the $T^0$ term in $H^0_{\rm eff}$ and the $i(\tilde r'_0 +\frac{4\tilde r'_a +2\tilde r'_e}{3})$ term from $\T^{1/2}$ gives:
\be
&&
\hspace{-1cm}
 i\left(\hat r'_t +\frac{4\hat r'_a +2\hat r'_e}{3}\right) \frac{1}{4} T^0\, \overline B_m H^{ik}_j (\Pi^{\rm out})^j_k (\Pi^{\rm out})^m_i 
[(\Pi^{\rm in})^a_b(\Pi^{\rm in})^b_a \eta_1^{\rm out}\eta_1^{\rm out}]
\non\\
&&=i\left(\hat r'_t +\frac{4\hat r'_a +2\hat r'_e}{3}\right) \frac{1}{4} T^0\, \overline B_m H^{ik}_j 
[\la (\Pi^{\rm out})^j_k (\Pi^{\rm in})^a_b\ra
\la (\Pi^{\rm out})^m_i (\Pi^{\rm in})^b_a\ra
\non\\
&&
\quad+
\la (\Pi^{\rm out})^j_k (\Pi^{\rm in})^b_a\ra
\la (\Pi^{\rm out})^m_i (\Pi^{\rm in})^a_b\ra]
\eta^{\rm out}_1\eta^{\rm out}_1
\non\\
&&= \frac{1}{2}i \left(\hat r'_t +\frac{4\hat r'_a +2\hat r'_e}{3}\right) T^0\,\overline B_k H^k \eta_1^{\rm out}\eta^{\rm out}_1
\non\\
&&=\delta(\tilde C+\tilde E+\tilde P +\frac{3}{2}\tilde {PA}-\frac{1}{3}\tilde P^C_{EW} -\frac{1}{3}\tilde P^E_{EW})(T^0)
\,\overline B_k H^k \eta_1^{\rm out}\eta_1^{\rm out}/3,
\en
which is similar to the is similar to the pairing of $T$ $ir'_t$ and leads to
\be
\delta(\tilde C+\tilde E +\tilde P+\frac{3}{2}\tilde {PA}-\frac{1}{3}\tilde P^C_{EW} -\frac{1}{3}\tilde P^E_{EW})(T^0)= \frac{3}{2}i \left(\hat r'_t +\frac{4\hat r'_a +2\hat r'_e}{3}\right) T^0.
\en

\subsection{Pairing $T^0\, \overline B_m H^{ik}_j (\Pi^{\rm out})^j_k (\Pi^{\rm out})^m_i$ and
$i\left(\check r^{(m)}_0+\frac{4\hat r^{(m)}_a+2\hat r^{(m)}_e+3\hat r^{(m)}_t}{6}\right)\frac{1}{2}\eta^{\rm in}_1\eta^{\rm out}_1\eta^{\rm in}_1\eta^{\rm out}_1$
}
Pairing the $T^0$ term in $H^0_{\rm eff}$ and the $i\left(\check r^{(m)}_0+\frac{4\hat r^{(m)}_a+2\hat r^{(m)}_e+3\hat r^{(m)}_t}{6}\right)$ term from $\T^{1/2}$ gives vanishing result.

The results of rescattering effects from $T^0$ are collected in Eqs.~(\ref{eq: delta TA1}), (\ref{eq: delta TA2}) and (\ref{eq: delta TA1}).
Rescattering effects from other TA are obtained and collected similarly.

\end{document}